
------------------------------------------------------------------------

The macros to actually print this file are not provided. Postcript files
for the figures are not provided either. If you want a hardcopy,
simply contact the author at the address given above.

\input jnl.tex
\input jnlfoot.tex
\input reforder.tex
\input eqnorder

\singlespace


\def\phys{Phys. Rev. B}
\def\lett{Phys. Rev. Lett.}

\def\dx2y2{\rm d_{x^2-y^2}}

\def\lacuo{\rm La_{2} Cu O_4}
\def\x214x{\rm La_{2-x} Sr_x Cu O_4}
\def\s2212{\rm Bi_2 Sr_2 Ca Cu_2 O_8}
\def\nco{\rm Nd_2 Cu O_4}
\def\xnco{\rm Nd_{2-x} Ce_x Cu O_4}
\def\ybco{\rm Y Ba_2 Cu_3 O_{6+x}}
\def\AAA{ A({\bf p}, \omega) }
\def\hbar{{\mathchar'26\mskip-9muh}}

\def\aphi0{\langle \phi_0 |}
\def\aphi1{\langle \phi_1 |}
\def\aphi2{\langle \phi_2 |}
\def\aphi3{\langle \phi_3 |}
\def\hH {\hat H}

\null
\vskip -.75in
\vskip .5in
{\singlespace
\smallskip
\rightline{ August, 1993}
}

\vskip 0.1in

\title CORRELATED ELECTRONS IN
       HIGH TEMPERATURE SUPERCONDUCTORS

\vskip 0.8in

\author  ELBIO ~DAGOTTO

\vskip 0.8cm

\affil Department of Physics,
       National High Magnetic Field Laboratory,
       and MARTECH,
       Florida State University,
       Tallahassee, FL 32306

\vskip 0.1in

\singlespace

\abstract { Theoretical ideas and experimental results concerning
high temperature superconductors are reviewed. Special emphasis is
given to calculations carried out with the help of computers applied to
models of strongly correlated electrons proposed to describe
the two dimensional ${\rm Cu O_2}$ planes. The review also includes
results using several analytical techniques. The one and three band
Hubbard models, and the ${\rm t-J}$ model are discussed, and their
behavior compared against experiments when available.
Among the conclusions of the review, we found
that some experimentally observed unusual properties of the cuprates have a
natural explanation through Hubbard-like models. In particular abnormal
features like the
mid-infrared band of the optical conductivity $\sigma(\omega)$, the
new states observed in the gap in photoemission experiments, the behavior of
the
spin correlations with doping, and the presence of phase separation
in the copper oxide superconductors may be explained, at least in part, by
these models.
Finally, the existence of superconductivity in Hubbard-like models is
analyzed. Some aspects of the recently proposed ideas to describe the
cuprates as having a $\dx2y2$ superconducting condensate at low
temperatures are discussed.
Numerical results favor this scenario over others.
It is concluded that computational techniques provide a useful unbiased
tool to study the difficult regime where electrons are strongly
interacting, and that considerable progress can be achieved by comparing
numerical results against analytical predictions for the properties of
these models.
Future directions of the active field of
computational studies of correlated electrons are briefly discussed.
}

\vskip 0.8truecm

\ \ \ Submitted to: Review of Modern Physics.

\vskip 0.8truecm

\vfill

\endtitlepage



\centerline{\bf CONTENTS}

\vskip 0.40cm

\singlespace

\def\leaderfill{\leaders\hbox to 1em{\hss.\hss}\hfill}
\line{\bf I. Introduction and experimental results\leaderfill 4}
\vskip 0.15cm
\line{\hskip .5truecm A. Structure and phase diagram of the cuprates\leaderfill
6}
\line{\hskip 1.0truecm  1. $\x214x$ \leaderfill 8}
\line{\hskip 1.0truecm  2. $\ybco$  \leaderfill 9}
\line{\hskip 1.0truecm  3. $\xnco$  \leaderfill 10}

\vskip 0.15cm
\line{\hskip .5truecm B. Normal state properties \leaderfill 11}
\vskip 0.15cm
\line{\hskip .5truecm C. Electronic models  \leaderfill 12}
\line{\hskip 1.0truecm 1. Three band model \leaderfill 12}
\line{\hskip 1.0truecm 2. One band models \leaderfill 13}

\vskip 0.24cm
\line{\bf II. Algorithms \leaderfill 15}
\vskip 0.15cm
\line{\hskip .5truecm A. Lanczos technique\leaderfill 16}
\line{\hskip 1.0truecm 1. Method \leaderfill 16}
\line{\hskip 1.0truecm 2. Dynamical properties \leaderfill 19}
\vskip 0.15cm
\line{\hskip .5truecm B. Quantum Monte Carlo technique \leaderfill 21}
\vskip 0.15cm
\line{\hskip .5truecm C. Sign problem\leaderfill 23}
\vskip .24cm
\line{\bf III. Correlated electrons at low hole doping\leaderfill 25}
\vskip 0.15cm
\line{\hskip .5truecm A. Results at half-filling \leaderfill 25}
\vskip 0.15cm
\line{\hskip .5truecm B. Properties of holes in antiferromagnets\leaderfill 28}
\line{\hskip 1.0truecm 1. String picture \leaderfill 29}
\line{\hskip 1.0truecm 2. Energy and momentum of a hole  \leaderfill 31}
\line{\hskip 1.0truecm 3. Dispersion relation of a hole \leaderfill 33}
\line{\hskip 1.0truecm 4. Dynamical properties of one hole\leaderfill 34}
\line{\hskip 1.0truecm 5. Binding of holes\leaderfill 36}
\line{\hskip 1.0truecm 6. Quasiparticles in models of correlated
electrons \leaderfill 38}
\vskip 0.24cm
\line{\bf IV. Comparing experiments with computer simulation results
\leaderfill 41}
\vskip 0.15cm
\line{\hskip 0.5truecm A. Magnetic properties in the presence of
carriers\leaderfill 41}
\line{\hskip 1.0truecm 1. Magnetic susceptibility \leaderfill 41}
\line{\hskip 1.0truecm 2. Antiferromagnetism at finite doping \leaderfill 42}
\line{\hskip 1.0truecm 3. Incommensurate spin order in doped
materials\leaderfill 44}

\vskip 0.15cm
\line{\hskip 0.5truecm B. Optical conductivity \leaderfill 45}

\line{\hskip 1.0truecm 1. Experimental results \leaderfill 46}
\line{\hskip 1.0truecm 2. Theoretical analysis of $\sigma(\omega)$ \leaderfill
48}
\line{\hskip 1.0truecm 3. Numerical results\leaderfill 51}

\vskip 0.15cm
\line{\hskip 0.5truecm C. Electron spectroscopy\leaderfill 55}
\line{\hskip 1.0truecm 1. Density of states (experiments)\leaderfill 56}
\line{\hskip 1.0truecm 2. Density of states (theory)\leaderfill 58}
\line{\hskip 1.0truecm  3. Angle-resolved photoemission\leaderfill 62}
\line{\hskip 1.0truecm  4. Fermi surface in models of correlated
electrons\leaderfill 64}

\vskip 0.15cm
\line{\hskip 0.5truecm D. Phase separation\leaderfill 65}
\line{\hskip 1.0truecm 1. Experimental results \leaderfill 65}
\line{\hskip 1.0truecm 2. Theoretical results\leaderfill 65}

\vskip 0.15cm
\line{\hskip 0.5truecm E. Superconductivity in models of strongly
correlated electrons\leaderfill 68}
\line{\hskip 1.0truecm 1. Superconductivity in the one and three band
Hubbard models\leaderfill 68}
\line{\hskip 1.0truecm 2. Superconductivity in the ${\rm t-J}$ model
\leaderfill 69}
\line{\hskip 1.0truecm 3. Phase diagram of the two dimensional ${\rm t-J}$
model
\leaderfill 72}
\line{\hskip 1.0truecm 4. Meissner effect and flux quantization\leaderfill 73}
\line{\hskip 1.0truecm 5. $\dx2y2$ superconductivity\leaderfill 75}

\vskip 0.24cm
\line{\bf V. Conclusions \leaderfill 76}
\vskip 0.24cm
\line{\bf VI. Acknowledgments\leaderfill 78}
\vskip 0.24cm
\line{\bf Footnotes \leaderfill 79}
\vskip 0.24cm
\line{\bf References \leaderfill 81}
\vskip 0.24cm
\line{\bf Figure Captions \leaderfill 95}

\vfill
\eject

\singlespace

\centerline{\bf I. Introduction and Experimental Results}

\vskip 1cm

\taghead{1.}

More than six years ago, J. G. Bednorz and K. A. M\"uller
(Bednorz and M\"uller, 1986)
announced the discovery of superconductivity in a pottery-like
copper-oxide material at a
temperature of about 30K. These compounds
are poor conductors, and thus their
result was unexpected and cautiously considered. However,
the confirmation of these experiments by Takagi et al. (1987)
generated a frenetic race for the preparation of materials with even
higher critical temperatures. Dozens of
``high-Tc'' compounds have been discovered in the last few years,
and currently a thallium-based
material has the highest confirmed critical temperature of about ${\rm
125K}$ (although very recent reports suggest an even higher Tc
of $133K$ for some copper-oxides
with ${\rm H_g}$ (Schilling et al., 1993)).
Since ${\rm 77K}$ is the boiling temperature of nitrogen, it is quite possible
that new technologies may emerge from the field of high-Tc, although
flashy early promises of levitating trains seem unlikely. The
critical current densities are still not high enough for most technological
applications, but
progress is very rapid on this front.

Although the initial frenzy has calmed down, the field still remains very
active and
large, and it is rapidly evolving.
On the experimental side results are being consolidated,
mainly due to a considerable improvement in the quality of the
samples compared with those used in the early studies of the high-Tc
materials. In particular, high quality
single crystals are currently available. Considerable work is being
carried out at temperatures above Tc, since
it is believed that the unusual normal state properties of these
materials may contain the key features to understand their superconductivity.
Much work has been devoted to some unexpected properties of the cuprates like
the linear
behavior of the d.c. resistivity with temperature,
a Hall coefficient which is temperature dependent, the presence of
short coherence lengths, and the energy dependence
of the relaxation rate $1/\tau$. The strong anisotropy of
the materials complicates the interpretation of the results.
These unusual
properties may suggest that the normal state cannot be described by
a Fermi liquid,
an issue which is currently under much discussion.
Actually, we know that organic superconductors, heavy
fermion superconductors, ${\rm Nb_3 Sn}$, and other materials are not
well described by the BCS model (Bardeen, Cooper, and Schrieffer, 1957).
However, the BCS or pairing $theory$ is a
broader concept and
describes fermions in which an effective attractive interaction produces
a condensate, with large overlaps between pairs. The source of the attraction
is not crucial for this theory to hold, and thus BCS ideas are not at all
excluded as a possible explanation of the behavior of the new superconductors.

The search for new materials has been mainly empirical since no
predictive theory is
currently known for the high-Tc compounds. The fact is that
no one knows why the cuprates behave as they do.
However, several theories have been $proposed$ to describe them.
One and three band Hubbard models, as well as the ${\rm t-J}$
model are believed to represent the gross features of
the electronic behavior of the new materials.
Unfortunately, most of the available experimental data are
not accurate enough to convincingly confirm or rule out most of the theories.
Antiferromagnetism, defects, phonons, dimensionality effects, and other
features
seem to conspire to hide the key properties of the normal state.
It is possible that theories that combine the pairing ideas with
the presence of strong antiferromagnetic correlations may properly describe the
high-Tc superconductors. In this family, consider for example
the spin-bag theories (Schrieffer, Wen and Zhang, 1988; Kampf and
Schrieffer, 1990),
antiferromagnetic Fermi liquid theories (Millis, Monien, and Pines,
1990), and
also the recently developed $\dx2y2$ theories that have attracted
considerable
attention (Bickers, Scalapino, and White, 1989; Monthoux, Balatsky, and Pines,
1991).
The
interchange of magnons may produce the attractive force needed to
pair the charge carriers (Scalapino, Loh, and Hirsch, 1986; Bickers,
Scalapino, and Scalettar, 1987). Note that
spin-bags are very different from Landau quasiparticles, since the
overlap Z between bare and dressed holes is very small (${\rm Z \sim
0.1}$ according to some calculations (Dagotto and Schrieffer, 1991)). This
effect produces considerable deviations from the $standard$ Fermi liquid
results, but these excitations still correspond to those of a Fermi liquid.
Among the Fermi liquid based theories we should include also the
so-called hole mechanisms (Hirsch and Marsiglio, 1989), and the
nested Fermi liquid (Virosztek and Ruvalds, 1990).

Some other theorists strongly believe that the BCS theory cannot work in
these new superconductors (Anderson, 1990b).
Instead, scenarios where the elementary excitations
in the new materials are spinons (zero charge, spin-1/2), and
holons (charge $e$, spin 0) have been proposed. As a toy model, the
one dimensional Hubbard model has been widely discussed (Anderson, 1990a).
However, it remains to be shown that one and two dimensional models have a
similar qualitative behavior.
Other non-Fermi liquid
theories that have attracted considerable attention are anyon
superconductivity (Laughlin, 1988; Chen et al., 1989; Wen, Wilczek, and
Zee, 1989), the
Marginal Fermi Liquid theories (Varma et al., 1989), gauge theories
(Nagaosa and Lee, 1990), and several others.
Clearly, we are still a long way from developing a $predictive$
theory to describe the new materials. We all expect that a
triumphant theory should indicate the
direction for developing new superconductors with even higher critical
temperatures.


It is interesting to remember that before 1986,
some theorists believed that condensed matter physics was a
mature field. Actually, few researchers were trying to find
high temperature superconductors by that time.
The discovery of the new superconductors has shown the weaknesses
of our area of research. After several years of intense theoretical studies, it
is clear that we still do not have the proper skills and
tools to deal with strongly correlated electrons.
The main problem is that real calculations are
not easy.
Most of the models proposed for the new superconductors
contain
interactions that are strong, and thus perturbative calculations of
bubble and ladder diagrams are questionable (nevertheless note that
there are other analytical methods that seem to be able to handle strong
correlations i.e. they can reproduce numerical results accurately. See
for
example, Fulde (1991) and Unger and Fulde (1993)).
Other approximations are self-consistent (mainly mean-field like) but
it is difficult to judge how close to the actual
properties of the model these results are.
It is interesting to observe that theories that start with the same
Hamiltonian, may arrive at completely different descriptions of their
properties.
The subject is so complicated that we need as much help as we can get.
The discovery of the new superconductors has clearly shown the need
for new, well-controlled approaches to the field of correlated electrons.

In an attempt to close the gap between a model, as defined by its
Hamiltonian, and, e.g., the actual properties of its ground
state, a large number of theorists have turned to the use of computers for some
help.
High-Tc physics has given theorists a strong motivation to work on
correlated electrons,
but it may be claimed that from a theorist's point of view
the particular model or material are not as
crucial as the development of the tools to handle them.
Computational results can contribute to the acceptance or rejection of
mean-field based theories, and can also indicate directions in which new
approaches should be developed. Much progress has been made in this
direction, with numerical calculations routinely being viewed as impartial
referees
that may eventually select the proper analytic description of a given
model. Actually, the cross-fertilization between computational
and analytical work in the area of correlated electrons is quickly growing.

As an additional motivation for carrying out computational work in the
context of high-Tc superconductors, note that the new cuprate compounds
differ markedly from the more conventional superconductors by having a
very small coherence length $\xi$. This length is usually associated
with the average size of a Cooper pair. For conventional superconductors
$\xi \sim 500 \AA$ to $10,000 \AA$, and thus the size of the pair is
larger than the average distance between pairs. This particular feature
allows a mean-field BCS-like treatment of the problem to be accurate and
reliable. However, the new superconductors have $\xi \sim 12 \AA $ to
$15 \AA$, and thus standard mean-field approximations are clearly
questionable. These results were obtained mainly via $H_{c2}$ (all
high-Tc materials are type II superconductors).
The cuprates are in the ``clean'' limit since the electron mean free
path ( $\sim 150 \AA$) is much smaller than $\xi$ (for more details see
Welp et al. (1989); Batlogg (1990), page 66; Batlogg (1991), page 48;
Burns (1992)).
The coherence length in the ${\bf c}$ direction is only $2$ to $5 \AA$
i.e. even smaller than the interplane distance, while $\xi$ in-plane
is only $\sim 3$ to $4$ lattice spacings
(since the distance between Cu ions in the plane is $3.8 \AA$).
Such small Cooper pair sizes indicates that studies on finite two
dimensional clusters,
as those reported in this review, may be of relevance to describe the
physics of the cuprates.


The field of high-Tc superconductors has generated several thousand
publications. It would be quite difficult to describe this field even if
only restricted to the theoretical aspects. In this review, we will
concentrate on the more modest goal of describing the progress achieved
in the study of models of correlated electrons using
computational techniques. However, several analytical techniques will
also be described, and their results discussed.
The review should be considered as a ``progress report'' in which we
have attempted to focus on some basic aspects of this rapidly
evolving field.
Another goal of the review is to provide to the reader a
simple overview of the experimental situation in high-Tc materials,
summarizing the basic agreements and discrepancies between theory and
experiment.
For additional literature, the reader should consult other review
articles or books like those by
Fukuyama, Maekawa, and Malozemoff (1989); Burns (1992);
Maekawa and Sato (1991); and others. For previous short reviews on
theoretical work
related to the subject of this paper see Dagotto (1991);
Fulde and Horsch (1993); and references therein.
It is also worth keeping in mind that all the vast literature related
with this subject cannot be mentioned in a single review article. The
author used his own judgment and prejudices to select the subjects that
he considered more relevant, and apologizes beforehand for any omission.

This review is organized as follows. In section I, a brief review of
the properties of some high-Tc superconductors is given. Models
to describe the behavior of electrons in these materials are presented.
In section II, the most important algorithms to study correlated
electrons are described. Results for the particular cases of
half-filling and a few holes doped into an antiferromagnet are
discussed
in section III.  A rough comparison of the numerical results with
experiments are discussed in section IV. Several observables like
the optical conductivity $\sigma(\omega)$, photoemission spectra $N(\omega)$,
magnetic
susceptibilities,
and others are considered and contrasted with theoretical predictions
coming mainly from numerical studies. It is concluded that some ``anomalous''
properties of the cuprates are not so unexpected once the electronic
models of correlated electrons are properly analyzed with powerful
unbiased tools like computational techniques. Regarding the final goal
of finding superconductivity in these models, we believe that $\dx2y2$
is the most likely channel where such a condensate may exist when
holes are immersed in an antiferromagnetic background.
Conclusions are presented in section V.


\vskip 1cm

\centerline{\bf A. Structure and phase diagram of the cuprates}

\vskip .5cm

In this section, we will
provide to the reader a short overview of the lattice structure and
phase diagram of the most widely studied high-Tc compounds
(note that in this review we will follow the convention
that a superconducting compound is said to belong to the
family of high-Tc superconductors $if$ it has ${\rm Cu O_2}$ planes).
In general, the high-Tc materials are tetragonal, and
all of them have one or more
${\rm Cu O_2}$ planes in their structure, which are
separated by layers of other atoms $({\rm Ba,O,La,...})$.
Most researchers in this field strongly
believe that superconductivity is related with processes occurring
in the ${\rm Cu O_2}$ planes,
with the other layers simply  providing the carriers (and thus they
are called ``charge reservoirs''). All high-Tc materials have such
charge reservoirs. In the
${\rm Cu O_2}$ planes, each copper ion is strongly
bounded to four oxygen ions separated by a distance of approximately $1.9 \AA$.
It is remarkable that
this distance is basically a local property of the ${\rm Cu-O}$ link, and it
is only weakly dependent on the actual compound being
considered.
Thus far, the material with the largest Tc reported and
confirmed (see below), has three
${\rm Cu O_2}$ planes at a short distance from each other. The fact that
Tc increases with the number of layers
has led to some theoretical speculations linking the number of
${\rm Cu O_2}$ planes with the critical temperature.

Another property common to these materials is the presence of
antiferromagnetic order at low temperatures in the undoped regime
i.e. when no free carriers exist
in the planes. Upon doping,
the spin order is destroyed, and
the superconducting phase appears.
However, this does not mean that spin correlations are unimportant
for superconductivity.
Even without strict long-range order, the spin correlation length can be
large in the superconducting phase
producing a local arrangement of magnetic moments that at
short distances
differs very little from that observed below the N\'eel temperature
in the insulating regimes. Actually, several ideas have been discussed in the
literature relating antiferromagnetism and superconductivity.
For example, the interchange of magnons instead of phonons may be
a possible mechanism of pairing in these materials, as will
be discussed in section IV.E.

A large number of compounds with the characteristic ${\rm Cu O_2}$ planes have
been
synthesized. This is not too surprising since it is possible to modify the
number of planes per unit cell, the atoms separating the nearby planes, as
well as the structure, composition, and size of the charge reservoir,
producing a huge number of combinations. In the table below, we present
a very short
list of the most widely studied compounds in this field, with their
critical temperatures ${\rm T_c}$
(for a more complete list see Harshman and Millis, 1992;
and Burns, 1992). For comparison we also show the critical
temperature
of some ``old'' superconductors like Nb, Pb, and ${\rm Nb_3 Ge}$. The
latter has the highest critical temperature known before 1986 (see
Testardi, Wernick and Roger, 1974). The Tc of a superconducting
heavy fermion material $({\rm UPt_3})$, and a fullerene are also given.

\vskip 0.4cm
\centerline{
\vbox{\offinterlineskip
\hrule
\halign{&\vrule#&\strut\quad\hfil#\quad\cr
\noalign{\hrule}
height2pt&\omit&&\omit&&\omit&\cr
&${\rm Material}$\hfil&&${\rm T_c}$\hfil&\cr
height2pt&\omit&&\omit&&\omit&\cr
\noalign{\hrule}
height2pt&\omit&&\omit&\cr
&${\rm Tl_2 Ca_2 Ba_2 Cu_3 O_{10} }$&&125&\cr
height2pt&\omit&&\omit&\cr
&${\rm Y Ba_2 Cu_3 O_{7}}$&&92&\cr
height2pt&\omit&&\omit&\cr
&${\rm Bi_2 Sr_2 Ca Cu_2 O_8 }$&&89&\cr
height2pt&\omit&&\omit&\cr
&${\rm La_{1.85} Sr_{0.15} Cu O_4 }$&&39&\cr
height2pt&\omit&&\omit&\cr
&${\rm Nd_{1.85} Ce_{0.15} Cu O_4 }$&&24&\cr
height2pt&\omit&&\omit&\cr
&${\rm Rb Cs_2 C_{60} }$&&33&\cr
height2pt&\omit&&\omit&\cr
&${\rm Nb_3 Ge}$&&23.2&\cr
height2pt&\omit&&\omit&\cr
&${\rm Nb}$&&9.25&\cr
height2pt&\omit&&\omit&\cr
&${\rm Pb}$&&7.20&\cr
height2pt&\omit&&\omit&\cr
&${\rm U Pt_3}$&&0.54&\cr
height2pt&\omit&&\omit&\cr
}
\hrule}
}

\noindent
In the remainder of this section, we will discuss the structure and
phase diagram of some particular high-Tc compounds in more detail.

\vskip 0.7cm

\noindent{\tt 1. ${ La_{2-x} Sr_x Cu O_4}$ }

\vskip 0.25cm






This compound was among the first high temperature superconductors discovered.
It crystalizes in a body centered
tetragonal structure which has been known
for several years from studies on ${\rm K_2 Ni F_4}$.
It is usually
called the ${\bf T}$ structure, and it is shown in Fig.1
(to visualize the tetragonal lattice, simply
concentrate only on the Cu atoms of the figure).
In $\x214x$, the ${\rm Cu O_2}$
planes are $\sim 6.6 \AA$ apart, separated by two ${\rm LaO}$ planes
which form the charge reservoir that
captures electrons from the conducting planes upon doping.
The atomic configurations of the elements forming this compound are
the following:
${\rm Cu: [Ar] (3d)^{10} (4s) }$,
${\rm La: [Xe] (5d) (6s)^2 }$,
${\rm O: [He] (2s)^{2} (2p)^4 }$, and finally
${\rm Sr: [Kr] (5s)^{2}}$.
In the crystal, oxygen is in a ${\rm O^{2-}}$ state that completes the
p-shell. Lanthanum loses three electrons and becomes
${\rm La^{3+}}$ which is in a stable closed-shell
configuration. To conserve charge neutrality, the copper atoms
must be in a ${\rm Cu^{2+}}$ state, which is obtained by losing
the ${\rm (4s)}$ electron (weakly bounded to the atom), and also
one d electron. This creates a $hole$ in the d-shell, and thus
${\rm Cu^{2+}}$ has a net spin of 1/2 in the crystal.
Each copper atom in the conducting planes has an oxygen
(belonging to the charge reservoir) above and below in the
{\bf c}-direction.
These are the so-called apical ${\rm O}$ atoms or just ${\rm O_z}$. Then,
in this compound the copper ions are surrounded by octahedra of
oxygens (shown in Fig.1). However, the
distance ${\rm Cu - O_z}$ is $\sim 2.4 \AA$ which is considerable larger than
the
distance ${\rm Cu-O}$ in the planes ( $\sim 1.9 \AA$).
Then, the ${\rm Cu-O_z}$ bond is much weaker
than the in-plane ${\rm Cu-O}$ bond, and thus considering the Cu
atoms as
immersed in perfect octahedra of oxygens is somewhat misleading.
The dominant bonds are those on the plane. We will see below that
many of the high-Tc materials have apical oxygens. Their distance to
coppers in the conducting planes
is remarkable similar in different compounds, being always about $2.4 \AA$.





Upon doping, ${\rm La^{3+}}$ are randomly replaced by
${\rm Sr^{2+}}$, and thus
electrons are taken from the ${\rm Cu O_2}$ planes. It will be shown
later that these electrons come from oxygen ions changing their configuration
from ${\rm O^{2-}}$
to ${\rm O^{-}}$ (and thus creating one $hole$ in their p-shell).
Metallic behavior is already observed for very small
doping concentrations, ${\rm x \ge 0.04}$. The sign of the Hall coefficient
shows
that indeed the carriers are holes as expected.
The actual  phase diagram of this material is shown in Fig.2, according to
results obtained by Keimer et al. (1992), and Birgeneau (1990).
Near half-filling, antiferromagnetic order is clearly observed which
theoretical studies have shown it is
well described by a simple Heisenberg antiferromagnetic Hamiltonian
representing the interactions between
the spin-1/2 holes located on the copper atoms (Chakravarty, 1990).
Experimentally, Aeppli et al. (1989), and Hayden et al. (1991b) showed
that spin-wave theory with only first neighbor interactions accounts for the
spin dynamics of ${\rm La_2 Cu O_4}$.
A small residual interaction between planes leads to a finite N\'eel
critical temperature of about $300 K$. The ``spin-glass'' phase of Fig.2
was considered by Harshman et al. (1988).
For Sr-dopings between ${\rm x \sim 0.05}$ and $\sim 0.30$,
a superconducting phase is found at low temperatures. The maximum
value of ${\rm T_c}$ is observed at the ``optimal'' doping ${\rm x \sim
0.15}$.
A structural phase transition was also found in this compound as
shown
in Fig.2. At high temperature the structure is tetragonal,
but at lower temperatures the copper atoms and the six
oxygens surrounding them, slightly deviate from their
positions forming an orthorhombic structure.
This small distortion is usually neglected in most theoretical studies
of this compound (for a description of the rest of the phase diagram
shown in
Fig.2, the reader should consult the original references by Keimer et
al., 1992; and Birgeneau, 1990. This phase diagram was also discussed
by Torrance et al., 1989).

\eject
\vskip 1cm

\noindent{\tt 2. ${Y Ba_2 Cu_3 O_{6+x}}$ }

\vskip 0.25cm

Superconductivity in
this material was discovered in early 1987 (Wu et al., 1987),
soon after the lanthanum compound was
reported by Bednorz and M\"uller.
The structure of this compound (sometimes called ${\rm YBCO}$)
is shown in Fig.3, and it is clearly
more complicated than the structure of $\x214x$. In ${\rm YBCO}$, there
are $two$ ${\rm Cu O_2}$ planes per unit cell approximately
$\sim 3.2 \AA$ apart, separated by yttrium ions.
The figure shows that these pairs of ${\rm Cu O_2}$
planes are themselves separated
by layers of atoms containing barium, oxygen and copper, which form the
``charge reservoir.''
As for the lanthanum compound,
the number of carriers in the conduction planes is controlled by the
amount of charge transferred between the conduction layer and the
charge reservoir. The distance between the pairs of conducting planes
is $\sim 8.2 \AA$.

The atomic configurations of Y and Ba are
${\rm [Kr] (5s)^2 (4d) }$, and
${\rm [Xe] (6s)^2 }$, respectively, while those of Cu and O have been
described before for the lanthanum compound.
In the crystal, yttrium is in the state ${\rm Y^{3+}}$, while
barium is in ${\rm Ba^{2+}}$. Copper on the planes is ${\rm Cu^{2+}}$, and
oxygen is ${\rm O^{2-}}$.
Note that in this compound
there are Cu atoms in the charge reservoir, contrary to what
occurs in $\x214x$. In combination
with oxygen atoms, they form
one dimensional structures along the {\bf b} direction
(shown in Fig.3)
which are called the ``Cu-O chains.'' The copper on the chain is in a state
${\rm Cu^{3+}}$.
Since not all cuprate
superconductors have chains, it is believed that this structure
does not play a
key role in the superconducting mechanism. However, their presence
affects other measurable properties of the material like the optical
conductivity, as will be
discussed in section IV.B. The distance ${\rm Cu-O}$
in the chains is $\sim 1.9 \AA$, as it is in the planes.
The chains are well-defined for
${\rm Y Ba_2 Cu_3 O_7}$, but at other oxygen concentrations they have
defects. Actually,
at the minimum oxygen concentration (x=0), no ${\rm Cu-O}$ chains
exist.





How can we control the amount of doping in the planes?
This is achieved by modifying the chemistry of the charge reservoir.
In this material,
the number of carriers depends on the oxygen content in the formula
${\rm Y Ba_2 Cu_3 O_{6+x}}$. In the case x=1, the oxygen atoms are
structurally ordered and form the ${\rm Cu-O}$ chains shown
in Fig.3, and when x is reduced, oxygen atoms are randomly
taken out from the chains. Adding oxygen to the compound is
believed to be equivalent to adding holes to the planes since the new oxygen
becomes ${\rm O^{2-}}$ trapping two electrons which presumably are
originated by creating two holes in the oxygens of the ${\rm Cu O_2}$ planes.
However, note that this scenario is still not fully supported by measurements
of the
sign of the Hall coefficient since this quantity is appreciably temperature
dependent
for this
compound. The possibility of having holes on the chains, rather than
only in the planes complicates this issue even further.
In this review we will not discuss these subtle details further, and we will
assume that increasing the oxygen content is equivalent to adding holes
to the conducting planes (for more details see Burns, 1992; and
references therein).

In Fig.4, the phase diagram of $\ybco$ is shown.
The range of defect concentrations (oxygen excess) in this compound is
large, allowing the properties to change from insulating to superconducting.
For x close to
0, an antiferromagnetic phase is observed with a N\'eel temperature
over 500K. This spin order is caused by the spin 1/2 holes in the
d-shell of the in-plane coppers, as in other compounds (note that
the copper of the chains are not magnetic).
At ${\rm x_c \sim 0.3}$ ($\delta_c \sim 0.7$ in the notation of the
figure),
antiferromagnetic long-range order disappears and
the superconducting phase starts developing. The ``optimal'' composition
(i.e. the one that gives the largest Tc) is close to ${\rm x \sim 1}
(\delta \sim 0.0)$.
Unfortunately, it is not possible to increase further Tc by
adding oxygen beyond x=1 since they have already completed the
${\rm Cu-O}$ chain structure at this composition.
Note also that a structural phase transition occurs in this material
near ${\rm x_c \sim 0.3}$ from a tetragonal to an orthorombic phase,
similar to that found for $\x214x$. In this structural
transition, the conduction
planes ${\rm Cu O_2}$ are unaffected.




\vskip 1.cm

\noindent{\tt 3. ${Nd_{2-x} Ce_x Cu O_4}$ }

\vskip 0.25cm

The structure of this compound is body centered
tetragonal
(shown in Fig.5), like
that of $\x214x$. The difference between the
two lies in the position of the oxygen atoms of the
charge reservoir. The structure
corresponding to $\xnco$ is usually called ${\bf T'}$ structure. It is
interesting to note that the ${\bf T}$ structure can only be easily
hole-doped, while the ${\bf T'}$ structure can be easily electron-doped, the
reason for this asymmetry being unknown.
The atomic configurations of the elements forming the compound is
${\rm Nd: [Xe] (4f)^4 (6s)^2 }$, and
${\rm Ce: [Xe] (4f) (5d) (6s)^2 }$, while Cu and O have been
described before.
In the crystal, copper becomes ${\rm Cu^{2+}}$, oxygen is ${\rm O^{2-}}$, and
neodymium is in a state ${\rm Nd^{3+}}$.  After doping, i.e. when a
${\rm Nd}$ ion is replaced by ${\rm Ce^{4+}}$, the
${\rm Cu O_2}$ planes get an excess of electrons. This is confirmed by
the experimentally
observed sign of the Hall coefficient.
It is believed that an added electron occupies a
hole in the d-shell of copper, producing a S=0 closed
shell configuration.





The phase diagram of this material is shown in Fig.6 comparing it with
a hole-doped compound.
The similarities between the two diagrams are remarkable. Both present an
antiferromagnetic phase with a similar N\'eel temperature
(although for electron-doped materials the antiferromagnetic phase
is more stable upon doping, since
${\rm x > 0.12}$ is necessary to destroy the spin long-range order).
Increasing x further, a
superconducting phase appears close to antiferromagnetism in both cases
with an ``optimal'' composition close to ${\rm x=0.15}$. The
electron-doped phase diagram clearly illustrates that superconductivity
is a relatively ``small'' effect compared with antiferromagnetism, and thus in
theoretical studies it is important to isolate the proper degrees of
freedom and energy scales of the pairing mechanism responsible for
superconductivity, from those causing the bulk
magnetic properties. Otherwise, the existence of
this phase can be hidden
in calculations which are dominated by other larger scales related with
the insulating state.




The family of high-Tc superconductors is very large.
A more complete list of superconducting materials and their
critical temperatures, can be found in a recent review article (Harshman and
Millis, 1992). In particular it is worth mentioning
the layered copper oxide superconductors that include bismuth and
thallium in the charge reservoir layers. These are more
complex compounds,
with the general formulas ${\rm Bi_m Sr_2 Ca_{n-1} Cu_{n}
O_{2n+m+2} }$ and ${\rm Tl_m Ba_2 Ca_{n-1} Cu_n O_{2n+m+2} }$ (where
${\rm m}$ and ${\rm n}$ are integers), and are typically identified
by the shorthand notation Bi(or Tl)m2(n-1)n, e.g. Bi2212.
In particular, Tl2223 has one of the highest confirmed critical temperature
${\rm T_c \sim 125K}$ of the high-Tc family.
(very recent results (Schilling
et al., 1993; Putilin et al., 1993) suggest the presence of
superconductivity at $133K$ in some copper oxides that include ${\rm H_g}$).
It contains $three$ ${\rm Cu O_2}$ planes per unit
cell, which are separated by Ca atoms $\sim 3.2 \AA$ apart, which
is very similar to the distance between planes in $\ybco$. The sets of
three planes are $\sim 11.6 \AA$ apart.
Details about the structure of these materials can be found in a review
article by Jorgensen (1991) (see also
Burns, 1992).

\vfill
\eject

\vskip 1.cm

\centerline{\bf B. Normal State Properties}

\vskip 0.5cm

It is widely believed that understanding the normal state properties of
the high-Tc cuprates will also shed light on the
superconducting mechanism. The basis for this expectation resides in the
unusual normal state properties of these materials. For example, strong
anisotropies are observed, mainly caused by the two dimensional nature of the
problem; and magnetic phases exist
close to the superconducting regions.
In addition, there are properties of the cuprates that have
raised the possibility of observing deviations from a Fermi liquid
description of the normal state.
For example, we know that
in a canonical Fermi liquid metal the magnetic susceptibility and the
Hall coefficient are temperature independent, the resistivity grows like
$T^2$ at low temperatures, and the NMR relaxation is proportional to
temperature ($1/T_1 \sim T$). These behaviors have $not$ been observed in
the cuprates, although a Fermi liquid description is still not ruled out
(see for example, Levin et al., 1992).

In this section, we will briefly describe the normal state property
of the cuprates that is  more frequently mentioned as indicative of an
unusual normal state, namely the experimentally observed
linear dependence of the resistivity $\rho$ with temperature
(for a short review see Batlogg, 1990 and 1991).
In conventional low-temperature ($T$) superconductors, it has been
experimentally observed that $\rho \approx a + bT^5$, at low $T$.
This temperature dependence arises from the scattering of
electrons with phonons. At higher temperatures, a linear behavior is
expected, and the interpolation between the two regimes is
given by the Gr\"uneisen-Bloch formula (Burns, 1992).
The residual resistivity $a$ at $T=0$ is caused by scattering
with magnetic impurities, point or line defects, other electrons, etc.
Several metals like Cu, Al, Ni, Na and others obey this behavior very
accurately (Burns, 1992, Chapter 4).
However, the behavior of $\rho$ observed in the high-Tc superconductors is
different. Single crystal measurements of $\rho$ in the ${\rm Cu O_2}$
planes carried out for several different
compounds have shown an approximate linear behavior over the
measured temperature range. In Fig.7, results for the in-plane
resistivity $\rho_{ab}$ are shown for ${\rm La_{1.85} Sr_{0.15} Cu O_4}$,
${\rm Y Ba_2 Cu_3 O_{7}}$, Bi2201 and Bi2212 compounds (the average between the
$a$ and $b$ crystal directions, i.e. $\rho_{ab}$, is measured
for multidomain or $twinned$ crystals although the preparation of untwinned
crystals is possible).
The linear behavior is
clear, and the slopes of the curves are very
similar for the various compounds suggesting a common scattering
mechanism for carrier transport in the ${\rm Cu O_2}$ planes.
The results for Bi2201 are particularly
interesting since the linear behavior is
observed in this material even at low temperatures close to
${\rm T_c = 7K}$, in a wide range of temperatures from 7K to
700K.





It is important to note that in those materials where the composition
can be changed easily by chemical doping, the linear behavior $\rho \sim T$ is
observed $only$ in a narrow carrier concentration window near
the ``optimal'' compositions i.e. those corresponding to
the highest critical temperatures
(Takagi et al., 1992; and Batlogg et al., 1992).
This detail is not sufficiently remarked in
the literature. To illustrate this result, $\rho_{ab}$ measured in single
crystals of
${\rm La_{2-x} Sr_x Cu O_4}$ for several compositions are shown in
Fig.8. Contrary to the linear dependence at ${\rm x=0.15}$, under and
overdoped samples follow a different power-law behavior
over a wide range. Then, the temperature dependence
of the resistivity is more involved than what theorists usually
believe. In addition, the electron-doped compounds do
$not$ show a linear behavior of the resistivity  with temperature but
rather a quadratic dependence $\rho
\sim T^2$ (see Hidaka and Suzuki, 1989; Tsuei, Gupta and Koren, 1989).
The $T^2$ dependence is
consistent with electron-electron scattering in a Fermi liquid, and thus
a lack of universality seems to exist between hole and electron doped
materials. This is another issue not sufficiently remarked in the
literature.
The behavior of the resistivity with temperature is not the only
``anomalous'' property of the normal state of the high-Tc
superconductors. ``Difficult to explain'' results have been observed in
the optical conductivity $\sigma(\omega)$, Raman scattering,
measurements of the Hall effect ($(R_H)^{-1} \sim T$ in the
${\rm YBCO}$ compounds), and several others. Some of these properties
will be discussed in the following sections.



\vskip 1cm

\centerline{\bf C. Electronic Models}

\vskip 0.5cm

\noindent{\tt 1. Three band model}

\vskip 0.3cm

After analyzing in some detail the structure and phase diagram of
the high-Tc superconductors, the next step is to write
a Hamiltonian to describe the behavior of electrons in these materials.
Due to the complexity of their structure it is important to make
some simplifying assumptions. To begin with, it is reasonable to construct
a Hamiltonian restricted to electrons moving on the ${\rm Cu O_2}$
planes. The very strong ${\rm Cu-O}$ bonds on the planes justifies
this assumption. Of course, some features of the phase diagram
can only be explained by adding a coupling between the planes (like the
existence of a
finite N\'eel critical temperature), but it is expected that those fine details
can be
studied $after$ the physics of the planes is understood. Even under
these assumptions,
the planar ${\rm Cu O_2}$
problem is still very difficult to analyze. The copper ions ${\rm Cu^{2+}}$
have 9 electrons in the five d-orbitals, while ${\rm O^{2-}}$ has the three
p-orbitals occupied. However, this complicated problem
can be further simplified as follows. All copper ions in the high-Tc
materials are surrounded by oxygens. As explained before in section
I.A.1,
in $\x214x$, a
${\rm Cu O_6}$ structure is formed around each ${\rm Cu^{2+}}$, which
corresponds to an elongated
octahedron. For other compounds like $\ybco$ the copper has five
oxygens in its vicinity; while in the electron doped material
$\xnco$, four oxygens form a square around the copper ion (see Figs.1,3,5).
In all these geometries, the degeneracy between the d orbitals produced
by the rotational invariance of isolated ions
is removed by the lattice structure.
After some calculations, it can be shown that the copper
and oxygen orbitals separate as schematically shown in Fig.9. The state with
the highest energy has mainly ${\rm d_{x^2 - y^2}}$ character, and it carries
the missing
electron (i.e. the hole) that gives the ion its spin 1/2.
Thus, in the absence of doping (i.e.
with one hole per unit cell on the plane) the material is well
described by a model of mostly localized spin-1/2 states that give to
these materials their antiferromagnetic character. The other
orbitals at lower energies
are occupied, and as a first
approximation they will be neglected in the construction of the
Hamiltonian.

What occurs upon doping? As an example, let us
consider $\x214x$, where an
additional
electron is removed from the ${\rm Cu O_2}$ plane by the substitution of
a ${\rm La}$ atom by a ${\rm Sr}$ atom. The energy levels
shown in Fig.9 may suggest that we simply have to take out another electron
from the ${\rm d_{x^2 - y^2}}$ orbital to describe the physics of the
doped compounds. However, in this picture the strong Coulombic
repulsion between holes in the same orbital is not taken into account.
Actually neglecting
interactions, one would have expected $\lacuo$ to be metallic with a
half-filled conduction band. However, the material is an
insulator with antiferromagnetic properties showing that correlations
are very strong. Double occupancy of the same orbital
must be energetically unfavored by the Coulombic interactions.
Based on this line of reasoning, it is possible to
construct a Hamiltonian for electrons in the copper oxide
planes (see Emery, 1987; Emery and Reiter, 1988a; Littlewood, Varma, and
Abrahams, 1987;
Varma, Schmitt-Rink, and Abrahams, 1987). Using the $hole$ notation,
where the vacuum is defined as all the orbitals shown in Fig.9 occupied, the
Hamiltonian is
$$\eqalign{
{\rm H =}
&{\rm -t_{pd}\sum_{\bf \langle i j \rangle} p^\dagger_{\bf j}( d_{\bf
i} + h.c.)
- t_{pp}\sum_{\bf \langle j j' \rangle} p^\dagger_{\bf j}( p_{\bf
j'} + h.c.)
+ \epsilon_d \sum_{\bf i}  n^d_{\bf i}
+ \epsilon_p \sum_{\bf j}  n^p_{\bf j} +} \cr
&{\rm U_d \sum_{\bf i} n^d_{{\bf i}\uparrow} n^d_{{\bf i}\downarrow} +
U_p \sum_{\bf j} n^p_{{\bf j}\uparrow} n^p_{{\bf j}\downarrow} +
U_{dp} \sum_{\bf \langle i j \rangle}
n^d_{{\bf i}} n^p_{{\bf j}} }.
\cr}
\tag {gg}
$$
\noindent ${\rm p_{\bf j}}$ are fermionic operators that destroy
holes
at the oxygen ions labeled ${\rm {\bf j}}$,
while ${\rm d_{\bf i}}$ corresponds to annihilation
operators at the copper ions ${\rm {\bf i}}$. ${\rm {\bf \langle i j
\rangle} }$ refers to pairs of nearest neighbors ${\rm {\bf i}}$ (copper) and
${\rm {\bf j}}$ (oxygen) sites.
The hopping terms correspond to the hybridization between nearest
neighbors Cu and O
atoms, and are roughly
proportional to the overlap between orbitals. For completeness,
a direct O-O hopping term is also included with amplitude ${\rm t_{pp}}$.
These hopping terms
allow the movement of the electrons on the lattice, providing their
kinetic energy.
${\rm U_d}$ and ${\rm U_p}$ are positive constants
that represent the repulsion between
holes when they are at the same d and p orbitals, respectively. ${\rm U_{pd}}$
has a similar meaning, i.e. it corresponds to the
Coulombic repulsion when two holes occupy adjacent ${\rm Cu-O}$.
In principle, interactions at larger
distances should also be included in the Hamiltonian, but they are
presumed to be screened by the finite density of electrons
(unfortunately,
the actual screening correlation length is difficult to calculate to
support this assumption).
The on-site energies $\epsilon_d$ and $\epsilon_p$ are shown in Fig.9. They
represent the difference in energy between the occupied orbitals
of oxygen and copper. In the strong coupling limit, and with one
particle per unit cell, this model reduces to the spin
Heisenberg model with a superexchange antiferromagnetic coupling
(Emery and Reiter, 1988a; Fulde, 1991).




The Hamiltonian Eq.(\call{gg}) shows that for $\Delta = \epsilon_p -
\epsilon_d > 0$, the first hole added to the system will energetically
prefer to occupy mostly the d-orbital of the copper ions. As explained
before, this is indeed
the observed situation in the ``undoped'' materials which have
one hole per unit cell. When another hole is added to this unit cell,
and working in the regime where ${\rm U_d}$ is larger than $\Delta$,
the new hole will mainly occupy oxygen orbitals.
This is in agreement with electron energy loss spectroscopy (EELS)
experiments (N\"ucker et al., 1987). From a band structure calculation
(Hybertsen et al., 1989)
we can roughly estimate the actual values of the parameters in the
Hamiltonian Eq.(\call{gg}). In eV's they are,

\vskip 0.4cm
\centerline{
\vbox{\offinterlineskip
\hrule
\halign{&\vrule#&\strut\quad\hfil#\quad\cr
\noalign{\hrule}
height2pt&\omit&&\omit&&\omit&\cr
&${\epsilon_p - \epsilon_d}$\hfil&&${\rm t_{pd}}$\hfil&&${\rm
t_{pp}}$\hfil&&${\rm U_{d}}$\hfil&&${\rm U_{p}}$\hfil&&${\rm U_{pd}}$\hfil&\cr
height2pt&\omit&&\omit&&\omit&&\omit&&\omit&\cr
\noalign{\hrule}
height2pt&\omit&&\omit&&\omit&&\omit&&\omit&\cr
&3.6&&1.3&&0.65&&10.5&&4&&1.2&\cr
height2pt&\omit&&\omit&\cr
}
\hrule}
}

\noindent showing that indeed we are in the strong coupling regime.
For a comparison between predictions for these parameters obtained by
different groups see Mila (1988).

\vskip 1cm

\noindent{\tt 2. One band models}

\vskip 0.3cm

The three band model has several parameters, and it is still somewhat
complicated. It would be desirable to reduce it to an even simpler
model. Zhang and Rice (1988) made
progress in this direction by the following argument. Consider
one copper ion surrounded by four oxygens. A hole at the oxygen can be in a
symmetric or antisymmetric state with respect to the central hole at the
copper ion.
These states can be combined with the Cu hole to form spin singlet
or triplet states. To second order in perturbation theory about the
atomic limit, Zhang and Rice showed that the spin singlet state has
the lowest energy, and assumed that it is possible to work in this
singlet subspace without changing the ``physics'' of the problem. Then,
the hole originally located at the oxygen, has been replaced
by a spin singlet state centered at the copper. This is equivalent to
removing one Cu spin-1/2 from the square lattice of copper spins, and
thus the effective model corresponds to spins and holes (absence of the
spin) on a two dimensional square lattice. The oxygen ions are no longer
explicitly present in the effective model. After some
calculations, their conclusion was
that the effective Hamiltonian describing the physics of the three band
model is the so-called ${\rm t-J}$ model (which was
previously introduced by Anderson, 1987)
defined as,
$$
{\rm H = J \sum_{\langle {\bf ij} \rangle}
( {\bf S}_{\bf i}.{\bf S}_{\bf j} - {{1}\over{4}} n_{\bf i} n_{\bf j})}
-t\sum_{{\bf \langle i j \rangle}\sigma}
[ c^\dagger_{{\bf i}\sigma} (1-n_{{\bf i}-\sigma})
  (1-n_{{\bf j}-\sigma}) c_{{\bf j}\sigma}
+ h.c. ],
\tag{hh}
$$
\noindent where
${\bf S}_{\bf i}$ are spin-1/2 operators at the sites ${\bf i}$ of
a two dimensional square lattice, and ${\rm J}$ is the antiferromagnetic
coupling
between nearest neighbors sites ${\bf \langle ij \rangle}$.
The hopping
term allows the movement of electrons without changing their spin,
and explicitly
$excluding$ double occupancy due to the presence of the projector
operators $(1-n_{{\bf i}-\sigma})$.
Then, this model has only three
possible states per site i.e. an electron with spin up or down, or a hole.
The rest of the notation is standard.

It is important to remark that the reduction of the three band model to the
${\rm t-J}$ model
is still controversial.
Emery and Reiter (1988b) have argued that the resulting quasiparticles of the
three band model have both charge and spin, in contrast to the ${\rm
Cu-O}$ singlets that form the effective one band ${\rm t-J}$ model. Their
result was based on the study of the exact solution in a ferromagnetic
background, and their conclusion was that the ${\rm t-J}$ model
is incomplete to represent the low-energy physics of the
three band model. Zhang and Rice (1990)
and Emery and Reiter (1990) continued their exchange of ideas on
this subject, and the issue is still unsolved. Other authors have
also contributed to the discussion (see for example Zaanen and Oles, 1988;
Stechel and
Jennison, 1988; Mila, 1988; Eskes and Sawatzky, 1988; Sch\"uttler
and Fedro, 1989; Ramsak and Prelovsek, 1989;
Belinicher and Chernyshev, 1993; and others).
Most of the results shown in the rest of
this review are for one band
models, and thus we will $assume$ that the reduction from the original
three band Hamiltonian to a one band is possible, although certainly
more work is needed to clarify this point.

In addition to the ${\rm t-J}$ and three band ${\rm Cu-O}$ models,
since the early days of high-Tc superconductivity theorists have been
extensively studying the two dimensional
$one$ band Hubbard model (Hubbard, 1963), mainly in the strong coupling limit.
This model is defined as,
$$
{\rm H = -t} \sum_{\bf {\langle ij \rangle},\sigma}
( c^\dagger_{{\bf i}\sigma} c_{{\bf j}\sigma} +
  c^\dagger_{{\bf j}\sigma} c_{{\bf i}\sigma} ) +
{\rm U \sum_{\bf i} (n_{{\bf i}\uparrow}-{{1}\over{2}}) (n_{{\bf i}\downarrow}
-
{{1}\over{2}})  },
\tag {oo}
$$
\noindent where, as usual, $c^\dagger_{{\bf i}\sigma}$ is a fermionic operator
that
creates
an electron at site ${\bf i}$ of a square lattice with spin $\sigma$.
${\rm U}$ is the on-site repulsive interaction, and ${\rm t}$ the
hopping amplitude. Although we know that the actual materials present
a band structure with three dominant bands (as shown in Fig.10),
the one band Hubbard model tries to mimic the presence of the
charge transfer gap $\Delta$ by means of an $effective$ value of the
Coulomb repulsion ${\rm U_{eff}}$,
and thus it presents only two bands (see Zaanen, Sawatzky, and Allen,
1985).
The ``oxygen'' band becomes the lower
Hubbard band of this model. Note also that in the strong coupling limit,
it can be shown that the Hubbard model reduces to the ${\rm t-J}$ model
Eq.(\call{hh}) with the addition of terms involving three sites. These terms
have not received much attention, and are usually excluded from the
numerical studies described in the rest of the review. Their importance
is unclear. Also note that the term ${\rm -1/4 n_{\bf i} n_{\bf j}}$
appears spontaneously in the strong coupling expansion of the Hubbard
model. Again, its actual importance compared with the rest of the terms
is not obvious.

Why is this Hamiltonian so much studied?
The dimensionality
of the problem is easy to understand since the model attempts to
describe electrons in the ${\rm Cu O_2}$ planes. The restriction of
working in strong coupling can also be understood easily. For example,
at half-filling this model reduces to the Heisenberg model when the
on site Coulomb interaction is large, and we know that the Heisenberg model
describes well the spin dynamics of the undoped cuprates. However, the
particular form of the Hamiltonian, where the oxygens are not
considered and the interactions are restricted to an on-site term, is
more difficult to justify.
The reader may find enlightening the article by Anderson and Schrieffer
(1991) where this issue is discussed. There are also interesting
calculations on finite clusters
by Hybertsen et al. (1990) and Bacci et al. (1991), where it was shown
that the one band Hubbard model (supplemented with a small next nearest
neighbor hopping term ${\rm t'}$) can reproduce the low-energy
spectrum of the three band model. Hybertsen et al. found that this is
achieved by taking the Hubbard model parameters as
${\rm U=5.4 eV}$, ${\rm t=0.43 eV}$ and ${\rm t' = -0.07 eV}$ i.e.
${\rm U/t \sim 12}$. These authors also found that the ${\rm t-t'-J}$
model with ${\rm J = 0.128 eV}$, and the same values of ${\rm t}$ and
${\rm t'}$ as for the one band Hubbard model, also reproduces well the
spectra of the more complicated three band Hamiltonian.

Of course, this is only a test of the short distance properties of the
models, but not of their long-range behavior. Thus, in this author's
opinion, there is no ``a priori'' clear justification for the enormous effort
carried out by hundreds of theorists to study this very particular
model (certainly including the author!).
However, as we will see below, several normal state properties of
the model qualitatively mimic those of the real materials and thus a ``a
posteriori'' justification for its use can be claimed. Unfortunately,
thus far there are no indications of superconductivity in
this model, although it is not clear whether that is a failure of the one
band approximation, or of the tools used to search for a superconducting
state (more details will be given in section IV.E.1).
Like with any simple model of a complicated material, we can only
justify its introduction after their low energy properties are known
with some accuracy. This is precisely the goal of most of the
computational work described
in the rest of the review.



\vskip 1.5cm


\centerline{\bf II. Algorithms}

\vskip 1.0cm

\taghead{2.}

The study of models of strongly correlated electrons is a difficult problem.
There are no well-controlled analytical techniques to analyze them.
Mean-field and variational approximations are self-consistent, but it is
difficult to judge if they actually describe properties of the ground
state of the system or of an excited state. As explained in the
introduction, these difficulties have led numerous groups to study
these models using computational techniques. In some sense the situation
is similar to that of particle physics in the 70's when it became
clear that an understanding of strongly correlated ``quarks'' in Quantum
Chromodynamics (QCD) would be difficult to achieve by the standard 1-loop
bubble
summation. In those days, the field of lattice gauge theory was
developed (for a review see Kogut, 1979; 1983),
and currently its predictions for the hadron spectrum of QCD have
reached a high level of accuracy. Currently, we may be facing similar
developments
in condensed matter, since it is clear that considerable progress can be
made with the help of computers in the study of correlated electrons.

A large number of techniques are currently being used to study
numerically the models of correlated electrons presented in
section I.C. However,
the vast majority of the papers in the literature can be
grouped into those where exact diagonalization (or Lanczos)
techniques were used, and those produced with Quantum Monte Carlo methods. Both
algorithms will be reviewed in this section.
A large number of physical results obtained with these
techniques will be discussed in the rest of the review.

\vskip 1cm

\centerline{\bf A. Lanczos technique}

\vskip .5cm

\noindent{\tt 1. Method}

\vskip .3cm

The basic idea of the Lanczos method is that a special basis can be
constructed where the Hamiltonian has a tridiagonal representation. This
is carried out iteratively as shown below. First,
it is necessary to
select an arbitrary vector $| \phi_0 \rangle$ in the Hilbert space of the
model being studied. If the Lanczos method (Lanczos 1950; Pettifor and
Weaire, 1985) is used to obtain the ground
state energy of the model, then it is necessary that
the overlap between the actual ground state $| \psi_0 \rangle$, and the
initial state $| \phi_0 \rangle$ be nonzero. If no ``a priori'' information
about the ground state is known, this requirement is usually easily satisfied
by selecting an initial state with $randomly$ chosen coefficients
in the working basis that is being used. If some other information of the
ground state is known,
like its total momentum and spin, then it is convenient to initiate the
iterations with a state already belonging to the subspace having those
quantum numbers
(and still with random coefficients within this subspace).

After $| \phi_0 \rangle$ is selected,
define a new vector by
applying the Hamiltonian $\hH$ to the initial state. Subtracting the
projection
over $| \phi_0 \rangle$, we obtain
$$
| \phi_1 \rangle = \hH | \phi_0 \rangle - { {\langle \phi_0 | \hH | \phi_0
\rangle}\over{\langle \phi_0 | \phi_0 \rangle} } | \phi_0 \rangle,
\tag {a}
$$
\noindent that satisfies $\langle \phi_0 | \phi_1 \rangle = 0$. Now,
we can construct a new state that is orthogonal to the previous two as,
$$
| \phi_2 \rangle  = \hH | \phi_1 \rangle -
{{\langle \phi_1 | \hH | \phi_1 \rangle}\over{\langle \phi_1
| \phi_1 \rangle}}| \phi_1 \rangle
- {{\langle \phi_1 | \phi_1
\rangle}\over{\langle \phi_0 | \phi_0 \rangle}} | \phi_0 \rangle.
\tag {b}
$$
\noindent It can be easily checked that $\langle \phi_0 | \phi_2 \rangle
= \langle \phi_1 | \phi_2 \rangle = 0$.
The procedure can be generalized by defining
an orthogonal basis recursively as,
$$
| \phi_{n+1} \rangle = {\hat H} | \phi_n \rangle - a_n | \phi_n \rangle
- b^2_n | \phi_{n-1} \rangle,
\tag {c}
$$
\noindent where $n=0,1,2,...$, and the coefficients are given by
$$
a_n = {{\langle \phi_n | {\hat H} | \phi_n \rangle }\over{\langle \phi_n
| \phi_n \rangle}}, \qquad
b^2_n = {{\langle \phi_n | \phi_n \rangle}\over{\langle \phi_{n-1} |
\phi_{n-1} \rangle}},
\tag {d}
$$
\noindent supplemented by
$b_0 =0$, $| \phi_{-1} \rangle = 0$.
In this basis, it can be shown that the Hamiltonian matrix becomes,
$$
H = \pmatrix{ a_0&b_1&0&0&\ldots\cr
              b_1&a_1&b_2&0&\ldots\cr
              0&b_2&a_2&b_3&\ldots\cr
              0&0&b_3&a_3&\ldots\cr
              \vdots&\vdots&\vdots&\vdots&\cr}
\tag {e}
$$
\noindent i.e. it is tridiagonal as expected. Once in this form the matrix
can be diagonalized easily using standard library subroutines.
However, note that to diagonalize completely the model being studied on
a finite cluster a
number of iterations equal to the size of the Hilbert space (or of the
subspace under consideration) are needed. In practice this
would demand a considerable amount of CPU time.
However, one of the advantages of this technique is that accurate enough
information about the ground state of the problem can be obtained after a
small number of iterations (typically of the order of $\sim 100$ or less). Thus
the method is suitable for the analysis of low temperatures
properties of the models of correlated electrons described in section I.C.

To understand the rapid convergence to the ground state which is
obtained using this
algorithm, it is convenient to consider a variation of this technique
known as the ``modified'' Lanczos method (Dagotto and Moreo, 1985;
Gagliano et al., 1986). In this method, the diagonalization proceeds
using ``$2\times 2$ steps'' i.e. first the Hamiltonian in the basis
$|\phi_0 \rangle$ and $| \phi_1 \rangle$ (defined before) is
diagonalized. The lowest energy state is always a better approximation
to the actual ground state than $| \phi_0 \rangle$.
This new improved state can be used as the initial state of
another
$2 \times 2$ iteration, and the procedure is repeated as many
times as needed, until enough accuracy has been reached. Then, it is
clear that the modified Lanczos method, or the original Lanczos, can be
described as a systematic way to improve a given variational state that
is used to represent the ground state of the system (Dagotto and Moreo,
1985; Heeb and Rice, 1993),
and thus it is not surprising that
ground state properties can be obtained accurately well before the rest
of the matrix eigenvalues are evaluated.

In spite of these advantages,
memory limitations impose severe restrictions on the size of
the clusters that can be studied with this method. To understand this
point, note that although
the lowest energy state can be
written in the $\{ | \phi_n \rangle \}$ basis
as $| \psi_0 \rangle = \sum_m c_m | \phi_m \rangle$, this expression is
of no practical use unless $| \phi_m \rangle$ itself is expressed in a
convenient basis to which the Hamiltonian can be easily applied. For
example,
in spin-1/2 models it is convenient to work in the basis where $S_z$ is
defined at every site, schematically represented as $| n \rangle = | \uparrow
\downarrow \uparrow ... \rangle$.
For the models described in section I.C,
the size of this type of
basis set grows exponentially with the
system size. For example, the dimension of the Hilbert space of a
Hubbard model (four states per site) on a ${\rm N}$ site cluster is
in principle ${\rm 4^N}$,
which for ${\rm N=16}$ corresponds to
${\rm \sim 4.3 \times 10^9}$ states. Such a memory requirement is
beyond the reach of present day computers. In practice this problem can
be considerably alleviated by the use of symmetries of the Hamiltonian
that reduce the matrix to a block-form. The most obvious symmetry is the
number of particles in the problem which is usually conserved at least
for fermionic problems. The total
projection
of the spin ${\rm S_{total}^z}$, is also a good quantum number. For
translational invariant
problems, the total momentum ${\rm {\bf p}}$ of the system is also conserved
introducing a reduction of
${\rm 1/N}$ in the number of states (this does not hold for models with open
boundary conditions or explicit disorder). In addition, several Hamiltonians
have
additional symmetries. On a square lattice, rotations of $\pi/2$
about a given site, spin inversion, and reflections with respect to the
lattice axes are good quantum numbers (although care must be taken in
their implementation since some of these operations are combinations of
others and thus not independent). Introducing these symmetries the
linear size of the largest block that is necessary to diagonalize for a Hubbard
model on the $4 \times 4$ square cluster is $\sim 1,350,000$ at
half-filling and zero momentum (see Fano, Ortolani, and Semeria, 1990;
Fano, Ortolani, and Parola, 1992. In these papers interesting group
theory tricks have been used to make the study of this cluster possible
in an efficient way. See also Dagotto et al., 1992b).
Then, it is clear that the use of
symmetries is very important to carry out Lanczos calculations on large
enough clusters. Currently, the one band Hubbard model can be studied on
clusters only slightly larger than the $4 \times 4$ lattice at least
near half-filling, while at low electronic densities much larger
systems
can be dealt with. The three band Hubbard model can be
analyzed on the cluster ${\rm
Cu_4 O_8}$ ($2 \times 2$ cells), but not much bigger.
The ${\rm t-J}$ model has been studied on clusters of
up to $26$ sites at low hole density (see section III.B), and perhaps
lattices of $32$ sites will be reachable in the near future. Note that
this model reaches a maximum in the
dimension of its Hilbert space at an intermediate hole density.
At this point, it is also convenient to clarify that it has become
common
place in the Lanczos literature to diagonalize not only clusters with
$M \times M$ sites, but also other square clusters that completely cover the
two dimensional square lattice, which have axes forming a nonzero
angle with the lattice axes. Examples can be found in Oitmaa and Betts
(1978). Some of the ``magic'' number of sites that admit such a covering of the
bulk lattice with ``tilted'' squares are $N = 8,10,16,18,20,26,32,...$.
The
general rule is $N = n^2 + m^2$ where the positive integers
$n,m$ are both even or odd. For
example,
$ 10 = 1^1 + 3^2$, $20 = 2^2 + 4^2$, $50 = 5^2 + 5^2$, etc.

How can we obtain explicitly the actual ground state of the problem?
Each element $| \phi_n \rangle$ of the basis is represented by a
large set of coefficients, when it is itself expanded in the basis
selected to carry out the problem (like the ${\rm S^z}$ basis).
Thus, in practice it is not convenient to store each one
of the $| \phi_m \rangle$ vectors individually since such a procedure
would
demand a memory requirement equal to the size of the Hilbert space
multiplied
by the number of Lanczos steps (typically $\sim 100$). However,
there is a simple solution to this problem, and it consists of running
the Lanczos subroutine $twice$. In the first run, the coefficients $c_m$
are obtained, and in the second  the vectors $| \phi_m \rangle$ are
systematically reconstructed one by one and stored in the vector
$| \psi_0 \rangle$. Another procedure to get the ground state is to
use the modified Lanczos method described before. In the $2 \times 2$
steps the ground state is always explicitly at hand. While this technique
converges more slowly to the ground state than the standard Lanczos method,
the latter needs to be run twice to get the ground state explicitly.
Thus, in some cases it is easier to use the modified Lanczos approach
which is somewhat simpler to program. An even more pedestrian technique
is the power method which consists of applying the Hamiltonian $n$-times
to the initial state until all excited states are filtered out, and only
the ground state remains. This procedure is the slowest in speed of
convergence,
but in simple problems is enough and easy to program.

To end this section about the Lanczos method, we will describe a recent
attempt to increase the size of the clusters
that this technique can reach. The idea is that for some particular
cases it may occur that
the wave function of the ground state expanded in some working
basis
that is selected for the problem (schematically
$| \psi_0 \rangle = \sum_m c_m | m \rangle$),
may contain states with very
small weight $c_m$. Then, it could be possible to neglect those states in
the basis, and still get accurate enough results for the ground state
properties. These types of ideas (that we call ``truncation'' method)
have been recently used in Quantum
Chemistry by Wenzel and Wilson (1992) and by Riera and Dagotto (1993a-b)
in the context of correlated electrons
(see also Knowles and Handy, 1989;
De Raedt and von der Linden, 1992; Kovarik, 1990; Prelovsek and Zotos,
1993; and references therein). For the particular case of the ${\rm
t-J_z}$ model the approach works very well, and clusters of 50 sites can
be easily studied keeping only a few hundred thousands states in the
basis (which is a negligible percentage of the total basis set size).
Physical results obtained with this approach will be described elsewhere in
this
review. However, when the method is applied to the ${\rm t-J}$ model its
convergence to the ground state energy becomes slow (logarithmic) when the
size of the basis is increased (Prelovsek and Zotos, 1993).
Probably to describe properly
the strong quantum fluctuations of the spin background most of the $S^z$
basis is needed. Then, the truncation technique is very accurate for
particular Hamiltonians while for others it only provides a rough
estimation of the ground state properties.
This approach should be
seriously considered every time a new problem
that needs computational work appears.

\vskip 1.cm

\noindent{\tt 2. Dynamical properties}

\vskip .3cm

One of the most appealing features of the Lanczos method is that it
allows the calculation of $dynamical$ properties of a given Hamiltonian
(Mori, 1965; Haydock, Heine, and Kelly, 1972;
Gagliano and Balseiro, 1987).
As shown below, the Quantum Monte Carlo technique is, unfortunately, not
suitable to
extract this information since the simulations are carried out in
imaginary time. Then, currently the Lanczos approach is the only
reliable technique to evaluate dynamical responses
in a controlled way (of course, with the restriction of working on
small clusters). Here, we will set up the main formalism. In general, we
are interested in calculating quantities like,
$$
I(\omega) = -{{1}\over{\pi}} Im [ \langle \psi_0 | {\hat O}^\dagger {{1}\over{
\omega + E_0 + i\epsilon
- {\hat H}}} {\hat O} | \psi_0 \rangle ],
\tag {f}
$$
\noindent where ${\hat O}$ is the operator that we are analyzing (which
depends on the actual experimental set up under consideration), $| \psi_0
\rangle$ is the ground state of the Hamiltonian ${\hat H}$
whose ground state energy is $E_0$,
$\omega$ is the frequency, and $\epsilon$ is a small
(real) number introduced in the calculation
to shift the poles of the Green's function
into the complex plane.
Introducing a complete basis, $\sum_n | \psi_n \rangle \langle \psi_n | = 1$,
and
using the well-known distribution property
${{1}\over{ x + i\epsilon}} =
P({{1}\over{ x}}) - i \pi \delta(x)$, valid when $\epsilon \rightarrow 0$
(where $x$ is real, and $P$ denotes the principal part),
we arrive to
$$
I(\omega) = \sum_{n} | \langle \psi_n | {\hat O} | \psi_0 \rangle |^2
\delta(\omega - ( E_n -E_0)),
\tag {g}
$$
\noindent which is another way to express the spectral
decomposition of a given operator. $| \psi_n \rangle$ are eigenvectors
of the Hamiltonian with eigenvalues $E_n$.
In practice, the $\delta$-functions
are smeared by a finite $\epsilon$ i.e. they are replaced by Lorentzians
according to $\delta(x) \rightarrow {{1}\over{\pi}} {{\epsilon}\over{x^2 +
\epsilon^2 }}$.

In order to evaluate numerically Eq.(\call{g}), it is convenient to write the
Hamiltonian
matrix in a special basis. As before, we will apply the Lanczos method
to write ${\hat H}$ in a tridiagonal form but instead of starting the
iterations with a random state, we choose
$$
| \phi_0 \rangle = {{{\hat O} | \psi_0 \rangle} \over
{\sqrt{\langle \psi_0 | {\hat O}^\dagger {\hat O} | \psi_0 \rangle}}},
\tag {h}
$$
\noindent as the initial configuration for reasons that will become
clear soon.
Following Fulde (1991), consider the matrix $(z - {\hat H})$ and
the identity $(z - {\hat H}) (z - {\hat H})^{-1} = I$, where $z = \omega
+ E_0 + i \epsilon$.
Decomposed in the basis ${ | \phi_n \rangle }$ defined before in
Eq.(\call{c}),
with $| \phi_0 \rangle$ as given by Eq.(\call{h}), we arrive to $\sum_{n}
(z - {\hat H})_{m n} (z - {\hat H})^{-1}_{n p} =
\delta_{m p}$. For the special case $p=0$ we obtain,
$\sum_{n}
(z - {\hat H})_{m n} x_n =
\delta_{m 0}$,
where $x_{n} = (z - {\hat H})^{-1}_{n 0}$.
This
represents a system of equations for the unknown $x_0$. The
particular case of $n =0$ corresponds to
$\langle \phi_0 | {{1}\over{z-{\hat H}}} | \phi_0 \rangle$ which is
the quantity we want to study. Then, we need to solve this linear
system of equations.

For this purpose we use Cramer's rule i.e.
$x_0 = {{detB_0}\over{det(z - {\hat H}) }}$, where the
matrices in the $\{ | \phi_n \rangle \}$ basis are given by,

$$
z - {\hat H} = \pmatrix{ z-a_0&-b_1&0&0&\ldots\cr
                                -b_1&z-a_1&-b_2&0&\ldots\cr
                                0&-b_2&z-a_2&-b_3&\ldots\cr
                                0&0&-b_3&z-a_3&\ldots\cr
                                 \vdots&\vdots&\vdots&\vdots&\cr},
\tag {i}
$$
\noindent and
$$
B_0                            = \pmatrix{ 1&-b_1&0&0&\ldots\cr
                                0&z-a_1&-b_2&0&\ldots\cr
                                0&-b_2&z-a_2&-b_3&\ldots\cr
                                0&0&-b_3&z-a_3&\ldots\cr
                                 \vdots&\vdots&\vdots&\vdots&\cr},
\tag {j}
$$

\noindent where the coefficients $a_n,b_n$ were defined before when the
Lanczos method was introduced. The determinants of these matrices are expanded
as $det(z - {\hat H}) = (z - a_0) detD_1 -
b^2_1 detD_2$,
and $detB_0 = detD_1$, where in general
the matrix $D_n$ is obtained from Eq.(\call{i}) by removing the first
$n$ rows and columns.
Then, it can be easily shown that
$$
x_0 = {{1}\over{z - a_0 - b^2_1 {{detD_2}\over{detD_1}}     }}.
\tag {k}
$$
\noindent The ratio of determinants on the r.h.s. of Eq.(\call{k}) can also
be expanded as
$$
{{detD_2}\over{detD_1}}
= {{1}\over{z - a_1 - b^2_2 {{detD_3}\over{detD_2}}     }},
\tag {l}
$$
\noindent and the procedure can be repeated until a full continued fraction
is constructed. Recalling the definition of the spectral intensity
$I(\omega)$, finally it can be shown that
$$
I(\omega) = -{{ 1
}\over{\pi}} Im[{\strut{\langle \psi_0 | {\hat O}^\dagger
{\hat O} | \psi_0 \rangle}\over\displaystyle{ z - a_0 -
{\strut{b^2_1}\over\displaystyle{z-a_1 -
{\strut{b^2_2}\over\displaystyle{z - a_2 - ...}}
}}             }}],
\tag {mm}
$$
\noindent which establishes the relation between Eq.(\call{f}) and a continued
fraction expansion. Recalling that $z = \omega + E_0 + i \epsilon$, then
for any value of the frequency $\omega$, the width $\epsilon$,
and knowing the ground state energy
of the system, we can obtain the spectral function.
{}From the eigenvalues of the Hamiltonian in the
special Lanczos basis obtained by iterating with the initial state
Eq.(\call{h})
we can get very accurately the positions of the poles in the spectral function.


In practice the best way to proceed in order to get the dynamical response
of a finite cluster is in two steps. First, run the Lanczos subroutine using
Eq.(\call{h}) as the initial state. It is clear that with this procedure
we are testing the subspace of the Hilbert space in which we are
interested, and thus all the states found in the Lanczos step will
contribute to the spectral function (there will be as many poles
as iterations are carried
out, assuming that this number is smaller than the total size of the
subspace being explored).
Secondly, in order to find the intensity of each pole
it is useful to recall that any energy
$eigenvector$ $| \psi_n \rangle$ of the tridiagonal representation of the
Hamiltonian
can be written as $| \psi_n \rangle = \sum_m c^n_m | \phi_m \rangle$,
where $| \phi_m \rangle$ are the orthonormalized vectors defined in the
Lanczos procedure, with $| \phi_0 \rangle$ given by Eq.(\call{h}).
Then, it can be easily shown that
$$
| \langle \psi_n | {\hat O} | \psi_0 \rangle |^2
= | c^n_o |^2 \langle \psi_0
| {\hat O}^\dagger {\hat O} | \psi_0 \rangle,
\tag {m}
$$
\noindent and thus the intensity can be written in terms of the first
component of each eigenvector obtained when the tridiagonal Hamiltonian
matrix is diagonalized. In summary, the whole process simply amounts to
a Lanczos run with a very particular initial state. To test the
convergence
of the procedure it is generally enough to plot the spectral function with a
particular $\epsilon$, and test by eye how the results evolve with the
number
of iterations. Other more sophisticated methods to terminate the
iterations can be implemented (Petiffor and Weaire, 1985).

Sometimes it is necessary to calculate moments of the distribution
$I(\omega)$. This can be done very easily. For example,
the integral over frequency of the spectral function gives,
$$
\int^{\infty}_0 d\omega I(\omega) =  \sum_n | \langle \psi_n | {\hat O}
| \psi_0 \rangle |^2
= \langle \psi_0
| {\hat O}^\dagger {\hat O} | \psi_0 \rangle,
\tag {n}
$$
\noindent where we have assumed that the eigenvectors of the Hamiltonian
are normalized to one, i.e.
$\sum_n | c^n_0 |^2 = 1$. Eq.(\call{n}) is a generic expression for some of
the ``sum-rules'' frequently mentioned in the literature for various
operators ${\hat O}$ (as shown in some examples in the remainder of the
review). If higher moments of the distribution are
needed, the following relation holds,
$$
\int^{\infty}_0 d\omega \omega^p I(\omega) =  \langle \psi_0
| {\hat O}^\dagger {\hat O} | \psi_0 \rangle
\sum_n | c^n_0 |^2 (E_n - E_0)^p,
\tag {o}
$$
\noindent where all the necessary information to calculate it was
obtained before in getting the spectral function (poles and intensities).


\vskip 1.0cm

\centerline{\bf B. Quantum Monte Carlo technique}

\vskip .5cm

The Monte Carlo method is well-known in the context of statistical
mechanics and condensed matter physics (for a recent review see Binder
and Heerman,
1992). Here, we will briefly
describe an application of this general algorithm to the quantum
mechanical many-body problem of interacting electrons on a lattice, working
in the grand-canonical
ensemble. The basic idea of this approach was presented some time ago
by Blankenbecler, Scalapino
and Sugar (1981). Suppose we want to evaluate the
expectation value of a physical observable ${\hat O}$, at some finite
temperature ${\rm T=1/\beta}$. If ${\hat H}$ is the Hamiltonian of the model,
this expectation value is defined as,
$$
\langle {\hat O} \rangle = { { Tr({\hat O} e^{-\beta {\hat H}}) }
\over{ Tr(e^{-\beta {\hat H}}) } },
\tag {p}
$$
\noindent where the notation is the standard.
{}From now on, let us concentrate on the particular case of the
one band Hubbard model which
was defined in Eq.(\call{oo}). The Hamiltonian of this model, with the
addition of
a chemical potential,
can be naturally separated into
two terms as,
$$
{\hat K} = -{\rm t} \sum_{\langle {\bf i j} \rangle, \sigma}
(c^\dagger_{{\bf i}\sigma}  c_{{\bf j}\sigma} +
c^\dagger_{{\bf j}\sigma}  c_{{\bf i}\sigma}) -
\mu \sum_{\bf i} (n_{{\bf i}\uparrow} + n_{{\bf i}\downarrow}),
$$
$$
{\hat V} = {\rm U} \sum_{\bf i} ( n_{{\bf i}\uparrow} - {{1}\over{2}})
(n_{{\bf i}\downarrow} - {{1}\over{2}}).
\tag {q}
$$
\noindent Discretizing the inverse temperature interval as
$\beta = \Delta \tau L$, where $\Delta \tau$ is a small number,
and $L$ is the total number of time slices,
we can apply the well-known Trotter's formula to rewrite the
partition function as,
$$
Z = Tr(e^{-\Delta \tau L {\hat H}}) \sim Tr( e^{-\Delta \tau {\hat V}}
e^{-\Delta \tau {\hat K}} )^L,
\tag {r}
$$
\noindent where a
systematic error of order $(\Delta \tau)^2$ has been introduced,
since $[{\hat K},{\hat V}] \neq 0$. In order to integrate out
the fermionic fields
the interaction term ${\hat V}$ has to be made quadratic in the
fermionic creation and annihilation operators by introducing a
decoupling Hubbard-Stratonovich transformation. At this stage,
we can select from a wide variety of possibilities to carry out
this decoupling i.e. we can choose continuous or discrete, real or
complex fields, belonging to different groups.
In particular, and for illustration purposes, here we use a simple
transformation using a discrete ``spin-like'' field (Hirsch, 1985),
$$
e^{-\Delta \tau {\rm U} ( n_{{\bf i}\uparrow} - {{1}\over{2}})
(n_{{\bf i}\downarrow} - {{1}\over{2}}) } =
{ {e^{-\Delta \tau {\rm U/4}}}\over{2}} \sum_{s_{{\bf i},l}= \pm 1}
e^{-\Delta \tau s_{{\bf i},l} \lambda
(n_{{\bf i}\uparrow} - n_{{\bf i}\downarrow})},
\tag {s}
$$
\noindent which is carried out
at each lattice site ${\bf i}$, and for each temperature
(or imaginary-time) slice $l$. The constant $\lambda$ is defined
through the relation $cosh(\Delta \tau \lambda) = exp(\Delta \tau {\rm U/2})$.
The transformation Eq.(\call{s}) reduces the four-fermion self-interaction of
the Hubbard model to a  quadratic term in the fermions coupled to the
new spin-like field $s_{{\bf i},l}$. Thus, in this formalism the
interactions between electrons are mediated by the spin field.
Now we can carry out the integration of the fermions. While this
is conceptually straightforward, and for a finite lattice of
${\rm N} \times {\rm N}$ sites it gives determinants
of well-defined matrices, arriving to the actual
form of these matrices is somewhat involved, and beyond the scope of
this review. Then, here we will simply present the result of the
integration (more
details can be found in Gubernatis et al.,1985; and White et al.,1989b).
The partition function can be exactly written as,
$$
Z = \sum_{ \{ s_{{\bf i},l} = \pm 1 \} } detM^+(s) detM^-(s),
\tag {t}
$$
\noindent where
$$
M^\sigma = I + B^\sigma_L B^\sigma_{L-1} ... B^\sigma_1,
\tag {u}
$$
\noindent and
$$
B^{\pm}_l = e^{\mp \Delta \tau \nu(l)} e^{-\Delta \tau {K}}.
\tag {v}
$$
\noindent $I$ is the unit matrix, $\nu(l)_{{\bf ij}} =
\delta_{\bf ij} s_{{\bf i},l}$, and $K$ is the matrix representation of
the operator ${\hat K}$.
Usually the physical observable ${\hat O}$, can be
expressed in terms of Green's functions for the electrons moving in
the spin field. Then, expressions similar to Eq.(\call{t}-23) can be derived
for Eq.(\call{p}).
Once the partition function is written only
in terms of the spin fields, we can use standard Monte Carlo techniques
(such as Metropolis or heat bath methods)
to perform a simulation of the complicated sums over $s_{{\bf i},l}$
that remain to be done. The probability distribution of a given spin
configuration is given in principle by ${{1}\over{Z}} detM^+ detM^-$
(unless it becomes negative, see next section).

A simple modification of the Blankenbecler, Scalapino
and Sugar algorithm allows the calculation of
ground state properties in the canonical ensemble i.e. with a fixed
number of electrons. This approach is called ``Projector
Monte Carlo.'' Consider the ground state $| \psi_0 \rangle$
of a system, and denote by $ | \phi_0 \rangle$ a trial state
with a nonzero overlap with the actual ground state. The expectation
value of a physical observable ${\hat O}$, can be exactly written as,
$$
{{\langle \psi_0 | {\hat O} | \psi_0 \rangle}\over{\langle \psi_0 | \psi_0
\rangle}} = \lim_{\lambda,\lambda' \rightarrow \infty}
{{\langle \phi_0 | e^{- \lambda' {\hat H}} {\hat O} e^{- \lambda {\hat H}} |
\psi_0
\rangle}\over{\langle \phi_0 | e^{- (\lambda' + \lambda) {\hat H}} | \psi_0
\rangle }}.
\tag {x}
$$
\noindent The steps necessary to simulate Eq.(\call{x}) using the Monte
Carlo method are
very similar to those discussed before in deriving Eq.(\call{t}). First,
$\lambda + \lambda'$ is discretized in a finite number of
slices, then the Trotter
approximation, as well as the Hubbard-Stratonovich decoupling, are
used. Fermions are integrated out, and all observables are finally
expressed in terms of the spin-fields which are treated using
a Metropolis algorithm (for details see White et al., 1989b; and Imada
and Hatsugai, 1989).
Results obtained using these techniques, as well as other modifications of
these methods, will be
discussed in several sections of the present review.

\vskip 1cm

\centerline{\bf C. Sign-Problems}

\vskip 0.5cm

For the one band Hubbard model,
the Quantum Monte Carlo simulations described before can be carried out
at half-filling  without trouble since the product $detM^+ detM^-$ is
positive (it can be shown that
$det M^+ = A \times det M^-$ for any configuration of the Hubbard-Stratonovich
spin fields, where $A$ is a positive number (Hirsch, 1985)). Results at
half-filling will be discussed below in section III.A. However,
in the case of an arbitrary density $\langle n \rangle \neq 1$ this is no
longer
true for the repulsive Hubbard model (other models like the attractive
Hubbard model do not have this problem and it
can be simulated at all densities). Then, the
``probability'' of a given spin configuration is no longer
positive definite. In this situation, to
obtain results using this technique
it is convenient to separate the product of the
determinants into its absolute value and its sign i.e.
$detM^+ detM^- = sign \times |detM^+ detM^-|$ for each spin
configuration.
Using this trick, the
expectation value of any operator ${\hat O}$ can be written as
$$
\langle {\hat O} \rangle = {{\langle\langle {\hat O} sign
\rangle\rangle}\over{\langle\langle sign \rangle\rangle}},
\tag {y}
$$
\noindent where $\langle\langle ... \rangle\rangle$ denotes an expectation
value obtained
using a probability proportional to
$|detM^+ detM^-|$. Similar tricks can be applied
to cases where the determinant becomes complex as it occurs in problems
of lattice gauge theory in the context of particle physics (Barbour et
al., 1986). Although Eq.(\call{y}) is an exact
identity, in practice the denominator can become very small
if the number
of spin configurations with positive and negative
determinants is similar. Unfortunately, this is the case for the Hubbard
model in some regime of couplings and densities, and at low
temperatures.
For example,
in Fig.11, $\langle\langle sign \rangle\rangle$
is shown as a function of density
$\langle n \rangle$, working at ${\rm U/t=4}$ and two temperatures, on
a $4 \times 4$ cluster (from White et al., 1989b)
The qualitative behavior is clear i.e. the sign is
decreasing rapidly when the temperature is reduced, especially
at densities close to half-filling. Similar trends have been observed for
larger clusters and couplings. Actually, it has been shown that
$\langle\langle sign \rangle\rangle$ converges exponentially to zero as the
temperature decreases (Loh et al., 1990).
This effect imposes severe constraints
on the temperatures that can be reached using Monte Carlo techniques in
simulations of the Hubbard model away from half-filling. This is the
well-known ``sign-problem.''




How does this complication affect the accuracy of the results?
In Fig.12, the expectation
value of the energy $E = \langle {\hat H} \rangle$ for the Hubbard model
is plotted as a function of temperature $T$,
on a $4\times 4$ cluster working at ${\rm U/t=4}$ and density
$\langle n \rangle = 0.875$ (from Moreo, 1993b), using the same number of
Monte Carlo sweeps
at all temperatures.
Note the rapid increase of the error bars as the temperature
decreases. For comparison, the exact result at zero temperature for the case of
two holes on a $4 \times 4$ cluster is also shown in the
figure (Dagotto et al., 1992b). It would have been difficult to
accurately obtain this zero temperature energy from the Monte Carlo data
alone.
Of course, increasing considerably the number of sweeps in Fig.12 and reducing
$\Delta \tau$, better results can be obtained, and thus this figure is
just to roughly illustrate the trends in the sign-problem. Actually, with some
effort,
the energy of the ground
state can be obtained with small error bars even away from half-filling
(remember also
that the Monte Carlo results need additional corrections to take into
account the systematic $\Delta \tau$ errors).
However,
results for other quantities like spin-spin correlations show a
similar qualitative behavior, but typically with larger error bars than
in the case of the energy, and it is difficult to improve these results
even with long Monte Carlo runs.




The study of the ``sign-problem'', and the possibility of
finding a cure for this
malice, is a very important subject in the context of simulations of
correlated electrons.
Some time ago, considerable excitement was generated by
a paper by Sorella et al. (1989) where it was claimed
that using a projector Monte Carlo algorithm, and an appropriate trial
wave function $| \phi_0 \rangle$, the mean value of the sign would converge
to a nonzero constant as $\beta \rightarrow \infty$. In such a case it was
argued that some physical quantities could be calculated simply by
neglecting the signs of the determinants. Unfortunately, these conclusions were
somewhat premature
as discussed later by Loh et al. (1990) and Sorella (1991), where it was shown
that
the expectation value of the sign actually decreases exponentially with
$\beta$.
Then, neglecting the signs of the determinants
leads to an uncontrolled approximation. Loh
et al. (1990) showed that some physical quantities related with superconducting
correlations present a $qualitatively$ different behavior with and
without the signs included in the averages.

It is also important to clarify that the ``sign-problem'' is not
only caused by the sign that appear due to fermionic
anticommutations. For example, consider the case of
the spin-1/2 Heisenberg model with nearest and
next-nearest neighbors interactions, which can be simulated using
Random Walk Monte Carlo methods (Barnes, 1991. See also Dagotto, 1991). In this
technique, matrix elements of the interactions
are used as probability in the Monte Carlo algorithm. Unfortunately,
it is not possible to write these matrix elements in
a positive definite way for an arbitrary value of the couplings in the
Hamiltonian. Then, the sign-problem is a widely extended plague that
affects several areas of quantum simulations, not only
strongly correlated electrons.
The study of the sign-problem continues attracting considerable
attention.
Some recent attempts to fight it can be found in Hamann and Fahy, 1990;
Batrouni and Scalettar, 1990;
Assaad and De Forcrand, 1990; Dagotto et al., 1990a; Zhang and Kalos,
1991; Fahy and Hamann, 1991; Furukawa and Imada 1991a;
and references therein.


Finally, we will briefly describe a recently proposed technique
to alleviate the sign-problem. The method
is based on the possibility
that the operators used to describe, e.g., hole excitations in Hubbard and
${\rm t-J}$ models are ``poor", in the sense
that they are a bad approximation to the actual ``dressed'' quasiparticle
operators that create real holes in these models.
Having proper quasiparticle operators alleviates the sign problem since
in Projector or Green's function Monte Carlo methods an initial state is
selected upon which $e^{- \Delta \tau {\hat H}}$ acts repeatedly till
convergence is reached ($\Delta \tau$ being a small number), and thus
if the initial Ansatz is very good, it may occur
that the sign problem destroys the statistics only after a good convergence is
observed (at least in the ground state energy).
A method to systematically
construct better operators was discussed by Dagotto and Schrieffer
(1991), Boninsegni and Manousakis (1991), and Furukawa and Imada (1991b);
and actually implemented
by Boninsegni and Manousakis (1991, 1993), and Furukawa and Imada
(1991b), with good results for the
cases of one and two holes in the ${\rm t-J}$ model, and the weak
coupling Hubbard model. In this technique the information gathered using
Lanczos methods is very useful to guide construction of
the variational states.

We would also like to mention a class of numerical work which is closely
allied with traditional diagrammatic calculations and so takes a rather
different approach than the Monte Carlo and exact diagonalization
methods we shall focus on. Pioneered by Bickers and Scalapino (1989), and
further developed by Serene and Hess (1991), these techniques rely on a
numerical solution of self-consistent equations for the one and two
particle interacting electron Green's functions and interaction
vertices.
These approaches are able to study the competition between particle-hole
and particle-particle instabilities, and estimate transition
temperatures. Unfortunately, an adequate treatment of this technique is beyond
the
scope of this review.


\vskip 1.5cm

\centerline{\bf III. Correlated electrons at low hole doping}

\vskip 1cm

\taghead{3.}

\centerline{\bf A. Results at half-filling}

\vskip 0.5cm

As explained before,
in the one band Hubbard model at half-filling $\langle n \rangle =1$,
particle-hole symmetry arguments can be used (Hirsch, 1985) to show
that $detM^+ = A \times detM^-$ ($A>0$)
for any configuration of the Hubbard-Stratonovich
spin field in the Quantum Monte
Carlo (see Eq.(\call{t})). Then, the product of determinants
cannot be negative, and a simulation where the probability of the
spin configuration $\{ s_{{\bf i}l} \}$ is proportional to $detM^+ detM^- $
can proceed without problems. For this particular
density,
strong numerical evidence suggests that the ground state
has long-range spin order for any nonzero value of the coupling. For
example, in
the strong coupling limit ${\rm U/t >> 1}$, the Hubbard model is
equivalent to the spin-1/2 Heisenberg model defined by the
Hamiltonian,
$$
{\rm H = J }\sum_{\langle {\bf ij} \rangle}
( {\bf S}_{\bf i}.{\bf S}_{\bf j} - {{1}\over{4}} ),
\tag {yy}
$$
\noindent where ${\rm J = 4t^2 / U}$, and the rest of the notation
is standard. This model has been extensively
studied using several different analytical and numerical methods.
According to these results the ground
state has antiferromagnetic long-range order at zero temperature
(for details see Oitmaa and Betts, 1978; Reger and Young, 1988; Chakravarty,
1990;
Manousakis, 1991; Barnes, 1991; and references therein). Exotic scenarios
like those described by the RVB states or flux phases do not seem to be
realized
in this model (at least from the point of view of computational studies
the flux phases do not seem stable in the ${\rm t-J}$
model either (Bonesteel and Wilkins, 1991; Dagotto et al., 1992b), and
thus they will not be addresed in this review).
Even including next-nearest neighbor  spin-spin
interactions no indications of such states have been found
numerically (Dagotto, 1991; and references therein).

At small and intermediate coupling ${\rm U/t}$ in the one band Hubbard model,
Quantum Monte Carlo simulations and Lanczos results
suggest that antiferromagnetic spin order is still present in the
ground state, even though double occupancy is allowed. The actual value of
the local moment decreases from 1/2 as the coupling is reduced. This state
is usually called Spin-Density-Wave (SDW) and is analytically connected to
the N\'eel-like ground state at large ${\rm U/t}$.
For example,
Fig.13 shows the exact spin-spin correlation function defined as ${\rm C(r) =
C({\bf i} -{\bf j}) =
{{1}\over{4}} (-1)^{i_x + i_y + j_x + j_y}\langle
(n_{{\bf i}\uparrow} - n_{{\bf i}\downarrow} )
(n_{{\bf j}\uparrow} - n_{{\bf j}\downarrow} ) \rangle}$ in the ground
state of the Hubbard model
as a function
of distance on a $4 \times 4$ cluster and at ${\rm U/t =4}$ (we used the
notation
${\rm {\bf i} = (i_x,i_y)}$ and the sign was introduced to make the
correlations
positive if staggered order exists). This result was obtained with the
Lanczos technique (Fano, Ortolani, and Parola, 1992; Dagotto et al., 1992b).
The X point corresponds to
a site located at (2,0) lattice spacings from the origin, while M
is located at (2,2) from the origin. Indications of long-range order
are clear in this figure. For comparison, we also include in Fig.13 results
obtained using Quantum Monte Carlo techniques applied to the same
cluster. It is interesting to note the convergence of
the QMC results to the Lanczos results
as the temperature is reduced.
Note that even working at high
temperatures, it would have been possible to infer the tendency towards
spin long-range order from the results of this small cluster. This effect may
be caused by the rapid growth of the spin
correlation length $\xi$, when the temperature is decreased, since
in the regime where
$\xi $ is larger than the lattice size the results are qualitatively
similar to those at zero temperature. This example shows that
convergence to zero temperature results for a $finite$ cluster
can be reached at relatively high temperatures depending on the physics of
the problem.

In Fig.14a results for the spin-spin correlations on larger clusters
obtained using QMC are shown (Moreo, 1993b). The presence of spin
long-range order is clear in these studies, and the finite size effects
are very small. Temperature effects
do not alter significantly the qualitative behavior of the
correlations.
Fig.14b shows an alternative method to search for long-range order.
$S(\pi,\pi)$ is the structure factor at momentum ${\bf Q} = (\pi,\pi)$,
which corresponds to the sum of the correlation $C({\bf i} - {\bf j})$ over all
distances. If there is long-range order, this quantity has to increase
and diverge
with the size of the cluster. Such an effect is clearly observed
in Fig.14b at ${\rm U/t=4}$, showing the presence of
antiferromagnetic
spin order in this model. Actually, spin-wave theory makes specific
predictions for the finite size corrections to $S(\pi,\pi)$.
Corrections
of these forms have been observed by Quantum Monte Carlo and Lanczos
methods,
clearly showing that the lattice is ordered in the thermodynamic limit.
For what value of ${\rm U/t}$ does long-range
order exist?
Hirsch and Tang (1989) showed numerically that spin-density-wave
order exists even for ${\rm U/t}$ as small as $2$, and it is widely believed
that the ground state is ordered for all nonzero values of ${\rm U/t}$.












These results obtained at half-filling can be described
intuitively by a
mean field approximation to the one band Hubbard model (Schrieffer, Wen,
and Zhang, 1989). These type of approximations are self-consistent and it
is difficult to judge their accuracy unless contrasted against results
obtained using unbiased techniques, like the numerical methods described in
this
section. For the particular case of half-filling, the agreement between
numerics and mean-field results is very good, and thus the analytic
approximation seems to have captured the important physics of the problem.
Let us write
the number operator at a given site ${\bf i}$ as
$n_{{\bf i}\sigma} = \langle n_{{\bf i}\sigma} \rangle +
( n_{{\bf i}\sigma} - \langle n_{{\bf i}\sigma} \rangle )$, where
$\langle n_{{\bf i}\sigma} \rangle$ is the expectation value in the
ground state, and
the
second term will be assumed to be ``small''.
Within this approximation, the on site Coulombic
interaction becomes
$$
n_{{\bf i}\uparrow} n_{{\bf i}\downarrow} \approx
- \langle n_{{\bf i}\uparrow} \rangle \langle n_{{\bf i}\downarrow} \rangle
+ n_{{\bf i}\uparrow} \langle n_{{\bf i}\downarrow} \rangle
+ n_{{\bf i}\downarrow} \langle n_{{\bf i}\uparrow} \rangle.
\tag {z}
$$
\noindent As an Ansatz for the mean value
of the number operators we select a SDW state i.e. we choose
$ \langle n_{{\bf i} \uparrow} \rangle = {{1}\over{2}} [ 1 + S (-1)^{|{\bf
i}|}]$,
and $ \langle n_{{\bf i} \downarrow} \rangle = {{1}\over{2}} [ 1 - S
(-1)^{|{\bf i}|}]$,
where $S$ is a variational parameter to be fixed by
minimization of the mean field energy, and $|{\bf i}| = i_x + i_y$ where
${\bf i} = ( i_x, i_y)$. In the limit
${\rm U/t=0}$, the parameter $S$ vanishes and we correctly reproduce the
result that at any site $\langle n_{{\bf i} \uparrow} \rangle =
\langle n_{{\bf i} \downarrow} \rangle = 1/2$ if the particles are
non-interacting.
At large ${\rm U/t}$, $S$
converges to one and the Ansatz becomes a spin staggered
N\`eel state. This state is a good qualitative approximation to the
ground state of the Heisenberg model (which is the
effective model at large Coulombic repulsion and half-filling),
although spin fluctuations are not included. Then, both the limits of
large and
small ${\rm U/t}$ are properly described by the mean-field state.
To obtain results at intermediate values of the coupling we have to
solve the
mean-field Hamiltonian which is now quadratic in the fermionic fields.
After some algebra, and
working in momentum space using
$c_{{\bf i}\sigma} = {{1}\over{\sqrt{N}}} \sum_{\bf p} e^{-i {\bf p}.{\bf i}}
c_{{\bf p} \sigma}$, we arrive to the  Hamiltonian
$$
H_{MF} = \sum_{{\bf p}\sigma} \epsilon_{{\bf p}}
c^{\dagger}_{{\bf p}\sigma} c_{{\bf p}\sigma}
- {{U}\over{2}} \sum_{\bf p} S
[ c^{\dagger}_{{\bf p+Q} \uparrow} c_{{\bf p} \uparrow} -
  c^{\dagger}_{{\bf p+Q} \downarrow} c_{{\bf p} \downarrow} ]
+ {{U}\over{4}} S^2 N,
\tag {aa}
$$
\noindent where $\epsilon_{\bf p} = -2t (cosp_x + cosp_y)$, ${\bf Q} =
(\pi,\pi)$, and
the chemical potential $\mu$ is zero at half-filling if the Coulombic
interaction is written in a particle-hole symmetric form as in Eq.(\call{oo}).
Note that in the mean field Hamiltonian Eq.(\call{aa}) the spin index is
diagonal,
and the operators with momentum ${\bf p}$ only interact
with those of momentum ${\bf p+Q}$ (since ${\bf p+2Q = p}$). Thus
the diagonalization of $H_{MF}$ amounts to solving just a $2 \times 2$
matrix problem for each spin and momentum (restricted to only half the
Brillouin zone). The eigenvectors can be easily obtained following steps
very similar to those used in textbooks to study the BCS model (i.e.
attractive Hubbard model). They are given by
$$
\gamma^{(+)}_{{\bf p} \uparrow} = u_{\bf p} c_{{\bf p} \uparrow} -
v_{\bf p} c_{{\bf p + Q} \uparrow}, \ \ \ \
\gamma^{(+)}_{{\bf p} \downarrow} = u_{\bf p} c_{{\bf p} \downarrow} +
v_{\bf p} c_{{\bf p + Q} \downarrow},
$$
$$
\gamma^{(-)}_{{\bf p} \uparrow} = v_{\bf p} c_{{\bf p} \uparrow} +
u_{\bf p} c_{{\bf p + Q} \uparrow}, \ \ \ \
\gamma^{(-)}_{{\bf p} \downarrow} = v_{\bf p} c_{{\bf p} \downarrow} -
u_{\bf p} c_{{\bf p + Q} \downarrow},
\tag {ab}
$$
\noindent where the upper index indicates that there are two
eigenvectors per spin and momentum ${\bf p}$ in the reduced Brillouin
zone. $\gamma^{(+)}_{{\bf p}
\sigma}$ has eigenvalue $\lambda_{\bf p} =
E_{\bf p}$, while for $\gamma^{(-)}_{{\bf p} \sigma}$
it is $\lambda_{\bf p} = - E_{\bf p}$,
where $ E_{\bf p} = \sqrt{
\epsilon^2_{\bf p} + {\rm \Delta_{SDW}^2} }$, and the spin-density-wave
gap is given by ${\rm \Delta_{SDW}^2 = U^2 S^2/4}$.
The functions used in
the definition of the eigenvectors are
$$
u^2_{\bf p} = {{1}\over{2}} ( 1 + {{\epsilon_{\bf p}}\over{E_{\bf p}}}
), \ \ \ \
v^2_{\bf p} = {{1}\over{2}} ( 1 - {{\epsilon_{\bf p}}\over{E_{\bf p}}} ).
\tag {ac}
$$





\noindent
The self-consistent equation for the mean-field parameter $S$ is
$$
{{1}\over{U}} = {{1}\over{N}} \sum_{|{\bf p}| \leq |{\bf p_F}|}
{{1}\over{\sqrt{\epsilon^2_{\bf p} + {\rm \Delta_{SDW}^2}  }}},
\tag {ad}
$$
\noindent where the sum over momenta is restricted to half the Brillouin
zone (${\bf p_F}$ is the momentum at the noninteracting Fermi surface,
and at half-filling).
In this mean field approximation, the ground state is obtained by
populating
all the ``quasiparticle'' states with negative energy as shown in
Fig.15a, i.e.
$$
| \phi_{MF} \rangle = \Pi_{|{\bf p}| \leq |{\bf p_F}|}
(\gamma^{(-)}_{{\bf p} \uparrow})^\dagger
(\gamma^{(-)}_{{\bf p} \downarrow})^\dagger
| 0 \rangle.
\tag {af}
$$
\noindent where $| 0 \rangle$ is the empty state.
Solving the gap equation Eq.(\call{ad}), we obtain the result shown
in Fig.15b. The gap is finite for all nonzero values of ${\rm U/t}$. At
large coupling, it grows proportional to ${\rm U/t}$, while in the
very weak coupling region ${\rm U/t <1}$, it follows an exponential
behavior ${\rm \Delta_{SDW}} \sim e^{-{{\pi}\over{\sqrt{{\rm U}}} }}$. Then, in
this
approximation, spin-density-wave order exists for $all$ values of the
coupling in qualitative agreement with the numerical studies (for
additional details see Schrieffer, Wen, and Zhang, 1989). It is also
worth
remarking that
in two dimensional problems, the Mermi-Wagner theorem prevents the existence of
long-range order at finite temperature if the spontaneously broken
symmetry is continuous. Actually, the spin correlation length
becomes infinite only at zero temperature. In the
actual cuprate superconductors a small residual interaction in the
${\bf c}$-direction (i.e.
between ${\rm Cu O_2}$
planes) induces a $finite$ critical N\'eel
temperature as was shown in the phase diagrams of several compounds
(Figs.2,4,6). It is also worth remarking that spin long-range order
has been found numerically not only in the one band Hubbard model but
also in the three band case (Scalettar, 1989; Dopf, Muramatsu, and Hanke, 1990;
Scalettar
et al., 1991). Finally, note that calculations similar to those
described here but made explicitly for the ${\rm YBCO}$ material have been
reported
by Furukawa and Imada, 1992.

Summarizing, it is believed that the ``physics'' of
the half-filled limit is mostly understood in models of correlated
electrons with repulsive interactions. Then, the bulk of this review
is devoted to the more challenging and interesting, but
considerably less understood,
situation where carriers are added to the planes.

\vskip 1.cm

\centerline{\bf B. Properties of holes in antiferromagnets}

\vskip 0.5cm

In this section we will describe the present status of studies of a few
holes in an antiferromagnetic background. These studies were carried out mainly
using
numerical methods, but also with the help of some rough analytical
techniques, as shown below. The models we will use are the ${\rm t-J}$
and Hubbard models. Here, it is worth reminding the
reader that in the early days of these studies it was suggested that
the effect of doping could be mimicked by adding frustration to
the spin-1/2 Heisenberg model (Inui, Doniach, and Gabay, 1988).
This issue was studied by Nori, Gagliano,
and Bacci (1992); Bacci, Gagliano, and Nori (1991); and Nori and Zimanyi
(1991). These authors actually found that the effect of doping and
frustration are  quite different in models of correlated electrons.
Therefore, the original proposal mapping doping to frustration has currently
been
discarded.
Then, unfortunately our investigations below will be carried out
directly using the
rather complicated Hubbard-like models.

\vskip 0.5cm

\noindent{\tt 1.  String picture}

\vskip 0.3cm

To gain some intuition on the behavior of holes doped into an antiferromagnet,
we will start with the study of just one hole. In spite
of its apparent simplicity, this problem is highly non-trivial and
a considerable effort has been devoted to its analysis. The physics
of a hole arises
from a competition between the superexchange energy lost near the hole,
and its kinetic energy. It is reasonable
to expect that the antiferromagnetic order parameter will
reduce its magnitude near the hole, increasing the mobility of the carrier
inside such a ``spin-bag''
(Kampf and Schrieffer, 1990). In the regime of strong
coupling, where double
occupancy is suppressed, an interesting picture emerges. Consider a hole
added to an antiferromagnet at a given site.
As this initial state evolves in time,
the hole can move some distance $l$
away from its original position
by the action of the hopping term. However, in
such excursions the spins along the path of the
hole are incorrectly aligned with respect to the N\'eel background, as
is shown in Fig.16. Then, if the hole is moved a distance $l$
from the origin the energy paid
is proportional to $l$, and thus over the hole acts an effective
``confining'' linear potential that tends to localize it. Such a
confinement is not strict, since complicated paths have been found that
avoid the problem of having an energy that grows with $l$ and thus give
mobility to
the hole
(Trugman, 1988), but in general the effects of this so-called
``string'' linear potential strongly influence on the physics of
holes in antiferromagnets.




More formally let us consider
the problem of one hole moving in a staggered spin
background (Shraiman and Siggia, 1988a; and references therein. See also
Brinkman and Rice, 1970; Prelovsek, Sega, and Bonca, 1990; Eder, Becker,
and Stephan, 1990; Eder, 1992).
To simplify the problem we will take into account
only the Ising part of the spin interaction i.e. we will consider
the so-called ${\rm t-J_z}$ Hamiltonian defined as
$$
{\rm H = J_z} \sum_{\bf \langle ij \rangle} S^z_{\bf i} S^z_{\bf j}
- {\rm t} \sum_{\langle {\bf ij} \rangle,\sigma}
({\bar c}^\dagger_{{\bf i}\sigma} {\bar c}_{{\bf j}\sigma} +
 {\bar c}^\dagger_{{\bf j}\sigma} {\bar c}_{{\bf i}\sigma}),
\tag {ag}
$$
\noindent where no transverse spin fluctuations are included,
${\bar c}^\dagger_{{\bf j}\sigma} = {c}_{{\bf j}\sigma} (1 - n_{{\bf
i}-\sigma})$,
and the
rest of the notation was already introduced in the definition of the ${\rm
t-J}$
model (Eq.(\call{hh})). The
ground state in the absence of holes is a perfect N\'eel state with
energy ${\rm E_{0h} = -{{J_z}\over{4}} 2N}$ where ${\rm 2N}$ is the number of
links of a two dimensional square lattice with ${\rm N}$ sites. The state
obtained by
removing one arbitrary spin from the cluster will be denoted
as $|0 \rangle$. This state is like
a string of zero length, i.e. all the links, but the four around the
hole, are antiferromagnetic.
Applying $H' = H - E_{0h}$ (with
$t =1$) on $| 0 \rangle$ we obtain,
$$
H' | 0 \rangle = J_z | 0 \rangle - 2 |1 \rangle.
\tag {ah}
$$
\noindent The first term appears due to the four antiferromagnetic
links that are missing once the hole is created. The new state
$| 1 \rangle$
is defined as $| 1 \rangle = {{1}\over{2}}
\sum_{{\bf \tau}_1} | {\bf \tau}_1 \rangle$, where $\{ {\bf \tau}_1 \}$ denote
unit vectors in
the directions $\pm {\bf x},\pm {\bf y}$. The
state $| {\bf \tau}_1 \rangle$ represents a ``string'' of length one
which has been created from the state without strings by moving the hole
one lattice spacing in the direction
${\bf \tau}_1$. In general, we
can define states $ | {\bf \tau}_1, {\bf \tau}_2, {\bf \tau}_3, ...,
{\bf \tau}_l \rangle$
which are obtained after $l$ applications of
the hopping term starting from the no string state
$| 0 \rangle$.
The hole position is obtained
by adding as vectors ${\bf \tau}_1 + {\bf \tau}_2 + ... + {\bf \tau}_l$.
Now, let us apply the Hamiltonian to $| 1 \rangle$. The result is
$$
H' | 1 \rangle = {{5}\over{2}} J_z | 1 \rangle - 2 | 0 \rangle - \sqrt{3} |
2 \rangle.
\tag {ai}
$$
\noindent The diagonal term arises from a simple calculation of
the energy of a state with a string of length one (there are four
missing links caused by the hole in its original position, plus three links
that have become
ferromagnetic due to the movement of the hole one lattice spacing).
The second term corresponds to
moving the hole back to the origin. The new state in the third term
is defined as $| 2 \rangle =
{{1}\over{2\sqrt{3}} } \sum_{{\bf \tau}_1, {\bf \tau}_2 \neq -{\bf
\tau}_1} | {\bf \tau}_1,
{\bf \tau}_2 \rangle$, where the sum represents the twelve states
that can be obtained by moving a hole two lattice spacings from the
origin ($2 \sqrt{3}$ is its normalization).

These ideas can be generalized to $l$ applications of the hopping
term by defining an orthonormal ``string'' basis as
$$
| l \rangle = {{\sqrt{3}}\over{2}} (h^\dagger)^l | 0 \rangle,
\tag {25}
$$
\noindent where the operator $h^\dagger$ acting over a state with a hole
at a given position, creates $three$ new states where the hole has
moved one more lattice spacing (at least for $l$ larger than one lattice
spacing) (Shraiman and Siggia (1988a)).
More formally,
$h^\dagger | {\bf \tau}_1 {\bf \tau}_2 ... {\bf \tau}_l \rangle =
{{1}\over{\sqrt{3}}} \sum_{{\bf \tau}_{l+1} \neq -{\bf \tau}_l }
| {\bf \tau}_1 {\bf \tau}_2 ... {\bf \tau}_l {\bf \tau}_{l+1} \rangle$,
where $\sqrt{3}$ is the normalization of the three newly
created states. On the other hand, the application of $h$
retraces the path of the hole in one lattice spacing.
It can be shown that the hopping term in the original Hamiltonian
is equal to $\sqrt{3} ( h + h^\dagger )$, and that $h h^\dagger = 1$.
Then,
$$
H' = J_z ( {{3}\over{2}} + l) | l \rangle - \sqrt{3} ( | l-1 \rangle
+ | l+1 \rangle ),
\tag {aj}
$$
\noindent which is valid for $l \geq 2$ (note that some paths are
neglected
in this approximation i.e. we are implicitly working on a Bethe lattice).
This problem can be solved
numerically with high accuracy. According to Shraiman and Siggia (1988a), the
ground state energy of $H'$ is $e_{1h} = -2 \sqrt{3} + 2.74 J_z^{2/3}$, which
is very similar to the result obtained numerically in the fully
interacting problem on finite clusters as we will show later in Sec. III.B.2.
Eq.(\call{aj}) corresponds to a discretized version
of the problem of a Schr\"odinger particle in a linear potential, which in
the continuum limit becomes
$$
H' | l \rangle = (- a^2 \sqrt{3} t {{d^2}\over{dl^2}}
+ J_z {{l}\over{a}} ) | l \rangle,
\tag {ak}
$$
\noindent where $a$ is the lattice spacing, and a constant
energy term has been omitted (to derive Eq.(\call{ak}) it is
useful to formally expand in powers of the lattice spacing, which
is achieved using
$| l+1 \rangle = | l \rangle + a {{d}\over{dl}} | l \rangle
+ a^2 {{d^2}\over{dl}} | l \rangle + ...$). This
problem can be solved exactly and the result is expressed in terms of
Airy function eigenvalues. Actually, through the change of variables
$l/a = (J_z/t)^{-1/3} x$ ($x$ is dimensionless),
the coupling dependence can be extracted
explicitly, and Eq.(\call{ak}) can be written as,
$$
H' | l \rangle = t (J_z/t)^{2/3} ( - \sqrt{3} {{d^2}\over{dx^2}} + x ) | l
\rangle.
\tag {al}
$$
\noindent This result
clearly shows that the energy levels of a hole in a
N\'eel background without spin fluctuations behaves as $(J_z/t)^{2/3}$. This
characteristic
dependence is not restricted to the ${\rm t-J_z}$ model, but it
will also be found below in numerical studies of the ${\rm t-J}$ model that
take into
account all the hole paths, and the transverse spin fluctuations.
Thus, the string description of holes in antiferromagnets seems to have
captured  many of the important features of this complicated problem.

Finally, by dimensional analysis we can estimate
the characteristic ``size'' of the hole ground
state wave function in real space. Suppose this wave function decays at
large distances as $exp(-l^2/L^2)$, where $L$ is the typical scale we
are looking for (the results below are independent of the actual functional
form of the wave function). Changing variables in the exponent as before we
obtain $(l/L)^2 = ((ax)/L)^2 (t/J_z)^{2/3}$. In the dimensionless
variable $x$, the size of the wave function is also a dimensionless
number
$b$, and thus we arrive to the conclusion that
$ (a/L)^2 (t/J_z)^{2/3} =b$, which implies
$L \sim 1/J_z^{1/3}$.
Thus the hole
is able to explore a larger lattice when $J_z$ is reduced, which
is reasonable since the attractive potential is weaker. This also
tell us that finite cluster calculations will have stronger finite size
effects as the superexchange coupling is reduced.

\vskip 1cm

\noindent{\tt 2. Energy and momentum of a hole}

\vskip .3cm

Now, let us consider numerical results for the actual ${\rm t-J}$ model
in two dimensions, obtained without the approximations employed in
the string picture.
The
energy of one hole in the ${\rm t-J}$ model, $e_{1h}$,
measured with respect to the energy of the undoped system, is
shown in Fig.17. This result was obtained using
a $4 \times 4$ cluster and exact diagonalization techniques in the
subspace of one hole.
In the same figure, we also
show the energy of one hole in the ${\rm t-J_z}$ model (defined in
Eq.(\call{ag})) on a $8 \times 8$ cluster (Barnes et al., 1989). For this
particular case where the transverse spin fluctuations are switched off,
and
if only one hole is studied
it can be shown that there are no ``sign problems'' using a Guided Random
Walk approach developed by Barnes and Daniell (1988).
Thus, the study of one hole in the ${\rm t-J_z}$ model
can be carried out on relatively large clusters.\footto{1}
Fig.17 implies that the energy
of one hole is proportional to ${\rm J_z^{2/3}}$ with high accuracy.
Actually, the best Monte
Carlo fit $e_{1h}/t = -3.66 + 2.96 ({\rm J_z/t})^{0.655}$ is in excellent
agreement with the string picture explained previously.
More recently, the most accurate results available for the one hole energy in
the ${\rm t-J_z}$ model have
been obtained with the ``truncation'' Lanczos algorithm described
in section II.A (Dagotto and Riera, 1993).
The reported result is $e_{1h} = -3.620 + 2.924
({\rm J_z/t})^{0.666}$ obtained using clusters with up to 50 sites and five
digit accuracy ground state energies. The string picture is clearly
very robust for the ${\rm t-J_z}$ model.




However, in the ${\rm t-J}$ model it is not obvious that
the string picture should work. In principle,
it may occur that the spin fluctuations ``cut''
the strings restoring the N\'eel spin order.
Then, for some time it was assumed that this formalism
was not suitable for the more realistic ${\rm t-J}$
model. However, numerical results
on small clusters showed that the ground state energy of one hole in the
interval ${\rm 0.2 \leq J/t \leq 1.0}$ can be fit very accurately as
${\rm e_{1h}/t} = -3.17 +  2.83 ({\rm J/t})^{0.73}$, with an exponent close
to the 2/3 power-law. It may occur that in some region of parameter space, the
string typical time-scale is much faster than that of the spin fluctuations,
and
thus the strings cannot be easily erased (for a discussion see
Dagotto et al., 1990b). In
other words, the hole may ``emit'' a string and retrace it back in a
time
proportional to $1/t$, while the Heisenberg term needs a $1/J$ time to
cut the string. Similar
conclusions can be obtained by studying excited states of the hole
through the dynamical spectral function as shown below in section III.B.4.
Then, the string picture seems to work even in the presence of spin
fluctuations.

It is also interesting to note
that for the ${\rm t-J}$ model the momentum of the hole
ground state seems to be ${\bf p} = (\pi/2, \pi/2)$. The evidence for
this result comes from a combination of spin-wave, variational and
numerical methods carried out by several different groups
(Shraiman and Siggia, 1988a; Trugman, 1988;
Schmitt-Rink, Varma, and Ruckenstein, 1988; Kaxiras and Manousakis,
1988; Dagotto, Moreo, and Barnes, 1989; Sachdev, 1989;
Poilblanc, Schulz, and
Ziman, 1992; and references therein). These results are not too
surprising since in the one band Hubbard model at ${\rm U/t=0}$, the
Fermi surface is defined by the equation $cosp_x + cosp_y = 0$, and thus
${\bf p} = (\pi/2, \pi/2)$ belongs to this surface. In addition, we
expect a smooth connection between weak and strong coupling for one hole.
Although for a nonzero
coupling there is no symmetry argument requiring that all points on the
original Fermi surface will remain degenerate, it is reasonable to
expect
that it is one of those points that will be
emptied upon doping of a hole. For
the ${\rm t-J}$ model the selected momentum seems to be ${\bf p} = (\pi/2,
\pi/2)$.
However, note that the states
with momentum ${\bf p} = (0,\pi),(\pi,0)$ are very close in energy (as
discussed below) and thus small perturbations (like a $t'$ hopping
at distance of two lattice spacings) may change the hole momentum
(see Gagliano, Bacci and Dagotto, 1990).  The study of this near
degeneracy deserves more work.

For the particular case of one hole in an antiferromagnet, an
analytical approach has been developed that gives results
in good agreement with the exact diagonalization predictions. The
basic idea was introduced by Schmitt-Rink, Varma and
Ruckenstein (1988), and
is based on (i) the analysis of the Heisenberg term of the ${\rm t-J}$
Hamiltonian using the Holstein-Primakoff transformation and the 1/S
expansion; and (ii)
the replacement of the fermionic operators by the composition of a spin wave
and a spinless hole operators. The approach was used by
Kane, Lee and Read (1989) in the ``dominant pole approximation'' to study the
single-particle Green's function of one hole, assuming that the weight beyond
the
first pole is incoherent. This additional approximation may be
questionable since
the arguments based on the string picture suggest the presence of
sharp peaks at finite frequencies related with the excited
states of a particle moving in a linear confining potential.
More recently, Marsiglio et
al. (1991), Martinez and Horsch (1991), and
Liu and Manousakis (1991) have studied the self-consistent Born
approximation to this reformulated problem. This is equivalent to the
rainbow approximation for the holon propagator, where the spinon lines
are non-crossing.
A remarkable agreement
with the exact diagonalization results was found for small ${\rm J/t}$.
The inclusion of vertex corrections does not change the
zeroth-order results appreciably for the ${\rm t-J}$ model
at small ${\rm J/t}$ (Martinez and Horsch, 1991; Liu and Manousakis,
1991). Unfortunately, an extension of this approach to a finite density of
holes is difficult.

What is the shape of the hole ground state wave function in real space?
Hartree-Fock calculations
in the context of the spin-bag approach suggest that the hole ground state has
a cigar-like shape (Su, 1988), elongated along a diagonal of the
lattice. This result was numerically confirmed using Lanczos techniques
(Dagotto, Moreo and
Barnes, 1989) showing that this peculiar shape is a consequence of the
finite momentum ${\bf p} = (\pi/2, \pi/2)$ along the lattice diagonal.
The size of this ``spin polaron'' is proportional to
${\rm 1/J}^{1/3}$ (Barnes et al., 1989) in agreement with the string
picture,
and thus it increases as ${\rm
J/t}$ decreases. This explains the observed feature that finite size
effects become more severe on finite clusters as the superexchange is
reduced. Inside the spin polaron the antiferromagnetic order parameter
is depleted but there is no evidence that the spins are
ferromagneticaly aligned (unless a very small regime of ${\rm J/t}$
is reached where the Nagaoka ``phase'' is realized (Nagaoka, 1966;
Dagotto, Moreo and Barnes, 1989). Then, this is not a ferromagnetic polaron.
Finally, it is worth remarking that there are
no conceptual differences between holes in strong coupling
trapped in a ``string'' potential
created by the antiferromagnetic background, and the spin-bags
excitations introduced by Schrieffer, Wen and Zhang (1988) in a
spin-density-wave background.
In strong coupling the effective confining potential is mostly linear, while in
weak coupling it may have some other form, but one evolves
smoothly into the other by changing the coupling ${\rm U/t}$.

\vskip 1cm

\noindent{\tt 3. Dispersion relation of a hole}

\vskip 0.3cm

It is instructive to calculate the dispersion relation of one hole in
an antiferromagnetic background. Its total bandwidth ${\rm W}$, provides
information about the renormalization effects caused by the spin-waves
that are created and absorbed while the hole propagates.
Moreover,
if the normal state is $assumed$ to be formed by a gas of noninteracting
(but spin-wave renormalized) holes,
then some observables can be calculated (Trugman, 1990a) and compared
with experiments, once the dispersion relation is known.
In addition, the specific ${\bf p}$-dependence of the energy
provides information about anisotropies in the system.
In Fig.18a, early results
for the dispersion relation of one hole are shown. They
were obtained numerically on
a $4 \times 4$ cluster by fixing the momentum ${\bf p}$ in the initial
state used in the Lanczos approach (the subsequent iterations preserve
the quantum numbers of the original state).
${\rm W}$ is defined as the difference between the energy of the state
with the minimum energy (typically corresponding to momentum
${\bf p} = (\pi/2, \pi/2)$), and the state with the highest energy
which seems to correspond to ${\bf p} = (0,0)$ in Fig.18a (which becomes
degenerate
with ${\bf p} = (\pi,\pi)$ in the Born approximation). ${\rm W}$ can
also be obtained from the position of the first pole in the
hole spectral functions.
It can be observed that the total width is considerably smaller than the
bandwidth of a free electron which is ${\rm W=8t}$, and decreases as the
coupling ${\rm J/t}$ decreases.
The bandwith ${\rm W}$ seems proportional to
${\rm J/t}$, at least for small ${\rm J/t}$ (Dagotto et al., 1990b).
This result is in agreement with other calculations
by Kane, Lee, and Read (1989), Bonca, Prelovsek, and Sega (1990),
and Trugman (1990b)
(it is also interesting to note that studies for the 3-band model
at strong coupling reveal a dispersion relation similar to that
found for one band models (Ding, Lang, and
Goddard, 1992)). According to the string picture
discussed in section III.B.1, a hole
needs a considerable energy to move in the background of
antiferromagnetically aligned spins.
Due to this effect, the hole acquires a large effective mass $m^*$, which is
reflected in a bandwidth smaller than its bare value
(note that the proper definition of the
effective mass involves the dispersion relation energy vs. momentum only
near the bottom of the band, and relating $m^*$ with ${\rm W}$ is not strictly
correct).
In Fig.18b, a study of
${\rm W}$ for different cluster sizes using the Lanczos algorithm
is reproduced from Poilblanc et al. (1993).
Finite size effects are small. ${\rm W}$ is approximately linear in the
interval ${\rm 0.1 \leq  J/t \leq 0.5}$ for all the clusters considered.
On the other hand, in the Born approximation it was found that
${\rm W \sim 1.5 J^{0.79}}$
(Martinez and Horsch, 1991), and the difference may be due to the diagrams
neglected in this approach.











An interesting
feature of Fig.18a is the degeneracy between momenta
${\bf p} = (\pi/2,\pi/2)$
and ${\bf p} = (0,\pi),(\pi,0)$. The reader should be warned that
this is an artifact of
the $4 \times 4$ cluster which has a hidden symmetry making it
isomorphic to a $2^4$ hypercubic lattice (Dagotto et al.; 1990b).
However, it has been repeatedly shown that
results obtained with this cluster size are usually qualitatively accurate
when they are compared against predictions obtained on larger clusters, or
using other techniques
(as will be discussed extensively in this review).
Then, the $4 \times 4$ cluster results may be telling us that in the bulk limit
the energies of those
momenta are indeed very close to each other.
Analyzed from the point of view of the Hubbard model, this is not
much surprising since in the noninteracting limit
${\rm U/t = 0}$ both momenta belong to the
Fermi surface, and thus at least at weak coupling only a tiny splitting in
energy is expected. This issue has been studied in the opposite limit
(strong coupling) using the Born
approximation  by Liu and Manousakis (1991) at
${\rm J/t=0.2}$ on a $32^2$ cluster. The results are shown in
Fig.19 (see also Marsiglio et al., 1991, figure 6).
The difference in energy $\Delta$
between ${\bf p} = (\pi/2,\pi/2)$
and ${\bf p} = (0,\pi),(\pi,0)$ is only about $\sim 20\%$ of the total
bandwidth. Although this analytic result involves some approximations,
while the numerical calculations are carried out on finite clusters, their
similarity gives
support to the claim
that both momenta are indeed close in energy in the bulk limit.
One of the implications of this result is that numerical studies that
search for ``pockets'' of holes near ${\bf p} = (\pi/2,\pi/2)$
(Schrieffer, Wen and Zhang, 1988)
should be carried out at temperatures smaller than $\Delta$, to avoid
mixing with other states. This detail has not been sufficiently remarked
in the literature.

\vskip 1cm

\noindent{\tt 4. Dynamical properties of one hole}

\vskip 0.3cm

One of the main advantages of exact diagonalization algorithms is that
they provide information about $dynamical$ properties of the model under
consideration
(as was explained in section II.A). This is very important
since most of the
experimentally available information on superconductors is obtained
from dynamical response measurements as a function
of frequency $\omega$. Thus, Lanczos techniques provide theoretical results
that
can be compared directly with experiments.
In the particular case of carriers in an antiferromagnetic background,
the spectral function of one hole $A( {\bf p}, \omega)$ can be evaluated.
In the
approximation where holes behave like independent particles in the
normal state of the superconductors, this spectral function can be
contrasted against photoemission spectroscopy (PES) experiments.
Of particular importance is whether a
quasiparticle-like excitation exists
in the spectrum (i.e. a pole in the Green's function of the hole
with a finite residue). This issue will be studied in more detail
in section III.B.6.

The spectral function of one hole in the ${\rm t-J}$ model is defined
as
$$
A( {\bf p}, \omega ) = \sum_n | \langle \phi_n | {\bar c}^\dagger_{{\bf
p}\sigma} |
\phi^{gs}_{0h}
\rangle |^2 \delta(\omega - ( E_n - E_0)),
\tag {an}
$$
\noindent where the hole operator ${\bar c}^\dagger_{{\bf
p}\sigma}$ creates a hole
with momentum ${\bf p}$, and spin $\sigma$. $| \phi^{gs}_{0h} \rangle $
is the ground state of the undoped system, and $ | \phi_n \rangle$
are eigenstates of the problem in the subspace of one hole with
momentum ${\bf p}$ and spin $\sigma$. Their energies are $E_0$ and
$E_n$, respectively. The Lanczos approach can be straightforwardly used
to calculate this spectral function (section II.A).
In Fig.20, $\AAA$ is shown at momentum ${\bf p} = (\pi/2,\pi/2)$ on a
$4 \times 4$ cluster at several couplings ${\rm J/t}$ (from Dagotto et
al., 1990b. See also von Szczepanski et al., 1990). The
$\delta$-functions of Eq.(\call{an}) were plotted with
a (arbitrary) width $\epsilon = 0.1t$. The
number of iterations in the continued fraction necessary to
reach convergence is coupling
dependent, but typically only $\sim 100$
iterations are enough to get results with high accuracy (this number is
much smaller than the actual size of the one hole Hilbert space).\footto{2}
Note that the energies are measured with respect to
the ground state energy of the undoped (no holes) system, with energies
$\omega = E_n - E_0$ growing from left
to right (this is not the standard way to plot a photoemission spectrum
in the experimental literature, but in this case we will
simply follow the convention used in most of the papers on one hole results).

At a relatively large coupling, like ${\rm J/t=1}$,
the spectral function has a simple structure i.e. a
dominant peak at the bottom of the spectrum is clearly observed,
and a couple of spikes are present
at higher energies. Extensive studies (Dagotto et al., 1990b)
have shown that the dominant peak
at ${\rm J/t > 1}$ corresponds to a hole almost localized at a given
site with a large mass, while the higher energy
excitations correspond to
short string states of lengths one and two, respectively. The momentum
dependence of the energy of the lowest pole shows that the hole
quasiparticle is mobile, but with a large mass.
It is natural to relate this state with
a ``quasiparticle'' state corresponding to a hole
dressed by spin excitations.
If the coupling is reduced to more realistic values, the amount
of spectral weight at the bottom of the spectrum is also reduced but
remains finite. Reciprocally, more spectral weight appears at higher energies.
Let us consider the case of ${\rm J/t = 0.2}$ shown in Fig.20c. $\AAA$
still contains a large
peak at the bottom of the spectrum (quasiparticle), but now it is
followed by a lump of spectral
weight with some internal structure. In turn this is
followed by a pseudogap, and a second lump at higher energies.
In this type of numerical
study it has been empirically found that low energy structures are
less affected by finite size effects than the high energy ones, and thus
the pseudogap and second band may disappear as the lattice size is
increased. However, the large quasiparticle peak and the
satellite low energy peaks may survive the bulk limit, as discussed
below. Finally, at ${\rm J/t=0}$ the spectrum at ${\bf p} = (\pi/2,
\pi/2)$ becomes
symmetric with respect to $\omega=0$, and it contains two bands
separated by a pseudogap. As stressed before, the region of small
coupling is the one where finite size effects are more severe (since the
hole wave function size increases when ${\rm J/t}$ is reduced).
Thus, Fig.20d may be strongly affected by size effects.








It has been shown that peaks denoted by I, II and
III in Fig.20c (corresponding to ${\rm J/t=0.2}$) can be identified as the
ground state and two next excited states of the string problem described
before in section III.B.1 (Dagotto et al., 1990b). The main support to this
statement
is that the energies of the three states have a ${\rm (J/t)^{2/3}}$
power-law dependence with the coupling similar to that predicted by the string
picture.
This result has been nicely confirmed by
Liu and Manousakis (1991) in the self-consistent Born
approximation
discussed in section III.B.2. These authors have shown that for a $4 \times 4$
cluster they can reproduce very well the numerical results,
at least for small ${\rm J/t}$ (see Fig.21a
obtained at ${\rm J/t=0.1}$).
With this approximation it is possible to study
larger clusters, and check if the string excitations survive the bulk
limit. The results on a $32^2$
cluster (Fig.21d) have already converged to the bulk limit, and the main
structure
predicted by the string picture and the
small cluster approach is still observed, namely a large
quasiparticle
peak at the bottom of the spectrum (carrying a finite percentage of the
total spectral weight),
followed by satellite peaks at higher energies.
The pseudogap of the small cluster is filled
and becomes only a soft depression in the spectral weight. While both
methods (exact diagonalization on small clusters and the rainbow
technique on large clusters) are approximations to the full problem,
the good agreement in their predictions suggests that the results may
indeed describe the bulk limit behavior of a hole in the ${\rm t-J}$
model. Then, we arrive to the conclusion
that a quasiparticle exists in this model with ground state
and excited state energies well described by the string picture. More on
this issue will be discussed in section III.B.6.




To check for finite size effects in the results shown in Fig.20, recently
Poilblanc et al. (1993)
carried out an exact diagonalization study on square clusters with up to
$N=26$ sites.\footto{3}
Results are shown in Fig.22
corresponding to ${\rm J/t=0.3}$ i.e. in the physically realistic regime. The
general trend observed in these results is that {\bf (i)} the quasiparticle
peak is
robust and does not change much with the lattice size; and {\bf (ii)} the
structure
at larger energies is more affected by the size of the clusters.
However,
care must be taken in this study since the momentum at which the
calculation has been done is not the same for the four clusters.
Actually,
only the $4 \times 4$ cluster contains momentum ${\bf p} = (\pi/2,
\pi/2)$ while the rest have momenta close to, but never at, this same
value. Then, such a systematic error may distort the higher energy
features of the spectrum. In spite of this problem, it is reassuring
that the low energy quasiparticle seems quite robust, and may well
survive the bulk limit.




An interesting issue to discuss is the momentum dependence of $\AAA$.
In Fig.23, the spectral functions at several momenta are shown at
${\rm J/t=0.2}$ on the $4 \times 4$ cluster (Dagotto et al., 1990b).
Unfortunately, the small size of the cluster does not allow us to
obtain a high resolution result in ${\bf p}$-space. However, some qualitative
conclusions can be made. For example, the shape of the spectrum
for ${\bf p} = (0, \pi/2)$ and $(\pi, \pi/2)$ is
qualitatively similar to ${\bf p} =
(\pi/2, \pi/2)$ which corresponds to the actual hole ground
state. On the other hand, at momenta ${\bf p} = (0,0)$ and
$(\pi,\pi)$ there are
substantial qualitative changes since most of the
weight is concentrated at high energies.
An analysis for different clusters sizes (Poilblanc et al., 1993)
showed that this is not a peculiarity of the $4 \times 4$ cluster,
but indeed these two momenta seem to have a large
spectral weight at high energies. The only effect of increasing
the cluster size is to provide a finite width to the peaks, while the total
amount of spectral weight in its vicinity remains approximately
constant. This result is also in good agreement with the self-consistent
rainbow approximation (Martinez and Horsch, 1991).


\vskip 1cm

\noindent {\tt 5. Binding of holes}

\vskip 0.3cm

The ground state energy of two holes has been studied by several groups.
For clusters of different sizes
it has been found that the ground state belongs to
the ${\rm B_{1g}}$ irreducible
representation of the ${\rm C_{4v}}$ point group of the square lattice
(i.e. ${\rm d_{x^2 - y^2}}$ symmetry).
(Kaxiras and Manousakis, 1988; Bonca, Prelovsek and Sega, 1989;
Hasegawa and Poilblanc, 1989; Riera and Young, 1989;
Dagotto, Riera, and Young, 1990; Fehske et al., 1991; Poilblanc, Riera
and Dagotto, 1993; and references therein).
In Fig.24a, the average distance between the two holes, obtained from
a study of hole-hole correlations in the
exact ground state wave functions, is plotted as a function of
${\rm J/t}$ for different cluster sizes (Poilblanc, Riera, and
Dagotto, 1993; Barnes et al., 1992). At least in the region ${\rm
J/t \geq 0.5}$, it is clear that the distance between holes is
small (less than two lattice spacings) suggesting the formation of
a bound state. Although such a bound state of two holes in an
antiferromagnet is not sufficient evidence for the formation of a
condensate, it is nevertheless suggestive that attractive effective
forces are operative (at least in the ${\rm t-J}$ model). Thus, it is
important to carry out a detailed study of this two hole problem, and
in this section we review the present status of this subject.

Intuitively, it is clear that a bound state of two holes in an otherwise
undoped antiferromagnet will be formed
at large values of ${\rm J/t}$. The reason is that each individual hole
``breaks''
four antiferromagnetic (AF) links, which costs an energy of the
order of the superexchange coupling. At least in the small ${\rm t}$ limit (low
mobility), two holes minimize the lost energy
by sharing a common link. In this way they reduce the number of broken AF links
from
eight to seven. When the coupling ${\rm J/t}$ is reduced to more realistic
values,
this attraction may survive until some ``critical'' coupling is reached
where holes unbind . Of course, the picture of ``minimization of the
number of broken AF links'' as the origin of binding is very crude, and
probably  wrong at small ${\rm J/t}$ but no better intuitive picture is
available.
In Fig.24b the ``binding energy'' of two holes is plotted based on results
obtained
using the Lanczos approach on clusters of ${\rm N=16,18,20}$, and ${\rm
26}$ sites (Poilblanc, Riera and Dagotto, 1993a).
The binding energy is
defined as $\Delta_B = e_2 - 2 e_1$, where $e_n = E_n - E_0$, and
$E_n$ is the ground state energy of the ${\rm t-J}$ model in the
subspace of $n$ holes. If two holes minimize their energy
by producing a bound state, then $\Delta_B$ becomes negative. Note that
in the bulk limit we expect $\Delta_B$ to
vanish if the holes do not form a bound state, since $ e_2 \approx 2
e_1$ for two independent holes. However, on a finite
cluster, it can be positive due to hole repulsion.

As shown in Fig.24b, our expectation of finding a bound state of holes is
correct. $\Delta_B$ becomes negative (implying binding) very rapidly starting
at
small values of ${\rm J/t}$. However, note that the
convergence of the results increasing the lattice size is erratic.
The reason is that $\Delta_B$ is considerably
affected by finite size effects since it is defined as a difference
between large numbers.\footto{4}
In addition,
the energy of one hole
enters in the definition of $\Delta_B$ and, as discussed before, this
quantity carries an additional systematic error due to the absence of
momentum ${\bf p} = (\pi/2,\pi/2)$ in the discrete set of momenta of
the clusters with ${\rm N=18,20}$ and $26$ sites.
In spite of these problems, qualitative information
can be obtained from Fig.24b with some confidence.
The ``critical'' coupling, ${\rm J/t]_c}$
where two holes reduce their energy by forming a bound state,
slowly grows with increasing lattice size suggesting that it may
converge to a finite value in the bulk limit, since we know that at large
couplings there must be binding. In
Fig.24b, Green Function Monte Carlo results on $8 \times 8$
clusters obtained by Boninsegni and Manousakis (1992) are also shown.
With this approach, supplemented by the use of appropriate
variational guiding states obtained from the exact analysis of smaller
clusters, it has been possible to study $\Delta_B$ for
couplings ${\rm J/t}$ as small
as $0.4$, on large clusters. The dashed line in the figure shows an educated
extrapolation
suggesting that binding starts at ${\rm J/t]_c \sim 0.3}$, in
qualitative agreement with the exact results on smaller clusters. Then,
it seems quite possible that indeed a critical value of the
coupling ${\rm J/t}$ exists beyond which two holes form a bound state in an
antiferromagnetic background.





Note that calculations in the ${\rm t-J_z}$ model Eq.(\call{ag}),
(where transverse spin fluctuations are switched off) on cluster of up
to 50 sites have also shown that a
critical coupling exists for hole pair formation (Fig.25). The
technique used to study such a large cluster is based on the selective
truncation of the Hilbert space of the problem (Sec. II.A; Riera and Dagotto,
1993a;
and references therein; see also Inoue and Maekawa, 1992).
In this
model, the critical coupling is ${\rm J_z/t]_c \sim 0.18}$, in qualitative
agreement
with the results for the ${\rm t-J}$ model since in the absence of
spin fluctuations, a stronger tendency to pairing would be expected.




The analysis of $\Delta_B$ for Hubbard-like models is more difficult.
Studies of the three band case have been restricted to very small
systems, where the presence of hole binding has been reported (Hirsch et al.,
1988; Balseiro et al., 1988; Hirsch et al., 1989). Unfortunately, a finite size
analysis of $\Delta_B$ is not possible in this model. With respect
to the one band Hubbard model, an
analysis based on exact diagonalization and Monte Carlo simulations
for $4 \times 4$ clusters shows that there is
binding. However, $\Delta_B$ is considerably smaller than the binding energy
reported for the ${\rm t-J}$ model,
and thus finite size effects may be more
important (Fano, Ortolani, and Parola, 1990;
Dagotto et al., 1990a). Actually, studies of
one dimensional Hubbard chains on finite clusters
show also a negative $\Delta_B$ comparable in
magnitude to that of the two dimensional case. However, increasing the
size of the chain,
$\Delta_B$ seems to converge to zero in the bulk limit (Fye, Martins, and
Scalettar, 1990). Similar negative conclusions for binding were found by
Ding and Goddard (1993) in the one dimensional 3-band Hubbard model. Then,
currently there is no convincing numerical evidence that the two dimensional
Hubbard
model has hole binding near half-filling in the bulk limit.
More work is necessary to clarify this issue.

\vskip 1cm

\vfill
\eject

\noindent{\tt 6. Quasiparticles in models of correlated electrons}

\vskip 0.3cm

One of the most controversial issues in the
context of models of correlated electrons proposed to describe
the new superconductors, is whether a hole injected in the undoped
ground state behaves like a quasiparticle or not. While it is clear
that spin-wave excitations
will heavily dress the hole, increasing substantially
its mass, the central point is whether this renormalization
is so strong that the wave function renormalization
$Z$ at the Fermi surface vanishes. This scenario has been
proposed by Anderson (1990a) mainly based on results obtained in
the one dimensional Hubbard model where indeed $Z$ vanishes in
the bulk limit. However,
this is a very particular situation caused by the
dimensionality of the problem, and what occurs in two dimensions is
unclear. In Fig.26, we show a hole injected in
an antiferromagnetic chain ( e.g. large ${\rm U/t}$ Hubbard model)
that propagates due to the hopping term in the Hamiltonian.
As shown in the figure, the ``ferromagnetic'' link and the hole quickly become
separated.
The velocity corresponding to charge and
spin degrees of freedom are different in this model, and even if a wave
packet is constructed at time $t=0$ with spin and charge localized
near the same site, the time evolution of the packet will make charge and
spin decouple (see for example, Jagla, Hallberg, and Balseiro, 1993; see
also Song and Annett, 1992).
Then, it is natural that the overlap between the initial state and
the ground state is zero in the bulk limit. Of course, there is no
reason for this mechanism to work in two dimensions (2D).
In 2D the hole propagation costs an energy proportional to
the length of the walk, contrary to what occurs in 1D (remember the
string excitations described in section III.B.1).
Spin-charge separation is not obvious in the dimension of interest.

Other more conservative approaches, like the spin-bag or string
ideas, describe the hole as surrounded by a region where the
antiferromagnetic order parameter is reduced. The combination
of charge plus the depleted antiferromagnetic background moves
coherently, and
behaves like a particle with charge $Q=e$, and spin 1/2, i.e. $Z$ is
nonzero in this approach. Then, since
different theories drastically disagree on the nature of
quasiparticles in strongly correlated electrons,
the
important issue that needs to be clarified numerically is the following:
suppose we consider a
large but finite cluster of $N$ (even) sites, with $N-1$ spins and one hole.
The ground state of the system has spin-1/2 (unless
ferromagnetism is favored which only occurs in special cases). Where is
this spin-1/2 localized? Is it near the hole or spread all
over the lattice? In the first case, we are forming a spin
polaron and the hole is a dressed quasiparticle with a finite
$Z$ weight. In the second case, this quasiparticle is unstable, and
it basically decays into a holon and a spinon.
Semiclassical calculations by Shraiman and Siggia (1988b) support the
assumption that
$Z$ vanishes in the bulk limit in two dimensions.
The main idea is that an infinite range $1/r$ distortion in
the spin background occurs when a hole is injected in an antiferromagnet.
However, these calculations have been recently reanalyzed by Reiter (1992),
taking into account
quantum fluctuations, and it was found that $Z$ remains finite.
Similar conclusions were reached by Auerbach and Larson (1991)
suggesting that the hole is in fact a small polaron.

The study of $Z$ can be explicitly addressed using numerical techniques. In
particular, Lanczos methods that provide the hole spectral function
are especially suitable for this purpose. $Z$ is simply given
by the weight at the lowest pole in the spectrum, i.e.
$$
Z = {{| \langle \psi^{gs}_{1h} |
{\bar c}^\dagger_{{\bf p}\sigma} | \psi^{gs}_{0h} \rangle | }\over
{ [ \langle \psi^{gs}_{0h} |
{\bar c}^\dagger_{{\bf p}\sigma} {\bar c}^\dagger_{{\bf p}\sigma} |
\psi^{gs}_{0h} \rangle ]^{1/2} }},
\tag {ap}
$$
\noindent where $| \psi^{gs}_{nh} \rangle$ is the ground state in the
subspace of $n$ holes, and the rest of the notation is standard. With
the
normalization used in Eq.(\call{ap}), it can be shown that $0 \leq Z \leq 1$.
Results obtained for the ${\rm t-J}$ model using two dimensional
clusters of $16,18,20$ and $26$ sites are shown in Fig.27a.
The behavior of $Z$ suggests that the quasiparticle weight remains
finite in the bulk limit for all the explored
values of ${\rm J/t}$, although
work on larger lattices is necessary to confirm this result. A fit in the
interval ${\rm 0.1 \leq J/t \leq 0.4}$ suggests that $Z \sim J^{0.5}$,
which vanishes only at ${\rm J/t=0}$.
However,
care must be taken in the use of these clusters for calculations in
the one hole subspace since due to the geometry of the clusters, the
momentum ${\bf p} = (\pi/2,\pi/2)$ only exists for $N=16$. In the
other clusters the ground state has a momentum
${\bf p} = (\pi, \pi/3)$, $(4\pi/5, 2\pi/5)$, and $(9\pi/13, 7\pi/13)$,
for the $N=18,20$ and $26$ sites, respectively. In spite of this
difficulty, there are no major finite size effects in Fig.27a.
Nevertheless, it would be very
important to study clusters slightly larger than those currently
available (like $N = 32$ sites)
to carry out a finite size scaling study with the
proper momentum ${\bf p} = (\pi/2,\pi/2)$.
It would also be important to study numerically
the separation between spin and charge in two dimensions. Is the
spin-1/2 of ${\rm N-1}$ spins on an ${\rm N}$ site lattice (${\rm N}$
even) localized near the hole or spread over the entire cluster?
More work should be devoted to this issue.


The Born approximation to the spin-wave holon reformulation of
the ${\rm t-J}$ model can also be used to calculate the quasiparticle
weight (Martinez and Horsch, 1991; Liu and Manousakis, 1991).
Within this approximation there is a well
defined quasiparticle peak in the spectrum carrying a finite
percentage of the total weight in good agreement with
the Lanczos calculations. The Born approximation
predicts ${\rm Z = 0.63 J^{0.667}}$ in the
interval ${\rm 0.01 \leq J \leq 0.5}$, for ${\bf p}=(\pi/2,\pi/2)$
on a $16 \times 16$ cluster ($\sim$ bulk limit). The results
are actually very close to those shown in Fig.27a, in spite of the
different optimal power-law fit.
This is an excellent example of how
analytical and numerical techniques can complement each other
efficiently in the study of a given problem.
The results obtained in the Born approximation
are appreciably closer to the numerical
results than the dominant pole approximation
which instead predicts $Z \sim J$ (Kane, Lee and Read, 1989).
The remaining small
discrepancy between numerics and the Born approximation
may be due to higher order corrections in the former, or
to finite size effects in the cluster results.

Similar issues have been addressed for the one band Hubbard model, but
the results are less clear. The wave function renormalization
for one less particle than
half-filling on a $4 \times 4$ cluster has been obtained by Fano,
Ortolani and Parola (1992). The results are also shown in Fig.27a,
and they qualitatively resemble those obtained for the ${\rm t-J}$ model
at large ${\rm U/t}$, using the relation ${\rm J = 4 t^2/U}$ (however,
note that the definition of $Z$ of these authors is different from that
used in Eq.(\call{ap})!). Both in the ${\rm t-J}$ and Hubbard model, the
weight of the quasiparticle decreases as the strong coupling limit is
approached, and seem to converge to zero only as ${\rm U/t \rightarrow
\infty}$.
Unfortunately, clusters
appreciably larger than those with ${\rm N=16}$ sites are unreachable
using exact diagonalization techniques,
for the Hubbard
model with one less particle than half-filling .
Thus, the finite size
study performed for the ${\rm t-J}$ model has not been carried out for
this model.
However, $Z$
can also be obtained by studying $n ({\bf p})$.
The ``jump'' in this quantity at the Fermi surface is
proportional
to $Z$. Moreo et al. (1990a) studied $n ({\bf p})$ using Monte Carlo
techniques
finding a finite jump at the Fermi surface (see section IV.C.3 below).
On the other hand, Projector Monte Carlo calculations by Sorella (1992)
seem to suggest the vanishing of $Z$ as the cluster size is increased
working at ${\rm U/t=4}$. However, in the last calculation a small
magnetic
field was introduced to work with ``closed shell'' configurations in
order to alleviate the sign-problem. How $Z$ is influenced by this small
magnetic field in unclear, and thus we believe that more work should be
done in the one band Hubbard model to clarify these issues.

It is also instructive to obtain the weight of the lowest energy
peak in the
spectral decomposition of the operator that creates a $pair$ of holes over
an undoped spin background. As was discussed in section III.B.5, there
is a wide region of parameter space in the ${\rm t-J}$ model where
holes tend to bind in pairs, and thus isolated holes are unstable
against pair formation. In other words, using a grand-canonical
ensemble where the fermionic density $\langle n \rangle$
is regulated with a chemical
potential $\mu$, the state of one hole can never be reached in that binding
region.
Then, the
actual ``quasiparticles'' in this regime are the
hole pairs. They carry charge $Q=2e$, total spin 0 (if bound in
a singlet), and internal quantum numbers related
with the symmetry of the pair ($\dx2y2$ according to the results
discussed in section III.B.5).
While these arguments are suggestive, to verify
the picture that associates the bound state with a quasiparticle,
it is necessary to show that the spectral decomposition
of the pair operator that produces this state out of the undoped
system, contains a sharp $\delta$-function at the bottom of the
spectrum.
By complete analogy to the case of one hole, it is possible to define
the spectral function of the operator $\Delta_\alpha$ that creates
a pair of holes from the ``vacuum'' (antiferromagnetic state) as,
$$
P(\omega) = \sum_n | \langle \psi^{n}_{2h} | \Delta^{\dagger}_{\alpha}
| \psi^{gs}_{0h} \rangle |^2 \delta(\omega - (E^n_{2h} - E^{gs}_{0h})).
\tag {xx}
$$
\noindent We have defined
$\Delta_\alpha = {\bar c}_{{\bf i}\uparrow} (
{\bar c}_{{\bf i+x}\downarrow} +
{\bar c}_{{\bf i-x}\downarrow} \pm
{\bar c}_{{\bf i+y}\downarrow} \pm
{\bar c}_{{\bf i-y}\downarrow} )$, where $\alpha=s$ $(d)$ corresponds
to the $+$ $(-)$ signs
and represents a pair with extended s-wave symmetry ($\dx2y2$ symmetry).
$ \{ | \psi^n_{2h} \rangle \}$ are states of the two holes subspace with
energy $E^{n}_{2h}$, and the rest of the notation is standard.\footto{5}
The results
for a $4 \times 4$ cluster are shown in Fig.28 for ${\rm J/t=0.4}$
and both symmetries (Dagotto, Riera, and Young, 1990; see also
Poilblanc, Riera, and Dagotto, 1993; Chernyshev, Dotsenko, and Sushkov,
1993).
Let us consider
the first case: clearly for a d-wave operator, there is a sharp
$\delta$-peak at the bottom of the spectrum  containing an appreciable
amount of the total spectral weight. The rest of the spectrum seems
incoherent. This result supports the picture that the two holes form
a bound state that behaves like a quasiparticle.
On the other hand, the spectral function
corresponding to an extended s-wave operator is drastically different.
The weight at the bottom of the spectrum is negligible, and most of
the spectral weight is concentrated at high energy. The s-wave spectrum
clearly does not present a quasiparticle.

Poilblanc, Riera, and Dagotto (1993) have recently studied the two holes
``quasiparticle weight'' $Z_{2h}$ defined as,
$$
Z_{2h} = {{| \langle \psi^{gs}_{2h} |
{\Delta}_\alpha^\dagger | \psi^{gs}_{0h} \rangle | }\over
{ [ \langle \psi^{gs}_{0h} |
{\Delta}_\alpha^\dagger {\Delta}_\alpha
| \psi^{gs}_{0h} \rangle ]^{1/2} }}.
\tag {aq}
$$
\noindent on several
cluster sizes. Their results are shown in Fig.27b.
In agreement with the one hole study, no major
differences are observed between the currently available clusters,
suggesting that $Z_{2h}$ remains finite in the bulk limit (unless ${\rm
J/t}$ vanishes).\footto{6}
Finally, note that this self-consistent picture of bound states behaving as
quasiparticles
can be wrong if a multibody condensate
is formed when additional holes are added to the system. In other words, the
formation of a superconducting condensate (through a possible Bose
condensation of two hole bound states) or the possibility of
phase separation for small values of ${\rm J/t}$, will break down
the picture of independent pairs of holes at low temperatures. However,
it may still correspond to a good approximation to the normal state of the
problem above $T_c$.

\vskip 0.7cm

Summarizing the results of this subsection, currently available numerical
results for the ${\rm t-J}$ model suggest that $Z$ for one and two holes
is finite for
all finite values of ${\rm J/t}$. A similar result was obtained
using a self-consistent Born approximation. Results for the Hubbard
model are more controversial.





\vskip 1.5cm

\centerline{\bf IV. Comparing experiments with computer simulation results}

\vskip 1.cm

\taghead{4.}

\centerline{\bf A. Magnetic properties in the presence of carriers}

\vskip 0.5cm

In this section, the magnetic behavior of models of strongly
correlated electrons (away from half-filling) are compared against
experiments. The main conclusion of this analysis is that some of
the ``unusual'' magnetic properties of the cuprate materials
can be qualitatively
reproduced by simple one band electronic models. Before describing the
results, the reader should be warned
that the Raman scattering will not be reviewed in this
paper since not much computational work has been carried out at finite
hole doping. The reader should consult Dagotto and Poilblanc (1990), and
Gagliano and Bacci (1990) for an attempt to address this issue, and for
experimental references on this important subject.

\vskip 0.5cm

\noindent{\tt 1. Magnetic Susceptibility}

\vskip 0.3cm

Here, we will review experimental and theoretical results for
the magnetic susceptibility $\chi_M$ of doped compounds.
Fig.29 shows $\chi_M$ as reported by Johnston et al. (1988), and
Johnston (1989),
obtained in powder $\x214x$ and $\ybco$ (see also Lee, Klemm, and Johnston,
1989; and
Torrance et al., 1989). Common to both materials is the
pronounced S-shaped curve for the samples close or in the insulating
regime. At half-filling such a behavior is not surprising since
the appearance of a maximum in $\chi_M$
at temperatures of the order of the superexchange $J$ ($\sim
1500K$ for the cuprates)
is known to occur in the two dimensional spin-1/2
Heisenberg antiferromagnet on a square lattice (Gomez-Santos,
Joannopoulus, and Negele,
1989; Gagliano, Bacci, and Dagotto, 1991; and references therein).
Unfortunately,
the temperature where the maximum occurs is higher than the measurement
limit,
and thus this theoretical prediction has not been verified for the
undoped cuprates.
However, with increased  doping the results for $\x214x$ show that this
maximum becomes observable in the experimentally accessible temperature
range, and it was found to
shift towards lower temperatures with increasing ${\rm Sr}$ concentration. A
qualitatively similar trend seems to occur in $\ybco$ with increasing the
amount of oxygen.
In particular, for the
samples with the highest Tc (i.e. having ${\rm O_{6.96} }$), the
susceptibility $\chi_M$ is almost flat.
The same figures show that working at fixed temperature but
changing the amount of doping, the magnetic susceptibility
increases when the hole concentration is increased from the insulating
regime.

These results can be qualitatively understood  as follows: when the
antiferromagnetic parent compounds are doped, the effective ${\rm Cu-Cu}$
superexchange interaction ${\rm J}$
is reduced due to the presence of hole carriers. As a consequence, the
temperature at which the maximum of the magnetic susceptibility
is located should also
decrease with increasing doping, as observed experimentally. To
understand
the doping dependence at a fixed and low temperature, simply
remember that for an antiferromagnet
the susceptibility at zero temperature is inversely proportional to
the spin superexchange, i.e. $\chi_M {\rm  \sim 1/J}$. Thus, if J is reduced by
doping,
the zero temperature susceptibility should increase as
observed experimentally.  From these ideas
it can also be inferred that in the regime where $\chi_M$ is flat the
antiferromagnetic correlations governed by an effective superexchange
coupling
may no longer be very important. The fact that this occurs on materials with
large
Tc, like ${\rm Y Ba_2 Cu_3 O_{6.96}}$, is suggestive that
antiferromagnetism may not be crucial to describe superconductivity in
these materials after all.
But of course considerably more work is needed to clarify this issue.
Although the qualitative argument based on a smaller effective
superexchange seems correct, it is worth remarking that
in practice it is difficult to map $\chi_M$ to a universal curve
by a rescaling of the energy scale, especially since $\chi_M$ is not
known over a wide enough temperature range (Johnston et al., 1988), and
thus this argument
should only be considered as a first approximation to the problem.

Now, let us consider more elaborate
theoretical predictions for magnetic properties of models of the cuprates.
An analysis of $\chi_M$ using weak coupling resummation methods (mainly RPA)
in the one band Hubbard model has been carried out (Bulut et al., 1990;
Lu et al., 1990). Using
this approach the susceptibility is maximum at half-filling i.e. it
does $not$ reproduce the experimental behavior of the cuprates. Is this a
problem of the approximation or of the Hubbard model?
To clarify this issue,
the magnetic susceptibility of the one band
Hubbard model has been recently studied
using Quantum Monte Carlo techniques (Moreo, 1993a).
The results are shown in Figs.30a-b.
Since the sign-problem is especially severe at large ${\rm
U/t}$ and finite hole density, the temperature of this study is
relatively high ${\rm T=t/4}$ (although smaller than $J$, at large couplings!).
However, even at this high temperature
a qualitatively different
behavior is observed between ${\rm U/t=4}$ and ${\rm U/t=10}$. In weak coupling
$\chi_M$ is maximum at half-filling as predicted by the RPA
approximation, while at large couplings the
maximum is reached at $\langle n \rangle \sim 0.85$, in qualitative
agreement with experiments. Similar results in the strong coupling
regime have been obtained using
high temperature expansions for the ${\rm t-J}$ model (Singh and
Glenister,
1992a) (Fig.30c).
The series expansion results are valid directly in
the bulk limit, but are restricted to
couplings ${\rm J/t > 0.5}$, and to temperatures ${\rm T > J/2 \sim t/4}$
due to uncertainties in the analytic
continuations (involving Pad\'e approximants)
necessary to extrapolate the results from high to low temperatures
(smaller couplings can be studied but the results are reliable only for
even larger temperatures).
In spite of the limitations of the numerical and analytical results, it
is reassuring to find a nice qualitative agreement among them, and
also when they are compared with experiments.

In the regime of large ${\rm U/t}$, the
simple argument used
to explain the behavior of $\chi_M$ based on a reduction of the effective J
should be operative, and the numerical results of Moreo (1993a)
confirms it. In the
weak coupling regime it may occur that the numerical simulation did not reach
temperatures low enough to observe the growth of the magnetic susceptibility
away from half-filling, or that a genuine qualitative
difference exists between strong and weak coupling. Nevertheless, it is
interesting to observe that the regime of ${\rm U/t \sim 10}$ of the
one band Hubbard model is able to reproduce qualitatively the behavior
of the experimentally observed magnetic susceptibility. Later in this review
we will discuss that for this same regime of coupling, infrared
experiments for the optical conductivity,
as well as photoemission results, are also qualitatively
reproduced. The intermediate to large regime of ${\rm U/t}$
seems the most
promising to describe the cuprate materials using one band electronic models.

\vskip 1cm

\noindent{\tt 2. Antiferromagnetism at finite doping}

\vskip 0.3cm

One of the most distinctive features of the cuprate superconductors is
the presence of antiferromagnetism in the undoped compounds. The experimental
evidence for this behavior has been
widely discussed in the literature, and it is not
the purpose of the present review to further analyze this issue.
Instead, we will
center our attention into the more interesting, and less understood,
study of antiferromagnetism in $doped$ materials. Fig.31a shows single
crystal neutron
scattering results for the
antiferromagnetic spin correlation length as a function of ${\rm Sr}$
concentration
in $\x214x$ at low temperatures
(Birgeneau et al., 1988). The solid line is the function
${\rm {{3.8}\over{\sqrt{x}}} \AA}$ which roughly reproduces the
data. This function corresponds to the average separation distance
between holes
(assuming them to be static). Since the ${\rm Cu-Cu}$ distance in the
plane is about $3.8 \AA$, Fig.31a shows that at the optimal concentration
${\rm x \sim 0.15}$ for this compound,
the spin correlation is approximately 2.6 times the
${\rm Cu-Cu}$ distance. Then, antiferromagnetism
is still strong in this regime, and for this compound (while for
${\rm Y Ba_2 Cu_3 O_7 }$ it may be weaker). However, it is important to
remark that the presence of incommensurate correlations in these compounds
(to be discussed below) somewhat alters the results of Fig.31a.
Actually, the correlation length taking into account the spin
incommensurability
is shown in Fig.31b
(from Aeppli, 1992b). This correlation is larger than what was
previously found by Birgeneau et al. (1988).

Several theoretical works have studied the degree of suppression of
antiferromagnetism
by the addition of holes to the ${\rm t-J}$ or Hubbard models at half-filling.
As example, consider the dynamical spin structure function
$S({\bf Q}, \omega)$ calculated for the ${\rm t-J}$ model
on a $4 \times 4$ cluster at ${\rm J/t=0.4}$, and
several dopings (see Fig.32 taken
from Dagotto et al., 1992b. See also Bonca et al., 1989). The momentum is
${\bf Q} = (\pi,\pi)$.
The sharp peak at low frequencies and doping corresponds to
the spin-wave excitation which
at half-filling, and in the bulk limit, becomes massless.
The finite size of the cluster, plus the effect of doping opens a gap
in the spectrum corresponding to this momentum.
For doping ${\rm x=0.25}$ a considerable amount of
spectral weight is transferred to large energies. For the quarter-filled
system, $\langle n \rangle = 0.50$, the spin-wave peak has virtually
disappeared. Then, these rough numerical results
obtained on a small cluster show that the rapid reduction of
antiferromagnetism with doping can be mimicked by simple models of
spins and holes, and it is not at all surprising. The same result is
expected for most models of high Tc superconductors.


To make this discussion more quantitative, let us consider the
results shown in Fig.33 which were
obtained by Furukawa and Imada (1992) using the Projector
Monte Carlo technique. The inverse of the
static structure
factor $S({\bf Q})$
at ${\bf Q} = (\pi,\pi)$ is shown as a function of
the hole density (denoted by $\delta$ in the figure)
for the one band Hubbard model
at ${\rm U/t=4}$, on clusters with up to $10 \times 10$ sites. Let us assume
that the
mean value of the
spin correlation between spins located at the origin ${\bf o}$ and
at site ${\bf r}$ is given by $\langle {\bf S}_{\bf o}.{\bf S}_{\bf r}
\rangle \sim e^{i{\bf Q}.{\bf r}} e^{-r/\xi}$, where $\xi$ is the
spin correlation length. Under this approximation,
$S^{-1}({\bf Q})$ should be roughly
proportional to $\xi^{-2}$ in two dimensions, and it should vanish in the
bulk
limit if there is long-range order. The Monte
Carlo results of Fig.33 suggest
that as soon as doping is introduced,
the long-range order disappears, and the spin correlation length becomes
finite. The approximate linear dependence observed in Fig.33, implies
that $\xi \sim 1/\sqrt{x}$, in good agreement with experiments (Fig.31). It
would
be interesting to verify these results using a larger value of the
coupling ${\rm U/t}$ in order to
work in a regime closer to that of the cuprate superconductors, namely
${\rm U/t = 8 - 10}$. Unfortunately, Monte Carlo techniques at finite
doping and on large clusters, are not able to reach low temperatures
mainly due to the sign-problem.


Finally, it is interesting to note that the influence of
doping in
the antiferromagnetic correlations is non-universal
between hole and electron doped materials.
Single crystal
neutron scattering measurements on $\xnco$ by Thurston
et al. (1990) have shown that the 3D antiferromagnetic order
persists even  with ${\rm x}$ as high as 0.14, while on $\x214x$, a doping of
${\rm x \sim 0.02}$ is enough to destroy the long-range order. This difference
can
be attributed to several factors, but there is one that is very
important: carriers in electron-doped materials seem to reside in
${\rm Cu}$ ions, while carriers in hole-doped materials reside on oxygen ions.
As shown by Manousakis (1992) a study of the Heisenberg model
with $static$ randomly distributed
holes shows a dependence of the antiferromagnetic
spin correlation function that reproduces better the results for
$\xnco$ than those for $\x214x$, suggesting that the former loses
antiferromagnetism through a dilution process.

\vskip 1cm

\noindent{\tt 3. Incommensurate spin order in doped materials}

\vskip 0.3cm

Neutron scattering experiments have shown that $\x214x$ with ${\rm x=0.075}$
and ${\rm x=0.14}$ present incommensurate spin fluctuations. Fig.34 shows
the results reported by Cheong et al.(1991) as a function
of momentum for the two concentrations mentioned before (for a review
see Aeppli, 1992b).
The splitting of the peak at
$(\pi,\pi)$ indicates the presence of incommensurability. It is
important to note that the splitting was observed along the line
joining momentum
$(\pi,\pi)$ with $(0,\pi),(\pi,0)$ in a notation corresponding to
a two dimensional square lattice. Cheong et al. (1991)
also reported a $finite$ (although large) spin correlation length,
suggesting that the incommensurate correlations in this material are of
short-range
(while recent studies of the three dimensional ``Mott-Hubbard'' system
${\rm V_{2-y} O_3}$ show the presence of a
static incommensurate order).
Mason, Aeppli, and Mook (1992) showed that the incommensurate peaks in
the cuprates are suppressed
below Tc.
Is the presence of these incommensurable
correlations a universal feature of all high-Tc compounds?
Tranquada et al. (1992) have recently reported
results for $\ybco$. The claim is that separate peaks cannot be resolved,
but the data is consistent with four unresolved incommensurate peaks
similar to those reported in the lanthanum compounds. On the other hand,
electron doped
materials $\xnco$ analyzed by Thurston et al. (1990) showed $no$
indications of incommensurability  adding more evidence
against universality between hole and electron-doped materials, at least
with respect to their magnetic properties.

Can we understand the presence of these incommensurate correlations with
simple electronic models?
Hartree-Fock calculations have been presented for the one band
Hubbard model at large (Poilblanc and Rice, 1989), and small (Schulz, 1990a)
coupling ${\rm U/t}$. These studies predict the existence
of locally stable solutions of the self-consistent equations consisting
of line defects (solitons) in the antiferromagnetic order parameter.
These
solutions provide low energy modes for the holes which are trapped in
the solitons. For small ${\rm U/t}$, the ``charged'' lines of defects are
aligned along the crystal axes,
while for large ${\rm U/t}$ they align
along the diagonal (1,1) directions (contrary to experiments).
Then, weak coupling
Hartree-Fock calculations seem in agreement with experiments. A
``spiral'' state has been also discussed by Shraiman and Siggia (1988b),
which has incommensurate correlations produced by the movement of the
holes. Furukawa and Imada (1992) argue that perturbation theory can
explain the presence of incommensurability in the Hubbard model.
Tanamoto, Kohno, and Fukuyama (1992, 1993a, 1993b) have developed a mean-field
for the
${\rm t-J}$ model that reproduces the experimental results.

While Hartree-Fock results are suggestive, it is important to verify if they
correspond to the solution of minimum energy of the models under study using
more powerful and unbiased techniques. In Fig.35, results are shown for the one
band Hubbard model using Quantum Monte Carlo as discussed by
Moreo et al. (1990a).
The static spin structure
factor $S({\bf Q})$ is presented for several fermionic densities
(with $\langle n \rangle = 1$ corresponding to half-filling)
on a $8 \times 8$
cluster, ${\rm U/t=4}$ and ${\rm T=t/6}$. The Monte Carlo results suggest that
commensurate antiferromagnetism is suppressed when the fermionic
density is reduced away from half-filling, and that
weak short-range incommensurate correlations develop as the
doping increases.
Similar results were also obtained independently by Imada and
Hatsugai (1989), and by Furukawa and
Imada (1992), using a Projector Monte Carlo method.
Their results for $S({\bf Q})$ are shown in Fig.36a which was obtained at
${\rm U/t=4}$,
and $\langle n \rangle = 0.82$. The incommensurability is clearly observed
in the figure. It would be desirable to obtain results at
larger couplings ${\rm U/t}$, but
Monte Carlo methods have numerical instabilities in this regime.
Unfortunately,
${\rm t-J}$ model calculations
on large clusters are not available either since a good Monte Carlo
technique
for this model has not been developed. However, the Lanczos
algorithm can be applied to this model. Fig.36b shows $S({\bf Q})$
calculated by Moreo et al. (1990b) on
a $4 \times 4$ cluster at zero temperature, and at a realistic value of the
coupling
${\rm J/t=0.2}$.
Again a shift
from momentum $(\pi,\pi)$ is observed as a function of the number of
holes. Then, we conclude that the presence of incommensurate (short-range)
correlations seems
a general feature of several models of correlated electrons, although we
still do not have a clear intuitive understanding of its origin.


\vskip 1.cm

\centerline{\bf B. Optical Conductivity:}

\vskip 0.5cm

Since the early days of high-Tc superconductors, attempts
have been made to identify the superconducting gap and other special features
responsible for the pairing mechanism, using the same infrared
spectroscopy techniques
which were successful in the analysis of classical low temperature
superconductors.
However, also from the beginning it became clear that the infrared properties
of the new superconductors are extremely complicated. Actually, it is not
even clear if the superconducting gap has been properly
identified using these techniques. This section consists of two parts. First,
we will attempt to summarize the vast experimental
literature on infrared experiments closely following the excellent
reviews on this subject by
Timusk and Tanner (1989), and
Tanner and Timusk (1992).
Since different plots will be presented in different units, it is convenient to
remember that
$$
1eV = 8063 cm^{-1} = 11,600 K.
\tag {units}
$$
The main conclusions of this first part will be the following: {\bf i)} upon
doping
weight appears inside the charge-transfer gap of the undoped compounds,
defining the so-called ``midinfrared'' band;
and {\bf ii)} the conductivity decays as $\sim 1/\omega$ at low energies
instead of the
behavior expected for
free carriers namely $1/\omega^2$.
The origin of the midinfrared band is still unclear. Some experimental
authors attribute
it in part to trapped holes near dopant atoms, while others claim that the
${\rm Cu-O}$ chains of the $\ybco$ family originate most of this
weight, at least for these particular compounds.

After the experimental results are summarized, we will describe the
present status of numerical studies of the optical conductivity in electronic
models that may be of relevance for high-Tc materials. It is
claimed that after a considerable effort by several groups,
a consistent picture is emerging which suggests that the midinfrared
band may be caused, at least in part, by the spin excitations that
surround (i.e. renormalize) the
hole carriers. The anomalous $1/\omega$ behavior can be mimicked
by a perverse
combination of the oscillator strength at the far-infrared
$\omega \sim 0$, and the spectral weight located
at midinfrared energies, as some examples show. Finally, it is concluded
that the unusual
behavior of $\sigma(\omega)$ may not be correlated with the
presence of superconductivity in a given material or model,
and should be a general feature of strongly interacting electronic
systems.

\vfill
\eject

\vskip 1.0cm

\noindent{\tt 1. Experimental Results}

\vskip 0.3cm

As described before in section I.A.1,
$\x214x$ is perhaps the simplest of the  cuprate superconductors since
it has
only one Cu-O plane per unit cell. Its carrier concentration can be varied over
a wide range, $0 \le x < 0.3$, allowing a systematic
study of the transitions from an antiferromagnetic insulator to a
superconductor,
and then, increasing further the doping, to an ``anomalous'' metallic state.
While the many early experimental studies of optical properties of this
material
were carried out on ceramic samples (Doll et al., 1988; and references
therein),
good quality crystals
(large and homogeneously doped)
have been recently grown. In particular,
Uchida et al. (1991) have measured the
reflectance of $\x214x$ for several dopings between ${\rm x=0}$ and ${\rm
x=0.34}$,
at room temperature (see also Shimada et al., 1992a).
The optical conductivity  can be obtained after a
Kramers-Kroning analysis of the reflectance.
Their main result for the real part of the optical conductivity
$\sigma_1(\omega)$,
is shown in Fig.37a which was obtained at 300K.
The undoped crystal in the figure (${\rm x=0}$) shows a negligible conductivity
below 1eV, in rough agreement with the expectation of
a charge-transfer gap of about 2eV for this insulating compound.
With hole doping, the intensity above the gap is
reduced and new features appear in the region around and below 1.5eV,  i.e.
a transfer of weight from above the gap to lower energies seems to occur.
In the lightly doped case (say, ${\rm x=0.02}$) a clear feature
centered about $\omega \sim 0.5$ eV appears. This is the famous
``mid-infrared'' (MIR) band, that has been observed
in several other cuprate superconductors,
and was discovered in the early days of high-Tc in polycrystaline
samples (Orenstein et al., 1987; Herr et al., 1987).
At this doping, the far-infrared signal near $\omega \sim 0$
is small and difficult to see in the graph,
suggesting that the MIR and free carrier absorptions are independent features
in
this material.
Increasing further the doping, the far infrared conductivity quickly grows, and
at
dopings larger than ${\rm x \sim 0.20}$, it entirely masks the MIR band which
does not
change with doping as rapidly.\footto{7} At small frequencies the
conductivity decays much more slowly than the Drude-type
$1/\omega^2$ behavior expected for free carriers.

What happens in other materials?
We have already discussed that
electron-doped materials, like $\xnco$, are
structurally very similar to $\x214x$. It has been found that
their optical conductivities
are also qualitatively similar (see Fig.37b which is
taken from Uchida et al., 1991). A MIR
structure is clearly present near
$\omega \sim 0.4$eV for the ${\rm x=0.10}$ sample. Increasing
further the doping,
the MIR absorption merges with the low frequency free carrier absorption,
as it occurs in the lanthanum compounds.
Again, below 1eV
the results for $\sigma_1(\omega)$ cannot be fit with a free
carrier law $1/\omega^2$. Other compounds of the same family
can be obtained by replacing ${\rm Nd}$ by ${\rm Pr}$. The optical properties
of ${\rm Pr_{2-x} Ce_x Cu O_4}$ have been investigated (Cooper et al., 1990)
and the reported results are very similar to those of $\xnco$.

Now, let us consider $\ybco$. This material has
been widely analyzed experimentally since good crystals with a very sharp
superconducting
transition at Tc can be prepared.
However, it presents
complications that are absent in simpler compounds like
$\x214x$. In particular, $\ybco$ has ${\rm Cu-O}$ chains, that
contribute substantially
to $\sigma_1(\omega)$, and
two ${\rm Cu-O}$ planes per unit cell.
These features have
to be taken into account when a theoretical description of this particular
material is attempted.
Fig.38a shows $\sigma_1(\omega)$
for the stoichiometric compound ${\rm Y Ba_2 Cu_3 O_7}$ as measured
by Kamar\'as et al.(1990) at several temperatures.
The far infrared region depends
strongly on temperature especially below Tc, while the region above
$\sim 800 cm^{-1}$ is temperature independent. There is a clear minimum
at $\sim 400 - 500 cm^{-1}$ that can be easily seen
at low temperatures but also exists above Tc. It has been argued that it
may be caused
by the coupling between the midinfrared carriers and phonons
(Timusk and Tanner, 1989).
Fig.38b shows the conductivity at a fixed temperature of 100K, for
different oxygen compositions (and, thus, for samples with
different values of Tc) taken from Orenstein et al. (1990).
The 30K material has a  weak
midinfrared, and an onset of charge transfer absorption
at $\omega \sim 12,000 cm^{-1}$. Increasing the content of oxygen,
the MIR absorption increases and shifts to lower energies. The minimum
at $\sim 400 cm^{-1}$ mentioned before can also be seen in these crystals.

What is the behavior of the optical conductivity in other cuprate
superconductors? Let us now consider the so-called
BISCCO family. $\sigma_1(\omega)$
for $\s2212$ has been reported by Romero et al. (1993)
using transmission measurements and a Kramers Kroning analysis.
This is a two Cu-O layer material similar to YBCO,
but without the one dimensional chains. The results are shown in Fig.39a.
In this material it was observed that: 1) above $\sim 300 cm^{-1}$ the
decrease in the conductivity above Tc is closer to $1/\omega$ than the
$1/\omega^2$ expected from free carriers, and 2) at high frequencies the
temperature dependence is much weaker than at small frequencies.
Then, this compound also seems to display non-Drude behavior as the
other high-Tc materials.
It is interesting to remark that compounds of the same Bi family, but
which are $not$ high-Tc superconductors, like ${\rm Bi_2 Sr_2 Cu O_6}$
(Tc less than 5K), present also a behavior qualitatively
similar to that of BISCCO above Tc (Fig.39b). Then, from this study it can be
inferred that the non-Drude optical behavior is mainly
produced by holes in the ${\rm CuO_2}$ planes, and it is not necessarily
related with the appearance of a superconducting transition.
Several other materials present an analogous behavior. For example,
the compound
${\rm Tl_2 Ba_2 Ca Cu_2 O_8}$
has been analyzed (Foster et al., 1990; Shimada et al., 1992b) as well
as ${\rm Pb_2 Sr_2 L Cu_3 O_8}$ with L being a rare earth (Reedyk et
al., 1992).
Both present a robust MIR band absorption.

Photoinduced absorption experiments have also shown the presence
of a midinfrared band in the cuprate superconductors. Kim et
al. (1987) reported results for $\x214x$, while $\nco$ was
analyzed by Yu et al. (1992), and $\ybco$ by Nieva et al. (1993).
In all of them, the presence of spectral weight in the MIR region was observed.
Then, there is a qualitative agreement between experiments
carried out by adding holes chemically, or by photoexcitation.

How can we understand the presence of the MIR band in these
compounds? A detailed study by Thomas (1991) and Thomas et
al. (1992) on some insulating materials with vacancies suggests
that this band may be caused in part by trapped holes near dopant sites
(see Fig.40a). Let us concentrate on
the electron doped ${\rm Nd_{2} Cu O_{4-y}}$
compound. Since no Ce atoms have been added, there are no carriers
and the system is insulating. However, the removal of oxygen induces
vacancies which seem to produce an interesting structure in
$\sigma_1(\omega)$. There are two clear broad peaks in the experimental
results.
Thomas et al. (1992)
argued that the lower energy peak $E_J$, may be produced by the interaction
with light
of an electron captured in a bound state by the vacancy. Its
movement costs energy because it disturbs the antiferromagnetic
background of spins in a picture very similar to that of the ``string''
excitations
described in section III.B.1. Then, the peak $E_J$ is associated to an
excited state of a trapped hole.
The second peak $E_I$, may be produced by the
ionization energy of this trapped electron (perhaps there are other
bound states between $E_J$ and $E_I$ that are difficult to resolve).
Then, this analysis
suggests that the midinfrared absorption can be partially accounted for
by bound states processes, and is not all due to free carriers
(however, see
Millis and Shraiman, 1992).
Recent work by Cooper et al. (1993) supports these
claims. However, it is not clear whether 100\% of the MIR weight
is caused by trapped holes.
In the case of the $\ybco$ compounds, it has been
argued that chains produce a substantial contribution to the MIR
band (Schlesinger et al., 1990). Actually, Cooper et
al. (1993)
have recently shown that the MIR band can no longer be resolved in $\ybco$ with
$
{\rm x = 0.6}$ and $1.0$,
once the chain contribution was subtracted (see Fig.40b).
Again, it is not clear that 100\% of the MIR band is actually caused
by the chains.

Finally,
in this short summary of experimental results for the optical
conductivity of the cuprate materials,
we will $not$ discuss the important issue of whether the superconducting
gap has been observed using infrared techniques.
The reason is
that experimentally the answer is not clear. It is believed that
due to the small coherence length of the cuprates below Tc, the
materials are in the extreme clean limit, and thus the gap feature in
the conductivity should be difficult to observe. It has also been
argued that the signal may be lost in the strong midinfrared
conductivity region. Actually, the conductivity seems finite down to
$150 cm^{-1}$, and thus no definite answer to the presence of
a superconducting gap
in the high Tc materials has been reported thus far (Tanner and Timusk, 1992).

\vskip 1.0cm

\noindent{\tt 2. Theoretical Analysis of $\sigma(\omega)$}

\vskip 0.3cm

Here, we will review the present status of some theoretical studies of
optical properties of models related with high-Tc
superconductors. In particular,
we will concentrate on
the response to an external field of an interacting system of electrons
evaluated using numerical methods.
A closed formula for the intensity of the
Drude peak in $\sigma(\omega)$ will be derived as a special
case of a more general equation.
This derivation follows closely that given by Shastry and
Sutherland (1990), and by
Scalapino, White and Zhang (1992).\footto{8}
Most of the results for $\sigma(\omega)$
discussed later in section IV.B.3,
have been obtained using computational techniques due to the
difficulty in obtaining analytical information on models of
correlated electrons when the interactions are strong.


As an example, we will
consider electrons described by the Hubbard model coupled to an external
classical vector potential $A_{\bf l}({\bf r},t)$, where ${\bf r}$
denotes a site of the two dimensional lattice, $t$ is time, and ${\bf
l}$'s are unit vectors in the lattice axes directions.
The gauge-invariant way to couple particles on a lattice with a U(1)
gauge field is by
introducing phase-factors in the kinetic energy hopping term, i.e.
$$
{\hat H} = {\hat H_0} + {\hat V} = -t \sum_{{\bf r,l},\sigma}
[ c^{\dagger}_{{\bf r}\sigma} c_{{\bf r+l}\sigma} e^{ieA_{\bf l}({\bf r},t)} +
c^{\dagger}_{{\bf r+l}\sigma} c_{{\bf r}\sigma} e^{-ieA_{\bf l}({\bf r},t)} ] +
U \sum_{\bf r} n_{{\bf r}\uparrow} n_{{\bf r}\downarrow},
\tag {ar}
$$
\noindent where we have set $\hbar = c = a =1$ ($a$ is the lattice
spacing) and the rest of the notation is standard.
The phase factors ``live'' on the links of the lattice
defined by sites ${\bf r}$ and ${\bf r+l}$.
${\hat H}_0$ is defined
as the Hubbard Hamiltonian in the absence of a vector field
but including the electron-electron interactions, while
${\hat V}$ contains the field dependence and it vanishes when $A_{\bf l}({\bf
r},t)=0$.
Expanding in powers of the electric charge $e$, it can be easily shown that
$$
{\hat V} = -e \sum_{{\bf r,l}} {\hat j}_{\bf l}({\bf r}) A_{\bf l}({\bf r},t) -
{{e^2}\over{2}} \sum_{{\bf r,l}} {\hat K}_{\bf l}({\bf r}) A_{\bf l}^2({\bf
r},t)
+ ...,
\tag {as}
$$
\noindent where the ``paramagnetic'' current density
operator in the $l$-direction is defined as
${\hat j}_{\bf l} ({\bf r}) = it \sum_{\sigma}
( c^{\dagger}_{{\bf r}\sigma} c_{{\bf r+l}\sigma}  -
c^{\dagger}_{{\bf r+l}\sigma} c_{{\bf r}\sigma}  ) $, and the
operator ${\hat K}_{\bf l}({\bf r})$ is the kinetic energy
density also in the ${\bf l}$-direction, i.e.
${\hat K}_{\bf l}({\bf r}) = -t \sum_{\sigma}
( c^{\dagger}_{{\bf r}\sigma} c_{{\bf r+l}\sigma} +
c^{\dagger}_{{\bf r+l}\sigma} c_{{\bf r}\sigma} ) $.
The linear response current is thus given by,
$$
{\hat J}_{\bf l}({\bf r},t) =
- {{\partial {\hat H}}\over{\partial A_{\bf l}({\bf r},t)}} =
e {\hat j}_{\bf l}({\bf r}) + e^2 {\hat K}_{\bf l}({\bf r}) A_{\bf l}({\bf
r},t) + ...,
\tag {at}
$$
\noindent where the first term is the paramagnetic current density, and the
second corresponds to the diamagnetic contribution.

The next step in the calculation of $\sigma(\omega)$
for the Hubbard
model is to evaluate the mean value of the total current operator
${\hat J}_{\bf l} ({\bf r},t)$, in the ground state of the Hamiltonian.
As a starting point,
let us derive the expectation value of
an arbitrary time-dependent operator ${\hat O}(t)$, in the ground
state of a given system. Following
well-known steps described in several textbooks (see for example
Fetter and Walecka, 1971),
and working at first order in the external field contained in ${\hat
V}$,  it is possible
to show that the following approximation holds,
$$\eqalign{
\langle \psi(t)| {\hat O}(t)| \psi(t) \rangle =
\langle \phi_0 | {\hat O}   | \phi_0  \rangle +
i \int^t_{-\infty} dt_1
& [ e^{iE_0 (t_1 - t)} \langle \phi_0 |
{\hat V} e^{-i{\hat H_0} (t_1 - t)} {\hat O} | \phi_0 \rangle \cr
&-e^{iE_0 (t - t_1)} \langle \phi_0 |
{\hat O} e^{-i{\hat H_0} (t - t_1)} {\hat V} | \phi_0 \rangle ] + ... . \cr}
\tag {au}
$$
\noindent In the derivation of  Eq.(\call{au})
we made explicit use of the definition of an operator
in the interaction representation i.e. ${\hat O}(t) = e^{i{\hat H}_0 t} {\hat
O} e^{-i{\hat H}_0 t}$, and
the time evolution of a state which is given by
$|\psi(t) \rangle = {\hat T} e^{-i \int^{t}_{-\infty} dt_1 {\hat
V}(t_1) } | \phi_0 \rangle $
(where ${\hat T}$ is the time-ordering operator,
and $|\phi_0 \rangle$ is the ground state
of the fully interacting system
in the $absence$ of the external field, which has an energy $E_0$).

Let us now specialize Eq.(\call{au}) to our problem, i.e. consider
${\hat O}(t) = {\hat J}_{\bf l} ({\bf r},t)$.
Defining the Fourier transformation of the vector field as
$A_{\bf l}({\bf r},t) = \int^{+\infty}_{-\infty} d\omega A_{\bf l}({\bf
r},\omega)
e^{-i\omega t}$ (with a similar definition for the transformation of the
current $\langle {\hat J}_{\bf l} ({\bf r},\omega) \rangle$), and after tedious
but straightforward algebra, it can be shown that
$$\eqalign{
\langle {\hat J}_x (&{\bf r},\omega) \rangle =
e^2 \langle {\hat K}_x \rangle
A_x({\bf r},\omega) + \cr
&ie^2 \sum_{\bf r'} \int^{\infty}_0 d\tau
\langle \phi_0 |
[ {\hat j}_x({\bf r}) e^{-i({\hat H}_0 - E_0 - \omega)\tau} {\hat j}_x({\bf
r'}) -
{\hat j}_x({\bf r'}) e^{i({\hat H}_0 - E_0 + \omega)\tau} {\hat j}_x({\bf r})]
| \phi_0 \rangle A_x({\bf r'},\omega), \cr }
\tag {av}
$$

\noindent where 1) the gauge
field has been specialized to the x-direction ($A_{\bf l}({\bf r},t) =
\delta_{{\bf l},x} A_x({\bf r},t) )$; 2) the paramagnetic current in the ground
state
without external fields is assumed to vanish
($\langle \phi_0 | {\hat j}_{\bf l}({\bf r}) | \phi_0 \rangle = 0$);
3) the change
of variables $t-t_1 \rightarrow -\tau$, $t \rightarrow t$ was
carried out; and 4) the notation $\langle {\hat K}_x \rangle
= \langle \phi_0 | {\hat K}_x({\bf
r}) | \phi_0 \rangle$ was introduced, since this mean value is site
independent in the ground state of the
Hubbard model defined on a cluster with periodic boundary
conditions. Eq.(\call{av}) can be further simplified
by working in momentum space. Defining the spatial Fourier transformed
as $A_x({\bf r},\omega)$$= {{1}\over{N}} \sum_{\bf q}
A_x({\bf q},\omega)$$e^{i {\bf q}.{\bf r}}$, and applying the
operatorial identity
$\int^{\infty}_0 dx e^{i{\hat a}x - \epsilon x} = i/({\hat a}+i\epsilon)
$ (where $\epsilon$ is a constant, and ${\hat a}$ an arbitrary
operator), we arrive to a $general$ equation for the response of the total
current
to an arbitrary, but small, vector field in the x-direction,
$$\eqalign{
\langle {\hat J}_x({\bf q},\omega) \rangle
= e^2 \langle {\hat K}_x \rangle A_x({\bf q},\omega)+ e^2[
&{{1}\over{N}} \langle \phi_0 | {\hat j}_x(-{\bf q})
{{1}\over{{\hat H}_0 - E_0 + \omega
+ i\epsilon}} {\hat j}_x({\bf q}) | \phi_0 \rangle  + \cr
&{{1}\over{N}} \langle \phi_0 | {\hat j}_x({\bf q})
{{1}\over{{\hat H}_0 - E_0 - \omega
- i\epsilon}} {\hat j}_x(-{\bf q}) | \phi_0 \rangle ]  A_x({\bf q},\omega).
\cr}
\tag {ax}
$$
\noindent $\epsilon$ is a small parameter introduced to regularize the
poles that will appear at particular values of the frequency.

Now, let us study the special case of the response to an electric
field defined by
$A_x({\bf q}=0,\omega) =
E_x({\bf q}=0,\omega)/(i\omega - \delta)$, where $\delta$ is a
small number. We consider zero momentum since we are interested in
a $uniform$ electric field (a pure magnetic field will be analyzed in
section IV.E.4 when superfluidity is discussed). In linear response theory
the conductivity is defined through the relation
$\langle {\hat J}_x({\bf q}=0,\omega) \rangle = \sigma_{xx}(\omega)
E_x({\bf q}=0,\omega)$.
Using Eq.(\call{ax}) it can be shown that the $real$ part of the
conductivity for $\omega >0$ is equal to,
$$
\sigma_1(\omega) = Re \sigma_{xx}(\omega) =  D \delta(\omega) +
{{e^2 \pi }\over{N}} \sum_{n \neq 0}
{{| \langle \phi_0 | {\hat j}_x | \phi_n \rangle |^2}\over{E_n - E_0}}
\delta(\omega - (E_n - E_0)).
\tag {ay}
$$
\noindent where the so-called ``Drude weight'' $D$ is given by,
$$
{{D}\over{2 \pi e^2}} = {{\langle - {\hat T} \rangle} \over{4N}}
- {{1}\over{N}}
\sum_{n \neq 0}
{{|\langle \phi_0 | {\hat j}_x({ q_x}=0,q_y=0) | \phi_n \rangle |^2}\over{E_n -
E_0}},
\tag {az}
$$
\noindent and $\langle {\hat T} \rangle$ (=$2N\langle {\hat K}_x
\rangle$) is the total kinetic energy of the problem in two dimensions.
To derive Eq.(\call{ay}),
we have introduced a complete basis $I = \sum_n | \phi_n \rangle \langle \phi_n
|$,
and we also used the
well-known identity
$1/(u + i\epsilon) = P(1/u) - i\pi \delta(u)$ valid in the limit of
small $\epsilon$,
where $P$ denotes the principal part, and $u$ is a real number.
Note the important detail that
the real part of the
conductivity contains a delta function at zero frequency, which is
produced by the ``free'' acceleration of the quasiparticles.
For a system with
periodic boundary conditions, Kohn (1964) showed that $D$ can be used as an
order parameter for metal insulator transitions.
Actually, it can be shown
that for an insulator, $D$ converges exponentially to zero with
increasing lattice size, while for a metal it converges to a
nonzero constant which
implies $zero$ resistance in the
ground state. This is not surprising since in the Hubbard model there is
no dissipative mechanism (unless disorder or a finite temperature are
introduced). A model where
$D \neq 0$ can correspond to a perfect metal or a superconductor, showing that
the
vanishing of the resistivity in the ground state
is only a necessary condition to achieve
superconductivity, but is not sufficient.
This point will be discussed in more detail in section IV.E.4.
It is also interesting to note that only $interacting$ fermions on
a $lattice$ can have a metal-insulator transition, signaled by the
vanishing of the Drude weight. In the continuum or
for lattice free fermions, $[{\hat j}_x, {\hat H} ] = 0$, and thus all
the weight of the conductivity is concentrated at zero frequency.
Actually, for noninteracting carriers in the continuum, the well known Drude
formula
is recovered from Eq.(\call{az}) namely
$$
\sigma_1(\omega) = {{n e^2 \pi}\over{m}} \delta(\omega),
\tag {ba}
$$
\noindent where $n=N_e/N$ is the density of carriers, $N_e$ is the
number of particles,
and $m = 1/(2t)$ their mass. In deriving Eq.(\call{ba}) we have assumed
a $low$ density of particles, each carrying an energy approximately
given by $(-4t)$ as it occurs in the noninteracting Hubbard model at the bottom
of the band.

By integrating in $\omega$ both terms
of Eq.(\call{ay}), we can easily arrive
to the well-known sum-rule (Maldague, 1977) relating the total weight of
$\sigma_1(\omega)$, with the mean value of the kinetic energy in the
ground state,\footto{9}
$$
\int^{\infty}_0 d\omega \sigma_1(\omega) = {{\pi e^2}\over{4N}}
\langle - {\hat T} \rangle.
\tag {bb}
$$
\noindent At a given coupling, $\langle - {\hat T} \rangle$ is in principle a
function of the fermionic density, but near half-filling it  changes
smoothly with $\langle n \rangle$. Then, in several cases it is
a good approximation to assume that the spectral weight in
$\sigma(\omega)$ is conserved upon doping, and thus it can only be
redistributed (however, note that the larger the coupling ${\rm U/t}$,
the worse is this approximation as shown below in Fig.44a).


\vskip 1cm

\noindent{\tt 3. Numerical Results}

\vskip 0.3cm

After setting up
the formalism to calculate $\sigma(\omega)$
in models of correlated electrons, it is necessary
to find a reliable technique to
evaluate the complicated matrix
elements appearing in Eq.(\call{ay}).
In this review we will mainly describe calculations carried out with
the help of computers, since they can provide unbiased and fairly accurate
estimates of several physical quantities.  Unfortunately, the
computational analysis of $\sigma_1(\omega)$
is by no means simple.
Quantum Monte Carlo methods cannot handle the evaluation of
dynamical $\omega$-dependent quantities, since
in this technique calculations are carried out in imaginary time.
Analytic continuations from imaginary to real time have been
attempted but this approach is not yet fully developed for two dimensional
problems (see Silver et al., 1990; Jarrell et al., 1991).
Then, the Lanczos method applied to small clusters is
one of the few available tools to calculate
the optical conductivity of correlated electrons.
Several groups around the world have actively worked on $\sigma(\omega)$ using
this
technique. To apply the Lanczos formalism to the evaluation of
dynamical
quantities as discussed in section II.A.2, it is convenient to rewrite
Eq.(\call{ay}) as,
$$
Re\sigma_{xx}(\omega) = D \delta(\omega)
+ {{e^2}\over{\omega N}} Im[ \langle \phi_0 | {\hat j}_x
{{1}\over{{\hat H}_0 - E_0 - \omega - i\epsilon}} {\hat j}_x | \phi_0 \rangle
].
\tag {bc}
$$

\noindent which is precisely of the form of Eq.(\call{f}).
The second term on the r.h.s. (sometimes called the
``incoherent'' or ``regular'' part of the conductivity)
can be obtained using the Lanczos formalism of section II.A.2.
Typically a couple hundred Lanczos iterations
provides the incoherent part with high accuracy.
Thus, it is not
necessary to explicitly obtain all the excited states to calculate the
regular part of Eq.(\call{ay}), as would have been naively required.
The Drude contribution is calculated using
Eq.(\call{az}), since with
the Lanczos approach it is possible to obtain the mean value of
the kinetic energy operator also very accurately.
The effects
of electronic interactions are fully taken into account with this technique,
and
the approach works equally well for any coupling and doping
fraction. Of course, the intrinsic
problem of this method is the constraint to working on
relatively small clusters, since in the calculations vectors of size the
dimension of the Hilbert space of the cluster need to be used.
However, in recent years the
availability of supercomputers with large amounts of memory, like the
Cray-2, have allowed calculations in cluster sizes that are expected to
capture at least qualitatively the physics of models of correlated
electrons.

In Fig.41, $\sigma_1(\omega)$ is shown
for the one band Hubbard model at ${\rm U/t = 10}$,
evaluated on a $4\times 4$ cluster. The results are parametric
with the hole doping fraction in the interval
$0.0 \le {\rm x} \le 0.375$. The results of Fig.41 have recently been
discussed in the literature (Dagotto et
al., 1992a) but in this reference they were presented with
a very high resolution $\epsilon = 0.01t$ to
distinguish between the individual $\delta$-functions in the
spectrum. On the contrary, here in Fig.41, we give
to the $\delta$-functions a $large$ width $\epsilon = t$ to
simulate the several effects not considered in our idealized
Hamiltonian that contribute to the broadening of the peaks.
The results are very interesting (Fig.41).
Selecting appropriately
the coupling constant in the Hubbard Hamiltonian, and without
providing additional information,
a $\sigma_1(\omega)$ that resembles the experimental results for
hole and electron doped materials is obtained (compare Fig.41 against
Fig.37a,b). At half
filling ($x=0$), the weight is accumulated above a gap which is
about $6t$ for this coupling. If $t$ takes the value suggested in some
calculations (Hybertsen et al., 1990; Bacci et al., 1991)
i.e. $t \sim 0.3 - 0.4 eV$, then the gap is similar
to that observed in the high-Tc materials, namely about 2eV. As
explained before, with
the Hubbard model we can $mimic$ the charge-transfer gap of the real
materials by means of the Hubbard gap.
The weight in $\sigma_1(\omega)$ above the gap is produced by charge
excitations, and it is basically related to the upper Hubbard band
of the model.
The small weight below the gap at $x=0$ is produced by
the ``tails'' of the $\delta$-functions above the gap.

The more interesting results occur upon
doping, since a redistribution of the spectral
weight takes place. Since the kinetic energy in the sum-rule
Eq.(\call{bb}) is not strongly doping dependent, this basically
amounts to a transfer of weight from the charge excitations
band down to lower energies.
Consider ${\rm x=0.125}$, which corresponds to two holes on the $4 \times 4$
cluster.
In Fig.41 it is shown
that in the infrared region below the gap,
two main features dominate: i)
a sharp peak at $\omega=0$ which is precisely the Drude peak
with damping, and ii) considerable
weight centered about the middle of the gap
that hereinafter we will call the ``midinfrared'' (MIR) band in analogy with
the weight  observed experimentally
located inside the charge-transfer gap upon doping.
Increasing further the hole doping, it is observed that the Drude peak
grows rapidly with ${\rm x}$, while the weight at the midinfrared
band is only weakly doping dependent. In the scale of
the plot, results for ${\rm x=0.25}$ and ${\rm 0.375}$ are virtually
identical. Even the appearance of what
Uchida et al. (1991) called an ``isosbestic'' point (the point
around $\omega \sim {\rm 5t}$
where conductivities for different densities cross) is neatly
reproduced in the figure! It is important to remark that
results as those shown in Fig.41 can be obtained if ${\rm U/t}$ is selected in
the intermediate region of couplings, namely when ${\rm U}$ is of the order of
the bare bandwidth $8t$. If the coupling ${\rm U/t}$ is larger, then it
can be shown
that between the MIR region and the charge excitations there is an
empty region with no spectral weight. On the other hand,
if ${\rm U/t}$ is too small then, upon doping, the MIR band and charge
excitations
merge and they are difficult to distinguish. Note that the same ``optimal''
region in parameter space necessary to mimic experiments on cuprates also
appears in studies of photoemission (see section IV.C) and for the
magnetic susceptibility (section IV.A.1).

What is the origin of the MIR band in these numerical studies?
We know that this band also appears in
the ${\rm t-J}$ model, and thus it is not related to charge excitations.
It is tempting to associate the MIR band with
the considerable amount of spectral weight found above the quasiparticle
peak in the study of the spectral function of a hole
$A({\bf p},\omega)$ (sections III.B.4, and IV.C),
since both appear at similar
energies. That weight was caused by spin
fluctuations around the hole, i.e. the hole is a dressed quasiparticle
that carries a ``bag'' of reduced antiferromagnetism in its
neighborhood. At large ${\rm U/t}$, the properties of this spin polaron are
dominated by the string excitations (section III.B.1).
To support in part these claims note that
the presence of the MIR band is a two dimensional effect. Actually,
$\sigma_1(\omega)$ has also been calculated
on a 16-site ${\rm t-J}$ model chain (Stephan
and Horsch, 1990). $\sigma_1(\omega)$
has a robust Drude peak but, contrary to its two dimensional counterpart,
negligible
weight at finite frequencies. This result supports the previous interpretation
that the dressing of the hole by spin excitations is a key ingredient
in the explanation of the origin of the MIR band. In 1D, spin and
charge separation takes place (there are no ``string'' excitations
in one dimension), while in two dimensions the cloud of distorted
antiferromagnetic background seems to follow the holes.
This example also shows that it is dangerous to
naively relate 1D and 2D results.

To complete the analysis of the numerical results,
$\sigma_1(\omega)$ of Fig.41 with $x=0.25$ and $0.375$ in the range
$1t \le \omega \le 5t$ was fit with a power-law,
$\sigma_1(\omega) \sim A/\omega^{\alpha}$. It is remarkable that
the best fit is obtained for $\alpha \sim (1.00 \pm 0.05)$,
again in excellent agreement with experiments which consistently
suggest a $1/\omega$ decay of the conductivity at intermediate
energies. However, for this finite cluster we
know $exactly$ that there is a Drude peak that has to
decay as $1/\omega^2$ at small enough frequencies
(in practice $ \omega < 0.1$ $eV$ or less).
This peak carries
a considerable amount of the total spectral weight at $x=0.25$, and
it has been given the same large damping $\epsilon = t$ as the rest
of the $\delta$-functions. Then, the observed $1/\omega$ decay of
the numerical results is caused by a perverse combination of the
oscillator strength of the free carriers, and that of the MIR band.
This may be a possible explanation for the puzzling experimental results
observed in the cuprates.

At this point, it is convenient to clarify that
the purpose of this theoretical
exercise of analyzing in detail $\sigma(\omega)$ of Fig.41
is $not$ to claim that the Hubbard model
contains all the key ingredients to describe the superconductors. We only
want to remark that once a calculation can be performed under some
controlled approximations (in this case with the help of
computational techniques), and if enough damping is provided
to the otherwise sharp $\delta$-functions, then some of
the intriguing  ``anomalous'' features of the experimentally observed
optical conductivity can be qualitatively reproduced
using models of strongly
correlated electrons. To produce a band at mid-infrared energies with
electronic models, we only seem to need strongly dressed quasiparticles, and
thus
its presence may well be a generic feature of several theories.
Of course, the reader should have in mind that other processes may
substantially
contribute to the MIR band observed experimentally. As mentioned
before studies by several groups have shown that holes trapped
near dopant atoms, as well as chain contributions in the YBCO
family, can account for a large percentage of this weight. Thus,
we expect that both these effects plus that created by the
heavily dressed quasiparticles, will be operative in the cuprates.
It would be quite difficult to distinguish among them experimentally.
Note also that the relation between these unusual features
and the superconducting mechanism remains obscure, since measurements of
pairing correlations in the same cluster that presents a robust MIR band
do not show signals of superconductivity (as will be discussed in more
detail in section IV.E).

The MIR band in
$\sigma_1(\omega)$
was first observed numerically in the ${\rm t-J}$ model
independently by
Sega and Prelovsek (1990);
Moreo and Dagotto (1990); Stephan and Horsch (1990);
and Chen and Sch\"uttler (1991). For additional information see
also Rice and Zhang (1989).
Fig.42 shows some of those results which were obtained
on $4 \times 4$ clusters with one hole. In both cases
considerable
weight is observed at intermediate energies. It is interesting to
note that the results of Fig.42a were obtained using $open$ boundary
conditions (OBC). What happens with
the Drude peak for a perfect metal if the numerical study is
carried out on a finite cluster with OBC?
In this
case we do not expect to find a Drude peak at zero
frequency, since no current can propagate with OBC.
However, in the thermodynamic limit open and
periodic conditions should give the same result. The way in which this
``paradox'' is resolved is by the appearance of a Drude precursor peak
at very small frequencies when open boundary conditions are used as
shown in the inset of Fig.42a. The
position of this peak converges to zero increasing the size of the
cluster as was discussed by
Moreo and Dagotto (1990); and Fye et al., (1991).
Fig.42b shows $\sigma_1(\omega)$ obtained using periodic
boundary conditions instead. The Drude peak is not shown, and the large
weight between 0.5t and 2.0t is the MIR band.\footto{10}


Studies of $\sigma_1(\omega)$ using the three band Hubbard model
have also been carried out. Wagner, Hanke and Scalapino (1991)
obtained the result shown in Fig.43a, using a cluster with four copper
atoms and 10 oxygen atoms with periodic boundary conditions.
The structure at A indicates charge-transfer
excitations. Note the appearance of the spectral weight B at intermediate
energies upon doping. In this plot the Drude peak is not shown
explicitly. Results for the three band model have also been reported by
Tohyama and Maekawa (1991).
Finally, on Fig.43b results are presented for the so-called Kondo-Heisenberg
model with the particular values of parameters shown in the caption
(Chen and Sch\"uttler, 1991).
The Drude peak is not shown, and the large peak near
$\omega \sim 2$ should be considered part of the MIR band.
Then, all the models of correlated electrons that have been studied in
two dimensions present spectral weight inside the insulator
gap of the undoped system. The existence of the ``electronic'' MIR band
seems a generic feature of these models.

To complete this study, in Figs.44 and 45 we show the
expectation value of the kinetic energy operator, as
well as the Drude weight ${\rm D}$, as a function of doping obtained
on a $4 \times 4$ cluster at several couplings, for the one band Hubbard
and ${\rm t-J}$ models (Dagotto et al., 1992b). In both cases
the Drude weight seems to increase with
the number of holes at small hole doping
(with the only exception of the noninteracting
${\rm U/t=0}$ limit in
the Hubbard model). The doping dependence is approximately
linear, and thus the finite cluster approach roughly predicts
${\rm D \sim x}$. In the other limit of small electronic density,
${\rm D \sim 1-x = \langle n \rangle}$ like for a gas of weakly interacting
electrons. Similar results were obtained in the one dimensional Hubbard
model by Fye et al., 1991,1992;
Schultz, 1990b; Zotos, Prelovsek, and Sega, 1990;
and Kawakami and Yang, 1991.
Although the behavior of both models
is roughly similar, it is surprising that in the ${\rm t-J}$ model
the results are almost ${\rm J}$-independent. This effect can be traced back
to the three-site terms left aside in the derivation of the ${\rm t-J}$ model
from the Hubbard model, as recently discussed by Stephan and Horsch (1992).
The approximate ${\rm J}$-independence of the ${\rm t-J}$ model results has
also
been addressed recently by Poilblanc et al. (1993) in
a finite size scaling analysis of the optical conductivity using
clusters of up to 26 sites.

Note that the Drude weight ${\rm D}$ has a maximum located
near quarter filling for both the ${\rm t-J}$ and large-${\rm U/t}$ models.
The maximum
slowly moves towards half-filling when the interaction strength
${\rm U/t}$ is decreased in the one band case. Its position
may be used as a rough estimator of the doping fraction
at which the carriers turn from hole-like to electron-like (Dagotto et
al., 1992b).
Fig.44a also shows some of the weaknesses of the calculation.
The Drude weight of the Hubbard model at half-filling obtained
using Eq.(\call{az}) can actually become negative on a finite system
(Moreo, private communication; Fye et al., 1991;
Stafford, Millis, and Shastry, 1991).
This unphysical result is
a finite size effect that has been studied extensively in one
dimensional rings, where it was shown that ${\rm | D |}$ converges to
zero at half-filling as expected for an insulator,
but with alternating signs depending on the number of sites
of the chain (for more details about one dimensional calculations
of the optical conductivity
see Stafford and Millis, 1993; and Giamarchi and Millis, 1992).
A comparison between variational Monte Carlo techniques and exact
diagonalization has been presented by Millis and Coppersmith (1990).
Their conclusion was that finite size effects were important. However,
note that considerably better numerical results have been produced
after the publication of that paper (see for example Dagotto et al.,
1992b), and thus their results need revision.
Finally, note that
an attempt to study numerically the optical conductivity of the
Hubbard model in a three
dimensional cubic lattice was presented by Tan and Callaway (1992).

Some recent developments in the context of studies of $\sigma(\omega)$
are worth mentioning:
1) Poilblanc (1991) and Poilblanc and Dagotto (1991)
have claimed that by introducing different boundary conditions
on the clusters (through a phase factor),
and averaging over them, allows to obtain results closer to the bulk
limit. They have also studied the dependence of the
energy levels with an external flux in two dimensions. The Drude weight
can be obtained as the second derivative of the ground state energy
with respect to that flux; 2) The Drude weight can be evaluated by
Quantum Monte Carlo techniques (Scalapino, White and Zhang, 1992)
for the Hubbard model at those densities
where the sign problem allows calculations at low
temperatures. The Drude weight obtained by Quantum Monte Carlo
is approximately 0.3 on a $8 \times 8$ cluster
at quarter filling $\langle n \rangle = 1/2$, and ${\rm U/t=4}$,
in good agreement with
the Lanczos
result shown in Fig.44b for the same parameters. Finite size effects
seem to affect only weakly the Drude weight at least at that density;
3) Recently, interesting results have been obtained by
Tikofsky, Laughlin, and Zou (1992). These authors
used the anyon superconductivity formalism to calculate the
optical conductivity and total kinetic energy of the ${\rm t-J}$ model.
They observed an excellent quantitative agreement with the exact
diagonalization results discussed in this review.

\vskip 1.5cm

Summarizing, the main conclusions in the study of the optical
conductivity are the following:

\itemitem{$\bullet$} {\bf Experiments:}
all materials with ${\rm CuO_2}$ planes, regardless of their Tc,
seem to have MIR bands,
and in several of them (specially those at its ``optimal'' compositions)
a $\sim 1/\omega$ decay of the optical conductivity is observed
beyond some threshold in energy and
doping. Possible explanations include the presence of trapped
holes near dopant atoms, and the important contributions of
the chains in $\ybco$.

\itemitem{$\bullet$}
{\bf Computational Studies:} all proposed purely electronic
models of Cu-O planes studied numerically
predict a $\sigma_1(\omega)$ qualitatively similar to
that observed experimentally (see Fig.41). The weight in the MIR
region is created by the dressing of the hole quasiparticles with
spin excitations, and it is in correspondence with the substantial
spectral weight observed in the hole spectral function $A({\bf p}, \omega)$ at
similar
energies. Although
in these studies there is a Drude peak in the spectrum, the
conductivity can decay as $1/\omega$ if enough damping is provided,
 due to the combination of the
free carriers contribution with the MIR band. These results are
observed in a regime where superconductivity was not
numerically detected in the ground state, suggesting that the
MIR feature and superconductivity may not be related processes.


\vskip 1.0cm

\centerline{\bf C. Electron Spectroscopy}

\vskip 0.5cm

In a typical photoemission spectroscopy
(PES) experiment, photons of a known energy are directed to
a sample of the material being analyzed. An electron with initial energy
$E_i$ in the sample is ejected out with a kinetic energy $E_f$.
These electrons are collected, and their energy analyzed. Using energy
conservation, the initial energy is given by $E_i = E_f +
\phi - h \nu$, where $h \nu$ is the photon energy, and $\phi$ the work function
of the material. If the photon energies used are about 1KeV, the
technique is called x-ray photoemission spectroscopy (XPS); while for
the ``inverse'' experiment, where electrons are added to the system, we
use the name inverse photoemission spectroscopy (IPES).
In the simplest approximation, the photoelectron spectrum
provides the occupied density of states of the system. However,
one of the main problems of this technique lies in its surface
sensitivity, since photoelectrons mostly come from a narrow region near the
surface (deeper electrons are strongly scattered and absorbed in the
sample). Then, care must be taken to carry out the experiment on a
ultra-high vacuum  with clean crystal surfaces. In Fig.46, we show a comparison
between a XPS experimental result obtained by Ghijsen et al. (1988)
for the charge-transfer insulator
${\rm CuO}$, which is considered to have many similarities with the
high-Tc compounds, and numerical exact diagonalization results for
a ${\rm Cu_2 O_7}$ cluster using a multiband Hubbard-like Hamiltonian
(Eskes and Sawatsky, 1991).
The agreement at the scale of eV's is excellent, showing that both the
experimental method and the study of small clusters provides
interesting information about real materials (band structure
calculations applied
to the same compound do not predict the insulator gap, nor
some of the satellite peaks, although work by Svane (1992) has shown
that the self-interaction corrected local spin density approximation is
able
to reproduce the presence of antiferromagnetism in the undoped
compounds).  However, note that the relevant scale of physics
we expect for superconductivity in the cuprates is of order meV. Thus,
considerable
more work is needed to show that
both the experimental technique, and the theoretical approach are able to
provide
useful information in this subtle regime.
Another important technique that complements the PES results is the
x-ray absorption spectroscopy (XAS). In this method, an electron in an
occupied core-level (like the oxygen 1s level) is excited by the x-ray
to an empty state above the Fermi level, and thus here the unoccupied
states of the system can be analyzed. Of course, removing a deep core
electron may induce strong local distortions of the electronic density
that may affect the final results, and thus, like for PES, care must be
taken in the analysis of the data.

What PES and XAS results should we expect to observe in the high-Tc materials?
In Fig.47 a rough scheme of the electronic band structure of a charge
transfer insulator is shown. $Assuming$ that the bands do not change
with doping (rigid band approximation), then upon hole doping a PES experiment
should
observe that the Fermi energy is smaller than for the insulator, and
thus it will be located
below the top of the valence band. On the other hand,
for an electron doped material the Fermi energy is above the bottom of
the conduction band, and thus PES should give a two peak structure.
For XAS  experiments which test the $unoccupied$ states, the situation is
basically reversed compared with PES, i.e. for hole doped samples, a
double
peak structure should be observed, while for electron doping only one peak
is expected.

In this section,
experimental and theoretical results will be discussed. They systematically
show
that the naive rigid band picture is not correct for the high-Tc
cuprates. An apparent disagreement between PES and XAS experiments that were
supposed to probe the same electronic density of states will be discussed.
The experimental summary given in this section closely follows
review articles by Allen (1991a, 1991b), and Dessau (1992). See also
Fink et al. (1993) for a review on Electron Energy-Loss and XAS spectroscopy.

\vskip 1cm

\noindent{\tt 1. Density of States (experiments)}

\vskip 0.3cm

PES experiments on single crystals and ceramic samples
for ${\rm La_{2-x} Sr_x Cu O_{4+\delta}}$ have produced
interesting, and somewhat surprising, results (Allen et al., 1990;
Shen et al., 1987; Takahashi et al., 1990; Matsuyama et al., 1989).
Based on the rigid band picture
we would have expected that the Fermi
energy $E_F$ would move into the insulator valence band as ${\rm x}$ increases.
However, metallic samples with
${\rm x=0.15}$ produce (weak) PES intensity in the region of the gap of the
insulator, i.e. $E_F$ lies in $new$ states inside the insulator gap.
On the other hand, XAS experiments on ceramic samples suggest a different
picture for
the density of states of this material (Romberg et al., 1990; Chen et
al., 1991).
By exciting core electrons into unoccupied states near the Fermi energy,
XAS provides information about the empty multibody states of the
problem. XAS experiments on ${\rm La_{2-x} Sr_x Cu O_{4+\delta}}$ show
the presence of states in the gap like in PES, but also two
peaks: one with weak intensity at low energies which is associated with
empty states at the top of the  valence band for the ${\rm Sr}$ doped system,
and a peak at higher energies with a larger intensity which is
associated with the upper Hubbard band. Then, XAS results seem
approximately in agreement
with the naive rigid band picture described before, and in disagreement
with PES experiments.

A similar discrepancy was observed in electron doped compounds. The naive
rigid band picture discussed above would predict that PES experiments
in these materials should show a two peak structure, one corresponding
to the electrons added to the system at the bottom of the conduction
band, and a larger structure related with the O2p band (or the
lower Hubbard band in a one band description of the material). However,
actual PES studies of $\xnco$ consistently show that there is a steady
growth of spectral weight at energies $inside$ the insulator
gap (Allen et al., 1990; Namatame et al., 1990; Fukuda et al., 1989;
Suzuki et al., 1990)
as if electrons added to the system
would occupy new states created in the gap.
As example, consider the
recent experimental results obtained by Anderson et al. (1993)
which are shown in Fig.48 for single crystals with
${\rm x}=0.0$, $0.10$, and $0.15$. The ${\rm x=0}$ insulator spectrum has been
positioned on the energy axis relative to those for nonzero ${\rm x}$, by
aligning the higher energy main band and satellite
features.
These results suggest that the
Fermi energy, $E_F$, does not appear to change appreciably
with ${\rm Ce}$ concentration, and in
doped metals it lies in states that fill in the gap of the
insulator.\footto{11} In other words, $E_F$ does $not$ move into states present
in the
insulator, and does $not$ jump across the gap if the doping is changed
from holes to electrons, but has roughly the same position relative to
the valence band maximum for both holes and electrons. Such results are indeed
unexpected.
In contrast to PES data, the XAS experiments for electron
doped materials seem to
suggest that $E_F$ lies near the bottom of the upper Hubbard band upon
electron doping (Alexander et al., 1991).
In these experiments, only one peak is observed which
is assigned to the upper Hubbard band. This result is consistent with
the naive rigid band picture and inconsistent with PES experiments.
The fact that $\mu$ is observed directly in photoemission but must be
inferred after data interpretation in XAS, favors the PES result on this
question. However, the greater surface sensitivity of photoemission
relative to XAS can be mentioned as a reason against PES data.

What occurs in other materials?
Results for $\s2212$
(Olson, 1989; Olson et al., 1990; Dessau et al., 1992)
and $\ybco$ (Liu et al., 1992a and 1992b) studied with angle-resolved
photoemission
suggest
that in these compounds the hole doping causes a shift of $E_F$ together
with a non-rigid band variation in the intensity of the emission.
Inverse-photoemission studies of $\s2212$ also support these claims
(Watanabe et al., 1991; Takahashi et al., 1991).
Dessau et al. (1992)
claims that $\s2212$ samples may be in a doping regime where some
aspects of the Fermi liquid description are recovered (while the
previously described
materials $\x214x$ and $\xnco$, may lie much closer to the insulating
regime, disturbing their regular behavior).
Thus, at present Bi2212 and YBCO do not seem to be as paradoxical
as the single plane compounds.

Summarizing the experimental PES and XAS results for several high-Tc
compounds, there is an important feature that is novel and puzzling:
a discrepancy seems to exist between PES and XAS experiments
regarding the behavior of the chemical potential as a function of doping, which
does not seem to move in PES results while XAS data suggest that
$\mu$ crosses the insulator gap when changing from hole to electron doping.
Actually, recent results by Anderson et al. (1993)
have shown that PES and XAS data obtained from the
$same$ sample of $\xnco$ at the same
time, still show the same discrepancy. An interesting feature
that is not controversial is the appearance of weight in the insulator
gap upon doping. This result is also in agreement with the theoretical
analysis of models of strongly correlated electrons (shown below),
and in disagreement
with a rigid band picture of the problem. However, as recently remarked
by Fujimori et al. (1992) and Fujimori (1992), the formation of these gap
states is a very
general phenomenon that appear in several strongly correlated (but
non high-Tc) compounds, and
thus it might not be essential in understanding the mechanism of
superconductivity in the cuprates. A very interesting example is presented
in Fig.49 where
the inverse PES spectra of ${\rm La_{2-x} Sr_x Ni O_4}$ is shown. This material
is a hole doped charge transfer insulator. As seen in the figure, upon
doping, new states appear in the gap. This material is not even metallic
for small $x$, and the holes may be localized.
Then, gap states may not be crucial for superconductivity but
nevertheless it is a poorly understood general phenomena of strongly
correlated materials, that should be carefully analyzed.

\vskip 1cm

\noindent{\tt 2. Density of States (theory)}

\vskip 0.3cm

Can the experimental results for the cuprates
be understood using models of strongly correlated
electrons? As shown below, the answer is that the presence of gap
states is very natural in these models, at least in some region of
parameter space. On the other hand, the chemical potential $\mu$
moves across the gap when hole doping is changed into electron doping
in $all$ the purely electronic models of high-Tc considered in the
literature (of course, without explicit impurities). Thus, the
paradoxical behavior of $\mu$ still needs a theoretical explanation.

Let us consider the PES results in more detail. In order to analyze the
presence of gap states in the one band Hubbard model, it is necessary
to carry out a reliable calculation of the density of states
$N(\omega)$. This quantity can be evaluated using Lanczos methods
following the same approach that was described in detail in section
II.A.2. We use the definition of the density of states as
$N(\omega) = \sum_{\bf p} A({\bf p},\omega )$,
where the spectral function corresponding to a given momentum
${\bf p}$, and energies $\omega$ is given by,

$$
A({\bf p}, \omega) = \sum_n | \langle \phi^{N+1}_n |
c^{\dagger}_{{\bf p},s} | \phi_0 \rangle |^2 \delta(\omega -
E^{N+1}_n + E^N_0 ), \ \ (\omega > \mu);
$$

$$
A({\bf p}, \omega) = \sum_n | \langle \phi^{N-1}_n |
c_{{\bf p},s} | \phi_0 \rangle |^2 \delta(\omega -
E^{N-1}_n + E^N_0 ), \ \  (\omega < \mu).
\tag {bd}
$$

\noindent $| \phi_0 \rangle$ is the ground state in the subspace of
N particles (with energy $E^N_0$). $| \phi_n^{N \pm 1} \rangle $ are
eigenstates in the subspace of $N \pm 1$ particles with energy
$E^{N \pm 1}_n$. The fermionic operator $c_{{\bf p},s}$
destroys a fermion with a given momentum ${\bf p}$
and spin $s$
The results for $\omega > \mu$ corresponds to IPES, while
$\omega < \mu$ with energy equal to $|\omega - \mu|$ determines the PES
spectrum, when integrated over all momenta.
The results for $N(\omega)$ are shown in Fig.50.
They were obtained on a $4 \times 4$
cluster at ${\rm U/t=8}$ and for
different fermionic densities ${\rm \langle
n \rangle}$. The
$\delta$-functions appearing in the figure have been given an
(arbitrary) width $\epsilon = 0.2t$ to plot the results.
Fig.50a corresponds to half-filling.
A clear gap exists in $N(\omega)$ which is caused by the
antiferromagnetic order (or, equivalently, by the spin-density-wave order) in
the ground state. The large dominant structures located at energy $|\omega |
\sim 2-3t$
correspond to the ``quasiparticle'' band produced when one
electron is added $(\omega > 0)$ or removed $(\omega < 0)$
from half-filling. $\mu$ is located at
$\omega=0$,
and the symmetry under a reflection $\omega \rightarrow -\omega$ is
caused by the particle-hole symmetry of the one band problem at half
filling. The rest of the structure in $N(\omega)$ at larger $|\omega|$
has the same origin as
the ``incoherent'' part of the hole spectral function discussed before in
section III.B.4, namely the state $c_{{\bf p},\sigma} | \phi_0 \rangle $
obtained by the sudden anihilation of an electron is not an
exact eigenstate of the interacting problem. Decomposed in a complete
basis of eigenstates it has a projection in virtually all of the states
with the same quantum numbers, and thus only a $fraction$ of the total
weight resides on the quasiparticle peak. The rest of the weight is
distributed at higher energies.

What occurs at finite doping? For example, consider $\langle n \rangle
= 0.875$ that corresponds to two holes on the 16 site cluster (Fig.50b).
The PES spectrum (dashed line) resembles that
obtained at half-filling but it is shifted towards smaller energies.
Naturally, its total weight is reduced since the integral of the density
of states up to the chemical potential (or $E_F$)
has to be equal to the number of particles. $\mu$
is now located near $\omega \sim -2.4t$ as shown in the
figure. Consider now the IPES spectrum. At an
energy $\omega \sim 4t$ or larger, a considerable amount of weight is located,
which
corresponds to the upper Hubbard band (that also exists in the IPES
spectrum at half-filling). The interesting new feature is that
immediately after $\mu$, a considerable amount of weight is observed
which peaks near the chemical potential, and then slowly decreases covering
the
whole original antiferromagnetic gap of the half-filled case, and extending
into
the upper Hubbard band. In this respect, the result is qualitatively
similar to that found
experimentally, namely that states appear in the
gap upon doping. However, contrary to PES experiments the chemical
potential moves to the top of the valence band upon hole doping.
Reciprocally,
for electron doping it moves to the bottom of the conduction band. This
result is obvious from the particle-hole symmetry of this model, and
thus an explicit calculation is not needed.
Like for the optical conductivity $\sigma(\omega)$
(section IV.B), it is important to remark that other non electronic
processes may well be contributing substantially to the spectral
weight in the gap.
For example, impurities can produce localized gap states.
Then, the numerical
study of purely electronic models without impurities only
$suggest$ that $part$ of the weight may have electronic origin, but
does not claim to explain entirely the experimentally observed spectrum.
Finally, and for completeness, it is useful to compare the results
obtained using the exact
Lanczos approach Fig.50a-b, with those obtained with
a recently developed technique to produce real-frequency results from the
Quantum Monte Carlo data (White, 1990; Silver et al., 1990;
Jarrell et al., 1991) using the same cluster, coupling and density,
and temperature ${\rm T = t/4}$ (which cannot be lowered further
due to the sign-problem). Results are shown in Fig.50c, reproduced from
Scalapino (1991). $N(\omega)$ at half-filling is qualitatively
fine i.e. the gap appears in the spectrum with a magnitude similar to
that of Fig.50a. The structure of the lower and upper Hubbard bands is
washed out showing that this technique cannot reproduce the fine details
of $N(\omega)$, but at least part of the physics has been captured
by the approximation. However,
the results at finite doping are a bit more problematic since in this
case there is no trace of the upper Hubbard band and the pseudogap
that can be observed in the exact result Fig.50b.
On the other hand, the appearance of IPES spectral weight immediately after
the chemical potential is correctly reproduced. In the opinion of the
author, the use of this method is somewhat risky when applied to two
dimensional problems of correlated electrons since it is difficult to
judge ``a priori'' what features are real and which ones are spurious.
But in a combination with Lanczos results for smaller systems it may
be possible to tune the algorithm for particular cases, and improve its
predictive capabilities. More work should be devoted to this problem.

In the numerical result shown in Fig.50b the origin of
the states in the gap is easy to understand (Eskes, Meinders, and
Sawatzky, 1991; Dagotto et al., 1991).
Consider a ``snapshot'' of the doped ground state
at large ${\rm U/t}$, as shown in Fig.51a. Double
occupancy is strongly suppressed. The PES
spectral weight is obtained by destroying one electron, and that process
does not cost much energy. On the other hand, the IPES spectrum must
necessarily
consist of
two parts. The new electron added to the system can either
occupy an already occupied state (with opposite spin), and thus pay a
large energy
${\rm U/t}$,
or be created in an empty site, which costs less energy. The former
corresponds to the upper Hubbard band structure, while the latter is
the origin of the gap states. Adding suddenly a new electron to an empty site
does not correspond to an eigenstate of the problem. Decomposed into a
complete basis it has a projection over several other states in the
subspace without double occupancy, and thus its spectral function has
a typical finite width of order $W \sim
8t$, like the spectral function of one added hole has in
the ${\rm t-J}$ model (section III.B.4). This width is enough to fill the
entire gap at ${\rm U/t=8}$. Needless to say, this behavior cannot be
reproduced by a ``rigid band'' approximation.

Then, the explanation for the presence of weight in the
antiferromagnetic gap is fairly simple in electronic models. Note that
such a reasoning also
predicts that at large values of ${\rm U/t}$, where the gap is larger than
the typical total width of the spectrum of one added particle (of order
$t$),
then the gap cannot be filled completely. This prediction can be easily
verified by studying the results obtained at
${\rm U/t=20}$. Effectively it was found that
the gap is not filled in this case (Dagotto et al., 1991; 1992b). In the other
limit of
small ${\rm U/t}$, where the antiferromagnetic gap is much smaller than
${\rm 8t}$, a small doping fraction will wash out the details
of the gap entirely. This result has also been verified numerically.
Then, there is a window in parameter space, roughly centered around
${\rm U \sim 8t}$ in two
dimensions,  where the experimental results are qualitatively
reproduced by a one band Hubbard model,
i.e. the gap is filled entirely but a ``pseudogap'' remains
(here defined as a region with small spectral weight). In previous
sections of this review, we also found that a coupling ${\rm U}$
in the neighborhood of $8t$ to $10t$ systematically
reproduces several features of the experiments (see sections IV.A.1 and
IV.B.3).
For completeness, in Fig.51b the density of states of the ${\rm t-J}$
model is shown at ${\rm x=0.125}$ and ${\rm J=0.4t}$ obtained on a $4 \times 4$
cluster (Dagotto et al., 1992b).
Note the similarity between this figure and the spectrum of
the one band Hubbard model at the same doping for energies below the
pseudogap. Also note that results for the attractive Hubbard model have
been presented by Dagotto et al. (1991), and (1992a). In
this case the chemical potential remains inside the superconducting gap
upon doping.

Thus far, we have observed the appearance of spectral weight in the
insulator gap using $one$ band models. What happens with the more
realistic (and complicated) three band Hamiltonian defined in section
I.C.1? To gain some
intuitive understanding of this problem
it is convenient to study first the limit of zero
hybridization (i.e. all hopping amplitudes equal to zero). Consider the
Hamiltonian of the model written in this limit as
$$
{\rm H =} (\epsilon_d - \mu) {\rm \sum_{\bf i}  n^d_{\bf i} }
        + (\epsilon_p - \mu) {\rm \sum_{\bf j}  n^p_{\bf j} } +
{\rm U_d \sum_{\bf i} n^d_{{\bf i}\uparrow} n^d_{{\bf i}\downarrow}} +
{\rm U_p \sum_{\bf j} n^p_{{\bf j}\uparrow} n^p_{{\bf j}\downarrow}},
\tag {ggg}
$$
\noindent where $p$ and $d$ denote oxygen and copper sites,
respectively, and the rest of the notation was introduced in
Eq.(\call{gg})
(we use
hole operators, i.e. the ``empty'' ${\rm Cu O_2}$ cell has
6 electrons; and we also study the special case where the nearest neighbor
density-density
interaction ${\rm U_{dp}}$  is zero). Assuming that $\epsilon_p > \epsilon_d$,
the ground state energy of one hole is $E = \epsilon_d - \mu$; and in
the large $U_d$ limit the ground state energy of two holes corresponds
to having one
hole in a copper and another in an oxygen, and it has an energy
$E= \epsilon_d + \epsilon_p - 2\mu$. Then, in order to make stable the
state of one hole we need to tune the chemical potential such that it
lies in the interval $\epsilon_d \leq \mu \leq \epsilon_p$. As an
example, let us
consider one of the extremes of the interval, i.e. $\mu = \epsilon_d$.
In that case the energy of zero and one hole are zero, the energy of
two holes (one in copper and one in oxygen) is $\Delta = \epsilon_p
- \epsilon_d$, and two holes on a copper (oxygen) ion
have an energy $U_d$ ( $U_p + 2\Delta$).

Then, in the PES and IPES spectrum of a ${\rm CuO_2}$ Hamiltonian
in the atomic limit and
in the case of one hole in the ground state, we expect to observe weight at the
energies
mentioned before, namely $0$, $\Delta$, $U_d$, and $U_p + 2 \Delta$,
all referred to the chemical potential. When the hopping amplitudes
are turned on, each of these sharp energies will acquire a width that
grows with ${\rm t_{pd}}$.
The results of a numerical exact diagonalization study carried out
by Tohyama and Maekawa (1992), on a cluster
with four copper atoms and thirteen oxygen atoms are shown in Fig.52,
for both hole and electron doping (see also Horsch et al., 1989; and
Ohta et al., 1992).
As parameters these authors
considered ${\rm U_d =8.5eV}$, ${\rm U_p = 4.1eV}$, ${\rm t_{pd} = 0.966 eV}$,
and $\Delta = 3.255eV$. Also a direct oxygen-oxygen hopping was
included ${\rm t_{pp} = 0.395 eV}$.
Let us concentrate on the
results at the top of the figure. The inverse photoemission  structure for
$\omega < 0$ corresponds to the states with no holes (in the electron-like
language, this is the upper Hubbard band corresponding
to the copper atoms i.e. it is obtained by adding one more electron
to the system). Reciprocally, at large and positive $\omega$, we can
observe the lower Hubbard band that corresponds to two holes (four
electrons) distributed such that two holes are on the same copper
site ($d^9 \rightarrow d^8$). Near $E_F$ a gap of charge transfer origin exists
(but it is
difficult to see in Fig.52 since the $\delta$-functions have
been given a finite width to present the results). The arrow indicates the
position of $\Delta$ i.e. the place where weight related with the
case of two independent holes located one in copper and the other
in oxygen should start. We clearly see a considerable
spectral weight in that regime, in nice agreement with the atomic limit.

However,  note that in addition to these features an appreciable amount
of spectral weight exists at much lower energies, making the actual gap
at half-filling be considerably smaller than $\Delta$. These states are
the Zhang and Rice singlets (Zhang and Rice, 1988)
which correspond to a spin singlet combination
between the hole at the copper with a hole at the surrounding oxygen
ions (thus, Fig. 47 is actually incomplete!). The
energy of this correlated state is reduced by the formation of such a
singlet, and, according to Zhang and Rice, in the strong coupling limit
this energy becomes $E_{singlet} = \Delta - 8 t^2_{pd} [ 1/\Delta +
1/(U_d - \Delta) ]$. For the same reason there is a triplet in the
spectrum whose energy is equal to $\Delta$ in this limit.
Upon doping, the chemical potential shifts into the Zhang-Rice band,
and states fill the insulator gap, in a manner very similar to that
observed in the one band Hubbard model (see also S\'a de Melo and Doniach,
1990; and Wagner, Hanke, and Scalapino, 1991).


We have seen that one of the unconventional features of $N(\omega)$,
i.e. the presence of
states in the gap, has a possible explanation in the context of simple
electronic models of the superconductors. However, the second
paradoxical feature, namely the behavior of the chemical potential with
doping, cannot be explained using these models. This conclusion can be
inferred from the results for $N(\omega)$ discussed before, or
in more detail by considering the behavior
of the electronic density $\langle n \rangle$,
of the one band Hubbard model as
a function of the chemical potential $\mu$, as reported by Moreo,
Dagotto,
and Scalapino (1991).
In the
regime of small and intermediate ${\rm U/t}$, Quantum Monte Carlo
techniques are able to study large clusters, and the results are shown
in Fig.53a (where also Lanczos results on a smaller cluster are shown).
It is clear that in order to change the density from hole doped
($\langle n \rangle < 1$) to electron doped ($\langle n \rangle > 1$),
the chemical potential has to cross a gap which corresponds to the
``antiferromagnetic'' gap observed in the density of states. Similar
conclusions were reached by Furukawa and
Imada (1992). A Quantum
Monte Carlo study of the three band Hubbard model by Dopf, Muramatsu,
and Hanke (1990)
on a $4 \times 4$ Cu-O cluster (16 cells) at the particular couplings
shown in Fig.53b (i.e. in the proper charge-transfer regime) arrived
to similar conclusions.
i.e. the chemical potential needs to cross a gap to change the
density from hole to electron doping. See also Scalettar (1989); and
Scalettar et al. (1991).
Thus, it is clear that purely
electronic models of strongly correlated electrons $cannot$ explain the
strange pinning of the chemical potential observed in PES experiments.
We do not think that this qualitative conclusion would change as a
function of the couplings of the model as long as we have an
antiferromagnetic gap in the one band Hubbard model (or charge
transfer gap in the three band model).
Note that no work has been carried out including phonons and disorder
to explore their influence on $\mu$.



\vskip 1cm

\itemitem{$\bullet$} Summarizing, the study of $N(\omega)$ and
$\langle n \rangle$ vs. $\mu$ shows that models of strongly correlated
electrons predict the presence of new states in the insulator gap, when
doping is added to the half-filled ground state. This is in agreement
with several experimental results for the cuprate superconductors, and
other non-superconducting materials. On the other hand, the behavior
of the chemical potential with doping observed experimentally using
PES techniques remains a mystery since all theoretical models
consistently support the notion that $\mu$ needs to cross the insulator
gap when hole doping is changed into electron doping.
The presence of phase separation in the Sr and Ce doped compounds
may be a possible explanation for this problem (Moreo, 1993c).
The solution of this paradox may well be very important for our
understanding of correlated electrons and high Tc superconductors.


\vskip 1cm

\noindent{\tt 3. Angle-Resolved Photoemission}

\vskip 0.3cm

The photoemission data discussed in the previous section
provides information about the integral over all
momenta ${\bf p}$ of the spectral function $\AAA$ of electrons ejected from the
materials in the photoemission process.
However, it is experimentally possible to obtain explicitly $\AAA$ as a
function of ${\bf p}$. This technique is called
``angle-resolved photoemission spectroscopy'' (ARPES).
For example, ARPES results have been obtained for single crystals of
${\rm Y Ba_2 Cu_3 O_{6+x}}$ at several different
dopings in the interval
$6.2 \leq x \leq 6.9$. According to Campuzano et al. (1990, 1991);
Liu et al. (1992a, 1992b); and Veal et al. (1993), several
features of the fermiology of this material are by now
established using this method. In particular, band
dispersions and a Fermi surface have been observed. These authors claim
that predictions of band theory appear to be
quite reliable near the Fermi energy $E_F$ at least
in the oxygen range $x \ge 0.5$, where the
material shows metallic behavior.
They also interpret their results as giving support to a Fermi-liquid picture
of
this particular compound, although they clarify that no general agreement has
been
reached on what theoretical framework provides the best description of
their results.
Recently King et al. (1993) have studied the electronic
structure of ${\rm Nd_{2-x} Ce_x Cu O_{4-\delta}}$ using ARPES
techniques, with ${\rm x=0.15}$ and ${\rm 0.22}$.
They conclude that a Fermi surface is observed that agrees very
well with band structure calculations, and appears to shift with electron
doping
as expected by a band filling scenario.

${\rm Bi_2 Sr_2 Ca Cu_2 O_{8+\delta}}$
has also been carefully studied using ARPES techniques. Olson et
al. (1990) (see also Dessau, 1992) have
concluded that a Fermi surface exists in this material. Actually,
a band along the $\Gamma - Y$ direction in the Bi2212 notation
was observed to cross the Fermi surface in
rough agreement with predictions from band theory calculations (see Fig.54).
These authors also found that the spectra shows correlation effects in
the form of an increased effective mass, but the essence of the single
particle band structure is retained.
It may be convenient at this point to remind the reader that
the standard notation for ${\bf p}$-points of the
Brillouin zone of a ${\rm Cu-O}$ plane, is that the $\Gamma$ point is
at the center, the $M$ point is at the corner, and the $X (Y)$ points are
midway along the edges. However, Bi2212 has a different notation with
the mapping $\Gamma \rightarrow \Gamma$, $M \rightarrow X(Y)$, and
$X(Y) \rightarrow {\bar M}$. Fig. 55a clarifies this relation.
In Fig.55b, the experimentally observed band structure along various
high symmetry directions is shown (taken from Dessau, 1992).

Olson et al. (1990) claimed that
a good fit of their Bi2212 results can be obtained using a quasiparticle
lifetime linear
in $|E-E_F|$ as predicted in the Marginal Fermi Liquid (MFL) theory of
Varma et al. (1989), instead of being
proportional to $(E- E_F)^2$ as in a Fermi Liquid (FL) theory.
Additional
support for the MFL hypothesis came from neutron scattering measurements
by Hayden et al. (1991a) (see also Aeppli, 1992a; and B\'enard, Chen and
Tremblay, 1993).
However, Liu, Anderson and Allen (1991)
have recently shown that
once the difficult issue of the background in the ARPES spectra is handled
carefully, MFL and FL fits are equally good. Systematic
studies of ARPES experiments in materials with a well known many body
ground state are necessary before extracting conclusions from these
experiments for the high Tc superconductors (see for example,
Claessen et al., 1992). Then, current ARPES experiments
have $not$ been able to solve the important issue of whether there are
quasiparticles in the cuprates (i.e. whether the quasiparticle weight
$Z$ is a finite number). Actually different theorists extract quite different
conclusions from exactly the same ARPES data! More work remains to be done on
the
experimental side to reach a consensus about the correct description of
the normal state lineshape of the cuprates.

On the theoretical side, the study of $\AAA$ has been carried out only for
simple one and three band models of correlated electrons. A
comparison between theory and experiments is difficult, even at a
qualitative level, because Bi2212 and YBCO have a complicated structure
with two ${\rm Cu O_2}$ planes close to each other plus other
bands produced by charge reservoir ions. Then, only rough
qualitative details can be theoretically addressed like the presence of
dispersive bands in the spectrum and the existence of a Fermi surface
in the models under consideration.
These two issues will be briefly discussed in this section.
Let us first consider the undoped case. In the study of
spin-density-waves at half-filling in the one band Hubbard model, we
found in section III.A that a simple mean-field approximation was enough
to describe qualitatively the physics of the model. Using this
approximation, it can be easily shown (Dagotto, Ortolani, and Scalapino, 1992)
that
the spectral weight is given by,
$$
A({\bf p},\omega)_{MF} = {{1}\over{2}}
(1 + { {\epsilon_{\bf p}}\over{E_{\bf p}} } ) \delta(\omega - E_{\bf p}
) +
{{1}\over{2}}
(1 - { {\epsilon_{\bf p}}\over{E_{\bf p}} } ) \delta(\omega + E_{\bf p} ),
\tag {bf}
$$

\noindent where $E_{\bf p} = ( \epsilon_{\bf p}^2 + \Delta_{SDW}^2 )^{1/2}$,
$\epsilon_{\bf p} = {\rm -2t (cosp_x + cosp_y)}$, and the
spin-density-wave
gap $\Delta_{SDW}$ is given by the solution of the equation
$1 = {{U}\over{N}} \sum_{\bf p} {{1}\over{2 E_{\bf p}}}$. The mean-field
results are shown in Fig.56 for two different momenta, and at a coupling
${\rm U/t=8}$. At ${\bf p} =
(0,\pi)$ or $(\pi/2,\pi/2)$, the spectral function has sharp peaks, a
gap
in between,\footto{12} and it is symmetric around $\omega = 0$ (Fig.56a).
On the other hand, Fig.56b corresponds to zero momentum i.e. in the PES
spectrum an electron well below the Fermi
surface is destroyed. As expected, considerable spectral weight is observed in
the PES
spectrum, while that corresponding to the IPES is small. The situation is
reversed if a momentum ${\bf p} = (\pi,\pi)$ is used (not shown in the figure).
These mean-field results are in excellent agreement with numerical
results obtained on $4 \times 4$ clusters using exact diagonalization
methods, which are shown in Fig.56c,d (taken from Dagotto, Ortolani, and
Scalapino,
1992. See also Leung et al., 1992; Feng and White, 1992).
The structure away from the
dominant peaks in the numerical results shows that an important
percentage of the spectral weight is not in the quasiparticle peak
as was explained in section III.B.4
(in other words, in these models $Z < 1$ as discussed before).
The density of states
$N(\omega)$, can be obtained by summing over ${\bf p}$.

Away from half-filling, there is no universally accepted mean-field
approximation to describe models of correlated electrons. Then, we turn
to computational studies for some guidance. In Fig.57, $\AAA$ is shown
for the one band Hubbard model at ${\rm U/t=8}$, and $\langle n \rangle
= 0.875$. The chemical potential in this figure is located approximately at
$\mu = -2.4t$. Note the presence of a dominant peak
for ${\bf p} = (\pi,0)$ located right after the chemical potential on
the
IPES side of the spectrum (Fig.57a). This peak is followed by spectral weight
that
fills the original AF gap. At higher energies the remnants of the upper
Hubbard band can be observed. The effect of doping is to remove weight
from the lower and upper Hubbard bands, and create states in the gap as
we discussed in the previous section.
Moving in momentum space away from the non-interacting Fermi surface, in the
direction of the $\Gamma$ point, the
PES weight increases, while IPES weight decreases. On the contrary, increasing
the
momentum towards the $(\pi,\pi)$ point, the situation is reversed.
For momentum
${\bf p} = (\pi/2, \pi/2)$, the dominant peak seems split and thus
this momentum may be close to the new Fermi surface of the
doped system (Fig.57d). Coming back to the comparison between theory and
experiment, we can conclude that the existence of dispersive
features are observed both in ARPES results, and in studies of
simple models of correlated electrons.
Similar conclusions have been
obtained in the ${\rm t-J}$ model (Stephan and Horsch, 1991;
Dagotto et al., 1992b), and the three band Hubbard model (Dopf et al., 1992).
A more quantitative comparison
would be very difficult with the numerical methods currently available to
study these models.

\vskip 1cm

\noindent{\tt 4. Fermi surface in models of correlated electrons}

\vskip 0.3cm

The study of $\langle n_{\bf p} \rangle$ in the one band Hubbard model
using Quantum Monte Carlo techniques (Moreo et al., 1990a)
shows that working on a $16 \times 16$ cluster,
at doping $\langle n \rangle = 0.87$ and coupling ${\rm U/t=4}$,
the locus of points where $\langle n_{\bf p} \rangle =
0.5$ is close to that of a noninteracting system at the same filling.
These results are shown in Fig.58a. Although this simulation was carried
out at a finite temperature (due to sign problems), the conclusions are
similar to those reached by exact diagonalization studies on
$4 \times 4$ clusters (Dagotto, Ortolani, and Scalapino, 1992) at zero
temperature which are
schematically shown in Fig.58b (based on the results shown in Fig.57).
The ${\bf p} = (\pi,0), (0,\pi)$ states
do not seem populated while ${\bf p} = (\pm \pi/2, \pm \pi/2)$ are close to the
Fermi surface.
Studies of the ${\rm t-J}$ model using exact diagonalization techniques
at a doping of approximately $\sim 10 \%$ holes on a 20 site lattice, are
consistent
with a large electronic Fermi surface (see Fig.58c taken from Stephan and
Horsch, 1991).
High temperature expansion calculations arrived to similar conclusions
(Singh and Glenister, 1992b) i.e. the presence of a Fermi surface in
these models (note that a Fermi surface does not inevitably imply a
Fermi liquid as remarked by Anderson, 1990a).
A Lanczos study of
$\langle n _{\bf p} \rangle$ (Ding, 1992)
has produced a
Fermi surface very similar to that observed in Hubbard model
calculations, showing that holes doped into an antiferromagnet may
actually
prefer to be located in the vicinity of momenta ${\bf p} = (0,\pi),
(\pi,0)$ rather than at the ${\bf p} = (\pm \pi/2,
\pm \pi/2)$ (Fig. 58d). A similar ``non-standard''
result was discussed by Poilblanc and Dagotto (1990). The claim is based
on the following argument: if indeed two holes in an antiferromagnet form a
d-wave state (as argued before in this  review),
then the contribution of hole states at ${\bf p} = (\pm \pi/2,
\pm \pi/2)$ cancells since $f(k) = cosk_x - cosk_y$ vanishes, while ${\bf p} =
(\pm 0,
\pm \pi)$ makes $f(k) = 2$ maximum.
Note that all these results are in contradiction with
the picture that would have emerged from studies of single holes in
antiferromagnets. In this case, holes have momentum ${\bf p} = (\pm \pi/2,
\pm \pi/2)$ in the ground state, and thus assuming a rigid band picture, hole
$pockets$ should
appear in the neighborhood of these points in momentum space.
Then, once interactions are taken into account, the
rigid band approach does not seem a good approximation
to the present problem (for a different point of view see Eder and
Wr\'obel, 1992). Unfortunately, with currently available numerical
techniques it is difficult to study very low density of holes and
temperatures, to search for indications of hole pockets.

\vskip 1.cm

\centerline{\bf D. Phase Separation}

\vskip 0.5cm

\noindent{\tt 1. Experimental results}

\vskip 0.3cm

There is considerable experimental evidence that ${\rm
La_2 Cu O_{4 + \delta}}$ has a regime where
phase separation occurs (Jorgensen et al., 1988. See also Harshman et
al., 1989).
Using La NMR, Hammel et al. (1992) have shown that
the temperature at which phase separation takes place is $T_{ps} \sim 250 K$
(see also Hammel et al., 1990; and references therein) .
The separation occurs between a phase with a stoichiometry very
close to ${\rm La_2 Cu O_4}$, and one phase rich in oxygen that becomes
superconducting at about $40K$. In Fig.59 the phase diagram
temperature-$\delta$ is shown (Reyes et al., 1993; Hammel et al., 1992).
The regime of doping fractions
between $\delta \sim 0.01$ and $\sim 0.06$ is not thermodynamically
stable i.e. if a
sample is prepared with this nominal composition, it will
spontaneously separate into two regions with the densities shown in
the figure.
The oxygen poor phase exhibits long range antiferromagnetic order,
which is reasonable since $\delta$ is very small in this phase,
and we know ${\rm La_2 Cu O_4}$ is an antiferromagnet. Then,
phase separation may well be driven by the energy gained by
forming a magnetically ordered phase (Hammel et al., 1992).
Finally note that the more widely analyzed ${\rm La_{2-x} Sr_x Cu O_4}$
compound does not seem to phase separate, although a more careful work
should be carried out for this compound.

Given these experimental results, we may try to blame
oxygen ``chemistry'' for the
existence of phase separation. The dopant oxygen atoms may
cluster together for reasons that are unrelated with the physics of
electrons in the ${\rm Cu O_2}$ planes. If this would be the case, then
phase separation would just become one more curiosity of the
cuprate superconductors. However, it has been claimed that the
mobility of the oxygens is exceptionally large
at $T_{ps}$ (Hammel et al., 1992). Also
it is quite remarkable that $T_{ps}$ is very close to the N\'eel
temperature of the material where 3D antiferromagnetism develops.
These features open the
possibility, recently addressed by Emery and Kivelson (1993), that the effect
is actually produced by the physics of electrons in the
planes, and is due to phase separation into hole-rich and hole-poor
regions. In this scenario, the excess
oxygen would simply follow the holes of the plane in their
search for the minimum energy configuration. The important
physics would be contained in the planes.
Recently, other challenging concepts have been
introduced (Cho et al., 1993; Cho et al., 1992).
In a study of ${\rm La_{2-x} Sr_x Cu
O_{4+\delta}}$, changing both ${\rm x}$ and $\delta$, it has been
claimed that the oxygen poor phase has a novel segregation
of doped holes into walls of hole-rich material separating
undoped domains. This implies that the so-called ``spin-glass''
phase of the [214] material, may actually be formed by walls
of holes, that disrupt the long-range AF order, constraining
the spin correlation length to a maximum value given by the
distance between walls. In short, the presence of phase separation
may be an important issue in understanding the physics of the cuprates,
and it certainly deserves a careful theoretical analysis.

\vskip 1cm

\noindent{\tt 2. Theoretical results}

\vskip 0.3cm

Initially,
the study of phase separation in models of strongly correlated
electrons did not receive as much attention as the
search for superconductivity. However, after some effort
it was clear that even the mere existence of a
superconducting phase in these models was a subject considerably
more subtle than expected. Then, to gain more insight about the behavior
of correlated electrons it
became important to understand the full
phase diagram of some popular models of the cuprates, including
regimes where the phenomenon of phase separation takes place.

In the ${\rm t-J}$ model, it was rapidly realized that
at large ${\rm J/t}$, the model phase separated (Riera and Young, 1989;
Emery, Kivelson and Lin, 1990; Nori, Abrahams, and Zimanyi, 1990; and
references therein).
As explained
before in section III.B.5, adding a low mobility hole
to the undoped system amounts to removing four antiferromagnetic
links, thus increasing
the energy of the system. Then, in order to minimize
the number of ``broken'' AF links, an additional hole added to the system
will prefer to be located at a distance of one lattice spacing from the
first one. In this way, the number of broken AF links is minimized. When
more (low mobility) holes are added to the system, the configuration
that minimizes the energy is the one where they form a compact cluster.
Then, at large ${\rm J/t}$ the ground state of the ${\rm t-J}$ model
separates into hole-rich and hole-poor regions.
Of course, the regime of large ${\rm J/t}$ is not physically realized
in the high-Tc compounds (section I.C.2), and this problem seems only of
academic interest. However, as mentioned before, recent work by
Emery and Kivelson (1993) proposes a link between the experimentally
observed regime of phase separation, and that found in models of
strongly correlated electrons. Then, it
becomes important to study
whether the regime of phase separation at large ${\rm J/t}$ exists also
for small values of this coupling near half-filling, i.e. in the
physically interesting regime.

Unfortunately,
the answer to this problem is still controversial. Emery, Kivelson
and Lin (1990) claimed that the boundary of phase separation in the
${\rm t-J}$ model is schematically given by the diamonds shown in Fig.60a.
Their
result is based on variational arguments, and exact diagonalization
studies of the $4 \times 4$ cluster. Note that their phase separation
boundary seems to converge to zero as the undoped limit is approached.
However, precisely this behavior
near half-filling is the subject of controversy between different
groups.
The boundary of phase separation in the same model has been
studied by Putikka, Luchini and Rice (1992), using a
high-temperature series expansion up to tenth order in $\beta= 1/T$. The
series has a finite radius of convergence, and in order to get results
at small temperatures, an
analytic continuation based on Pad\'e and integrals
approximants is needed. At $J/t \sim 0.3$ and density $\langle n \rangle \sim
0.9$,
temperatures $T \sim t/5$ were reached by this procedure.
The result of the high
temperature expansion method is also shown in Fig.60a, with a solid
line. Note that the phase
boundary touches the half-filled axis at a finite coupling ${\rm J/t \sim
1.2}$, contrary to the results of Emery, Kivelson and Lin (1990).
Other numerical studies have also addressed this issue. For
example, Dagotto et al. (1992b) carried out a similar analysis as that
discussed by Emery, Kivelson and Lin i.e. exact diagonalization
on $4\times 4$
clusters. Dagotto et al. arrived to different conclusions since
they interpreted the points of Fig.60a as indicating the region where
$binding$ of holes takes place i.e. where individual holes are
unstable towards pair formation, but not necessarily towards phase separation
which
starts at a larger coupling similar to that given by the high
temperature expansions (and it is roughly signaled by the tendency of four
holes to
form a bound state). Similar conclusions using larger clusters were
obtained by Fehske et al. (1991), and by Prelovsek and Zotos (1993)
(in the last reference it is argued that a ``striped phase'', namely
holes forming domain walls, exists before the onset of hole clustering).
If phase separation would indeed start at a coupling ${\rm J/t > 1}$,
then it may only be of academic interest. Different would be the
situation
if in the realistic regime of ${\rm J/t \sim 0.2-0.4}$, phase
separation exists near half-filling. In short, the gist of the
disagreement between numerical studies and Emery's et al. results, is
whether small ${\rm J/t}$ and large ${\rm J/t}$ physics are related.
The numerical results suggest the presence of two regimes, while Emery's
et al. claim that the simplest assumption is phase separation for all values
of the coupling near half-filling.

What occurs in the one band Hubbard model? At large ${\rm U/t}$ this
model should be qualitatively equivalent to the ${\rm t-J}$ model at
small ${\rm J/t}$. Then,
it is worth studying the issue of phase separation directly in the
Hubbard model, where Quantum Monte Carlo techniques are available
for its analysis. Unfortunately, with this method it is difficult to
study large couplings, and thus results at low temperature have been obtained
only in the
intermediate region ${\rm U/t=4}$.
In Fig.53a, results given by Moreo, Dagotto, and Scalapino (1991) were
discussed in order to compare theoretical results with those
observed experimentally in PES experiments. But the same data can be used to
address the issue of phase separation in these models. The results of Fig.53a
show a study of the density $\langle n \rangle$, as a function of
the chemical potential $\mu$. This criterion is based on the following
idea:
$if$ a discontinuity is observed in $\langle n \rangle vs. \mu$, then
the densities inside the gap are unstable i.e. if a system is initially
prepared with such nominal densities, it will evolve in time into a phase
separated state, with the two regions having the densities corresponding to
the extremes of the gap. However,
the results of Fig.53a do not show signs of discontinuity. The
simulations have been carried out on cluster with up to $8 \times 8$ sites,
and thus finite size effects are not expected to change drastically
the results. However, it is possible that the finite temperature at
which
the simulations have been carried out may have some influence on the
results (due to the sign problem it is not
possible to work at temperatures as low as those obtained at half-filling).
The temperatures reached by Quantum Monte Carlo at finite
density are approximately similar to those reached by the high temperature
expansions.
However, Lanczos (Dagotto et al., 1992b), and perturbative results (Galan and
Verg\'es, 1991) support the results of Moreo, Dagotto, and Scalapino (1991).

The conclusions of Moreo et al. are in agreement with a
projector Monte Carlo simulation carried out
by Furukawa and Imada (1992). Using this algorithm there is no finite
temperature
contamination. However, the method is based on applying
the operator $e^{-\tau H}$ to an initial Ansatz for the ground state. To
obtain ground state properties, it is necessary to study the limit
$\tau \rightarrow \infty$ which can be
obtained numerically only up to some accuracy. Then, finite temperature
errors are traded for finite $\tau$ errors in this method.
The dependence of the chemical potential with
hole doping in their simulation is shown in Fig.61, where ${\rm U/t=4}$
and clusters of up to $12 \times 12$ have been used. After the chemical
potential crosses the antiferromagnetic gap at half-filling, Furukawa and Imada
(1992) concluded that the electronic density varies continuously as
$\langle n \rangle \sim \sqrt{\mu_c - \mu}$. This result is in agreement
with Moreo et al.'s conclusions, and also with the
behavior of the one dimensional Hubbard model. Then,
Monte Carlo studies suggest that the Hubbard model at intermediate
couplings does not phase separate. The behavior
at larger couplings is unknown.

The numerical results described in this section
are considered by Emery and Kivelson (1993) to be inconclusive.
The high temperature series may not be reliable for temperatures much
below $J$ since they are based on a Pad\'e analysis which is
difficult to control. Actually Singh and
Glenister (1992) claim that various Pad\'e approximants diverge from each
other for temperatures below $J/2$, and the uncertainty increases
with decreasing $J/t$ (however, the analytic continuation techniques
used by Putikka, Luchini and Rice (1992) based on integral approximants
allowed them to reach temperatures $\sim t/5$ for the free energy. These
approximants have more analytical information incorporated than
Pad\'e extrapolations).
The Monte Carlo simulations have also been
criticized since they are carried out at a
finite temperature $T \sim t/8$; while the Lanczos calculations on small
clusters can only work at a finite number of densities, and they may miss
small discontinuities in the $\langle n \rangle$ versus $\mu$ curve.
Then, more
work is necessary to further clarify the presence of phase
separation near half-filling at small ${\rm J/t}$. The effect of the
long-range Coulomb
interactions that tend to destroy phase separation deserves also more
theoretical work.


\vfill
\eject

\vskip 1.cm

\centerline{\bf E. Superconductivity in models of strongly correlated
electrons}

\vskip .5cm

\noindent{\tt 1. Superconductivity in the one and three bands Hubbard models}

\vskip 0.3cm

After our long journey through the several physical quantities that
characterize
the normal state of the cuprates, we have finally arrived to the study
of the superconducting phase. Leaving this subject to the end of the
review is not an accident. In spite of the
considerable effort that has been devoted to the search for
superconducting long-range correlations in models of strongly interacting
electrons, no clear indications of their existence have been found in the
realistic regime of parameter space. What became clear over the years is that
the
presence of superconductivity in Hubbard-like models is a subtle issue,
much more than originally believed.
Nevertheless, recent
results discussed below still leave open the possibility for the existence of a
superconducting phase in these models.

Early results by White et al. (1989a) suggested that
in the $\dx2y2$ mode, the pair-field susceptibility of the one
band Hubbard model was enhanced
at low temperatures with respect to the uncorrelated pair-field
susceptibility (Fig.62). Small finite size effects were observed in these
studies
between clusters
with $4 \times 4$ and $8 \times 8$ sites. These results were consistent with
the attractive $\dx2y2$ channel observed in the strong coupling
limit (${\rm t-J}$ model) for the case of two holes in an
antiferromagnetic background (as was shown in section III.B.5),
and led to considerable excitement.
However, we know that in a superconducting state in the bulk limit, the
expectation
value of the pair operator $\langle \Delta \rangle$ should be nonzero.
Although for any $finite$ system $\langle \Delta \rangle$ is identically
zero, the pair-pair correlation functions should indicate the presence of
superconductivity (if it exists in the ground state)
by converging to a finite number when the separation
of the pairs is sent to infinity. Do we see this effect in the one band
Hubbard model? Unfortunately,
studies carried out by several groups showed
that at the temperatures and lattice sizes currently accessible to
Monte Carlo simulations, there are no signals of superconductivity in
the ground state, as we will see below.

In Fig.63 the dependence with distance of the
equal-time pair-pair correlation function is shown for the $\dx2y2$ and
extended-s waves (Moreo, 1992b). The
operator that destroys a pair is defined as
$c_{{\bf i}\uparrow} ( c_{{\bf i+x}\downarrow} + c_{{\bf i-x}\downarrow}
\pm c_{{\bf i+y}\downarrow} \pm c_{{\bf i-y}\downarrow} )$, where the
$(+)$ sign corresponds to extended s-wave, and the $(-)$ to $\dx2y2$
wave.\footto{5}
The pair-pair correlation is defined as $P({\bf r}) = \langle
\Delta^\dagger({\bf 0}) \Delta({\bf r}) \rangle$, where ${\bf 0}$ and
${\bf r}$ are sites of the lattice,
and the susceptibility
is given by $\chi_{sup} = \sum_{\bf r} P({\bf r})$.
For both waves, the sign problem prevents the
simulation to be carried out at temperatures smaller than ${\rm T = t/6}$ but
this is the same temperature at which enhancement in the pair-field
susceptibility was observed by White et al. (1989a),
and thus indications of superconductivity
should be observed if they exist in the ground state.
Unfortunately, Fig.63 clearly shows that already at distances of
two lattice spacings or larger, the pairs are not correlated in any of
the two
channels, with a minimal finite size effect. Similar conclusions have
been obtained by Imada and Hatsugai (1989), and Imada (1991b).
The reason for the apparent contradiction between the results of Fig.62, and
63 is simple. A pair-field susceptibility contains
information about the pair-pair correlations at $all$ distances. Thus, a
susceptibility may be robust and actually increase with decreasing
temperatures,
if the short distance correlations are enhanced as the temperature is
reduced. Such an enhancement is $not$ related with long-range order and
if that occurs the susceptibility
should not increase like the number of sites, when the
cluster size is increased.
For the one band Hubbard model, it has been shown
(Scalapino, 1993) that
the equal-site $\dx2y2$ pair correlation satisfies the equality,
$$
\langle \Delta_d({\bf 0}) \Delta^\dagger_d({\bf 0}) \rangle =
1 - \langle n_1 \rangle - {{3}\over{4}} \langle m^z_1 m^z_2 \rangle
+ {{1}\over{4}} \langle n_1 n_2 \rangle,
\tag {bg}
$$
\noindent where $n_1$ is the number operator at site ${\bf 0}$, $n_2$ is the
number operator at the nearest neighbor site, while $m^z_1$ and $m^z_2$
are their respective spins in the z-direction.
Decreasing the temperature, the presence of
antiferromagnetic correlations induces a negative and large value for
$\langle m^z_1 m^z_2 \rangle$, and thus
also the pair susceptibility is enhanced, but this effect is
unfortunately unrelated with
superconductivity. Then, currently
in the one band Hubbard model there are no
indications of strong pairing correlations for the clusters and
temperatures available to numerical studies. This result has to be
contrasted against those obtained for the $attractive$ Hubbard model.
Using the same cluster sizes, temperatures and algorithm, clear
numerical indications of superconductivity were observed in this model.
Then, there is no doubt that
numerical methods can indeed detect this type of long-range order if
present in the ground state
(Scalettar et al., 1989; Moreo and Scalapino, 1991; Randeria, Trivedi,
Moreo, and Scalettar, 1992).
Another important issue is:
How do we know when a cluster is large enough to rule out the
presence of superconductivity in a given model? The only way is
to compare the lattice size with a typical correlation length $\xi$ in the
problem as predicted by other studies (typically based on
self-consistent approximations), at the temperature at which the
simulation was carried out. Unfortunately, this is a difficult task in
many of the proposed theories of the cuprates, and thus such a
comparison is difficult.

Results similar to those observed for the one band case, have
been obtained for the three band Hubbard model. In Fig.64, the equal
time pair-correlation function is shown for several channels as
described in the caption (Frick et al., 1990). The correlation is normalized
such that
superconductivity is signaled by the convergence of this quantity to a
nonzero value in the bulk limit. Since the pair correlation function
actually converges to zero as $N$ increases,
no indications of superconductivity in this model were detected in the
parameter region
analyzed by Frick et al. (1990) (for a similar study see
Dopf, Muramatsu, and Hanke, 1990). Scalettar (1989) found that in the
3-band model the pairing susceptibility indicated that the extended
s-wave channel was competitive with d-wave, and indeed, Scalettar et al.
(1991)
showed that the addition of an intersite copper-oxygen Coulomb repulsion
stabilized the s-pairing further. However, as in studies of the single
band model the equal-time correlations do not show long-range order.

\vskip 1cm

\noindent{\tt 2. Superconductivity in the ${\rm t-J}$ model}

\vskip 0.3cm

The results shown in the previous section suggest that the one and
three band Hubbard models do not superconduct, at least in the range
of temperatures and cluster sizes that are accessible to present day
numerical studies. Then, the natural question is: do any of the models of
correlated electrons that are currently widely studied present a
superconducting phase in some region of parameter space?
Trying to answer this question,
let us consider the ${\rm t-J}$ model in more detail. Lanczos studies of
this model near half-filling and ${\rm J/t < 1}$ by several groups, do not show
enhancement of the pairing
correlations. This is
not too surprising since, e.g.,
a hole density of ${\rm x=0.125}$ corresponds to
only one hole pair on a ${\rm 16}$ sites lattice, and just one pair cannot
produce long-range order. However, the Monte Carlo results for the Hubbard
model
shown in the previous section on larger clusters (and thus with more holes)
also do not show indications of long-range
pairing correlations, and we would expect some qualitative relation
between the Hubbard model and the ${\rm t-J}$ model at small ${\rm
J/t}$. Then, the tentative conclusion is that the ${\rm t-J}$ model does
not superconduct at small ${\rm J/t}$ and hole density.

However, we know that the ${\rm t-J}$ model presents hole
binding near half-filling approximately in the region ${\rm J/t > 0.3}$ (as was
observed in
section III.B.5). In addition, at large ${\rm J/t}$ it is well-established
that there is phase separation (section IV.D), and thus it is clear that
effective
attractive forces are operative in this model.
The pairs formed near
half-filling may actually condense at low temperatures or, in other words,
it can be argued that the attraction that leads to phase
separation may create mobile $pairs$ increasing ${\rm J/t}$
before that regime is reached. Actually,
this phenomenon explicitly occurs in the regime of low electronic density.
Emery, Kivelson and Lin (1990) have shown that a pair of electrons in an
otherwise empty lattice are bound
in a spin singlet in the region ${\rm J/t > 2}$, while phase separation
seem to occur at larger couplings. It is expected that these pairs may condense
at low
temperature into a superfluid phase. The appearance of superconductivity
near phase separation was also addressed using large-N techniques
by Grilli et al. (1991). These authors found an instability in the
$A_{1g}$ and $B_{2g}$ channels. Di Castro and Grilli (1992) studied the
relation between phase separation and superconductivity using slave
bosons. A study of the spectrum of the
${\rm t-J}$ model by Moreo (1992a) also suggested that superconductivity
may exist near phase separation.

Then,
according to these arguments, a numerical analysis of the two
dimensional ${\rm t-J}$ model
near phase separation may finally show the elusive
indications of superconductivity that we are looking for.
What doping fraction is the most favorable? As explained
before near half-filling (and also in the other extreme of small
electronic density) very few pairs are available to contribute to the
pairing correlations. Then, $\langle n \rangle = 1/2$
seems optimal since in this regime the maximum
number of pairs that can be formed grows
like ${\rm N/4}$, where N is the number of sites of the cluster.
An analysis in this region of parameter space
has been recently carried out
by Dagotto and Riera (1992; 1993). Indications of
superconductivity
in the ground state have been observed by these authors.
To discuss their results let us introduce the
pairing correlation function $C({\bf m}) = (1/N) \sum_{\bf i} \langle
\Delta^\dagger_{\bf i} \Delta_{\bf i+m} \rangle$ (where the operator
$\Delta$ has already been defined in the previous section), and the
pairing ``susceptibility'' $\chi_{sup} = \sum_{\bf m} C({\bf m})$,
as indicators of the presence of long-range superconductivity in the model.
Results obtained on a $4 \times 4$ cluster at density $\langle n \rangle = 1/2$
are shown in Fig.65a. The susceptibility has a large peak in the vicinity
of ${\rm J/t =3}$ suggesting strong pairing correlations. The sharp
decay
for larger values of the coupling is caused by the transition to the phase
separation regime as explained by Dagotto and Riera (1993). However, from our
discussion
for the one band Hubbard model, it is important to study the explicit
distance dependence of the pair correlations where the susceptibility is
enhanced. In Fig.65b the
correlations are shown for the extended s-wave and $\dx2y2$ symmetries. The
$\dx2y2$ channel seems enhanced, and appreciably large at the
largest distance available on this small cluster. Finally, Fig.65c shows
the
coupling dependence of the results. The pairing correlations have
maximum strength at ${\rm J/t \sim 3}$ as suggested by Fig.65a. Of
course, these results are not
final since a proper finite size study of these correlations has
not been carried out thus far. However, they are very suggestive that
indeed
the argument expressed above relating phase separation and
superconductivity may be operative.

To gain further intuitive insight into the superconducting region detected in
the
${\rm t-J}$ model, it is convenient to enlarge the Hamiltonian to
include a repulsive density-density interaction
${\rm V \sum_{\bf \langle i j \rangle} n_{\bf i} n_{\bf j}}$. This
${\rm t-J-V}$ model was studied near the atomic limit, ${\rm t \sim 0}$, by
Kivelson, Emery and Lin (1990). In the intermediate regime where
${\rm J \sim V}$, a ``liquid of spin dimers'' was reported, and the
possibility of s-wave superconductivity through a condensate of these spin
dimers was discussed. Recently, their results were
confirmed by Dagotto and Riera (1992) using exact diagonalization
techniques on a $4 \times 4$ cluster, at large ${\rm J}$ and ${\rm V}$
couplings
and for a density ${\rm \langle n \rangle = 1/2}$. These numerical results are
shown in Fig.66.
Three regimes were detected: i) at small superexchange, the system forms
a charge-density wave; ii) at intermediate couplings, spin singlets are
formed in a regular array (shown in Fig.66a); and, finally, iii) at large
${\rm J/V}$ there is phase separation. These results can be obtained by a
simple
minimization of the energy at $t=0$, or using numerical techniques
on a finite cluster in the large couplings regime.
At strictly $t=0$, the spin
singlets of the intermediate phase are not mobile, but including
corrections in powers of $t$, Kivelson, Emery and Lin (1990) argued that
the system would become superfluid. Similar conclusions were reached in
the numerical study by Dagotto and Riera (1992) where in addition it was
observed that the pairing correlations
are maximized at $intermediate$ values of the new coupling ${\rm V/t}$ as
it is shown in Fig.66b. The region of superconductivity observed at large
${\rm V/t}$, and for the pure ${\rm t-J}$ model $({\rm V=0})$ seem analytically
connected, and speculations linking their properties with those of the
attractive Hubbard model have been presented (Dagotto and Riera, 1992)
An interesting feature of the ${\rm t-J-V}$ model is that a spin-gap
appears in the spectrum. This is natural since in the large ${\rm
V,J}$ limit, the ground state is formed by short range dimers, and thus
an energy as large as ${\rm J/t}$ is needed to create a triplet
(however, note that the presence of the spin-gap suggests that
superconductivity
should appear in the s-wave channel at finite ${\rm V}$,
contrary to the previously described
claims at ${\rm V=0}$ that the $\dx2y2$ channel is dominant. An interesting
crossover may exist between the two regimes in the superconducting phase).
Recent
studies by Troyer et al. (1993) in the one dimensional version of this model
have shown that the spin gap opens very
rapidly when ${\rm V/t}$ is increased starting from the ${\rm t-J}$
limit (Fig.67).
Actually, the physics of the one dimensional chain ${\rm t-J-V}$ is very
similar to that of its two dimensional counterpart, with analogous
behavior of the pairing correlations and superconducting
susceptibilities (see Fig.68). Troyer et al. (1993) also studied the
influence of long-range interactions noticing that the superconducting
region is not enhanced by suppressing phase separation. Instead it seems
to follow phase separation forming a narrow strip in its
neighborhood.
For another study
including $1/r$ interactions see Barnes and Kovarik (1990).

Before addressing other issues related with superconductivity, note that
some of the usual models of correlated electrons defined
on particular geometries are also candidates to show superconductivity.
In particular, it is interesting to study ``ladders'' (i.e. two coupled
chains), and two coupled planes. The former has been explored to address
questions
concerning the stability of Luttinger liquids (Schulz, 1991; Anderson,
1990a), the existence of a spin-gap (Dagotto and Moreo, 1988;
Hida, 1991, 1992; Barnes et al., 1993), and superconductivity (Emery, 1986;
Dagotto, Riera and Scalapino, 1992).
There are real materials
like ${\rm (VO)_2 P_2 O_7}$ that consist of weakly coupled arrays of
one-dimensional metal oxide ladders (Johnston, 1987). Rice, Gopalan, and
Sigrist (1993) have recently also remarked the importance of studying
ladder systems for a possible description of ${\rm Sr_{n-1} Cu_{n+1}
O_{2n} }$ compounds.
The study of coupled
planes is also interesting since the structure of some of the high-Tc
superconductors like YBCO and Bi2212 have ${\rm Cu-O}$ planes
at short distance in the unit cell (Dagotto, 1992; Dagotto, Riera, and
Scalapino, 1992; Millis and Monien, 1993. See also Morgenstern et al., 1993).
Numerical studies of superconductivity have been carried out in both
the $ladder$ ${\rm t-J}$ and Hubbard models. In the former, there is a simple
argument that guarantees the presence of superconductivity in a
particular region of parameter space. Consider the limit of large
superexchange coupling along the $rungs$. In this limit, the undoped
ground state is formed by spin-singlets along these rungs. If two
holes are added to the system, energetically it is favorable to break
only one spin singlet, and thus they will share the same rung, leading
to short-range pairing. It can be shown that residual
interactions will favor a superconducting state in this limit, as was
effectively observed in a numerical study by Dagotto, Riera and Scalapino
(1992). Actually, this mechanism is very similar to that proposed earlier
by Imada (1991a) in a dimerized ${\rm t-J}$ model.
The Hubbard model has also been studied numerically on a ladder
geometry. Some (weak) indications of long-range superconducting order have been
detected in the ground state (Bulut, Scalapino, and Scalettar, 1992; Noack et
al., 1992).
The analysis of these models deserves further study.

\vskip 1cm

\noindent{\tt 3. Phase diagram of the two dimensional ${\rm t-J}$ model}

\vskip 0.3cm

Based on several calculations reviewed in the previous section and
others, it is possible to make an educated guess for the phase diagram of the
two dimensional ${\rm t-J}$ model. The result is shown in Fig.69a. At
large ${\rm J/t}$, there is a well-established region of phase
separation. At low electronic density (${\rm x \sim 1}$ in the figure), phase
separation starts near ${\rm J/t \sim 4}$ (Kivelson, Emery and Lin, 1990;
Putikka, Luchini and Rice, 1992). In the other limit, i.e. near
half-filling, it is not clear whether
phase separation starts at a finite coupling or not (see Sec. IV.D),
and
thus we prefer to leave that region
undefined in Fig.69a. We only know that in that regime
strong antiferromagnetic correlations (AF) are present, perhaps with
some slight modulation into an incommensurate state (section IV.A.3). For
very small ${\rm J/t}$ and close to half-filling, ferromagnetism exists
(FM). We have not discussed this phase in detail in the present review since it
is
not of obvious relevance to the understanding of the cuprates. For more
details
the reader is referred to the vast literature on this subject (of which
a small sample is
Kanamori, 1963; Mattis, 1981; Doucot and Wen, 1989;
Fang et al., 1989; Barbieri, Riera, and Young, 1990;
Trugman, 1990c; Putikka, Luchini, and Ogata, 1992).
At small electronic densities and couplings, the system numerically looks like
a
noninteracting gas of electrons, and thus tentatively we label this
regime as a Fermi liquid (FL), although more work should be done to
firmly establish this result.

The dashed line of Fig.69a represents the regime where holes
seem to bind in pairs, according to the criterion $\Delta_B <
0$ as was described in section III.B.5. At least in the low electronic
density region, it is well established that electrons form bound states
starting at ${\rm J/t = 2}$, but phase separation occurs at a larger
coupling (Kivelson, Emery and Lin, 1990). It is natural to assume that
these pairs may condense into a superconducting regime at
low temperatures. An explicit
analysis of this possibility was described before (Dagotto and Riera,
1993),
and the regime where strong pairing correlations were found is denoted
as ``superconductivity'' in Fig.69a. Certainly, more work is
necessary to quantitatively find the boundaries of this phase. It may
occur that in the region $\langle n \rangle \ll 1$, the actual symmetry
of the superconducting condensate is s-wave, since that is the channel
in which two electrons bind on an otherwise empty lattice, and ${\rm J/t >2}$.
Numerical indications of a crossover from s to d-wave condensates
increasing the fermionic density starting at $\langle n \rangle = 0$
have been found by Chen (1993).

Of
particular importance is the study of the superconducting regime away
from $\langle n \rangle \sim 1/2$, and closer to half-filling. Does this
phase follow  the phase
separation regime all the way to small hole density? In the one
dimensional
${\rm t-J}$ model, Ogata et al. (1991) have shown that indeed there is
a region where superconducting correlations are dominant near phase
separation, and for densities as close to half-filling as $\langle n
\rangle = 0.87$ (for results at low electronic density see
Chen and Lee, 1993; Hellberg and Mele, 1993).
Actually, in Fig.69b the regime
where the parameter $K_{\rho}$ is larger than one,
indicates that the superconducting correlations
decay the slowest in the ground state (for details about the definition
of this parameter see Ogata et al., 1991; Troyer et al., 1993; and
references therein).
For additional information about the one dimensional case see
Assaad and W\"urtz (1991); Imada and Hatsugai (1989); Imada (1990);
Bonca et al. (1992);
and Prelovsek et al. (1993).
Then, the one dimensional
results support the idea that superconductivity appears in electronic
models near the regime of phase separation even close to half-filling.
If these results can be extended to two dimensions, then
the ${\rm t-J}$ model would superconduct in the realistic regime
of couplings and densities, becoming a strong candidate to describe the
cuprates. Then, we may still have a chance of describing high-Tc
superconductors
with purely electronic models!
However, thus far these
are speculations that clearly need (and deserve) more work and thinking.

\vskip 1cm

\noindent{\tt 4. Meissner effect and flux quantization}

\vskip 0.3cm

In addition to the pairing correlation functions, it would be important
to develop a formalism to explore numerically the Meissner effect, and
the superfluid density. This issue can be addressed
following steps similar to those that led us to Eq.(\call{az}) for the Drude
weight. In other words, it is possible to express the ``superfluid weight''
${\rm
D_s}$, in terms of suitable current-current correlations.
First, let us recall the well-known basic ideas of London's theory. In this
approach the superfluid current
is proportional to the transverse gauge field,
$$
{\bf J}_s = -{{e^2 n_s}\over{m}} {\bf A},
\tag {bh}
$$
\noindent where $n_s$ is the carrier density of the superfluid, $e$ the
unit of
charge, and $m$ the mass of the particles. This
equation can be derived from the assumption that the carriers in the
superfluid do not collide as follows:
Let us define the supercurrent as
${\bf J}_s = - e n_s {\bf v}_s$, where ${\bf v}_s$ is the velocity of
the carriers in the superfluid. Taking time derivatives
on both sides, and replacing the acceleration by
${\bf a} = {\bf F}/m = -e {\bf E}/m$ (where ${\bf E}$ is an
external electric field), we arrive to
$$
{{d{\bf J}_s}\over{dt}} = {{e^2 n_s}\over{m}} {\bf E}.
\tag {bi}
$$
\noindent In the absence of external scalar potentials $\phi$, we can
derive Eq.(\call{bh}) by a time integration of Eq.(\call{bi}).
Now, we define $D_s$ through the proportionality constant
between the superfluid current and the gauge field, i.e.
$$
{\bf J}_s = - e^2 D_s {\bf A},
\tag {bj}
$$
\noindent and thus $D_s = n_s /m$. It can also be shown easily that
$D_s = 1/(e^2 \lambda^2)$, where $\lambda$ is the London's penetration
depth,
defined
by the  exponential decay of the magnetic field inside a superconductor
($B(x) = B(0) e^{-x/\lambda}$, where $x$ is perpendicular to the surface.
For details see Schrieffer, 1988).

Having set up these simple ideas
and definitions, now we need to evaluate the expectation value of the current
operator in the ground state of the many-body problem under
consideration in order to get $D_s$ (Scalapino, White and Zhang, 1992).
Using linear response theory, in section IV.B.2 we arrived to
a general relation between the current in the many-body interacting
ground state, and an
external gauge field in the $x$-direction, i.e. we showed that
$\langle {\hat J}_x({\bf q},\omega) \rangle
= K({\bf q},\omega)
A_x({\bf q},\omega)$,
where the function $K({\bf q},\omega)$ is given by,
$$\eqalign{
K({\bf q},\omega) =
e^2 \langle {\hat K}_x \rangle + e^2[
&{{1}\over{N}} \langle \phi_0 | {\hat j}_x(-{\bf q})
{{1}\over{{\hat H}_0 - E_0 + \omega
+ i\epsilon}} {\hat j}_x({\bf q}) | \phi_0 \rangle  + \cr
&{{1}\over{N}} \langle \phi_0 | {\hat j}_x({\bf q})
{{1}\over{{\hat H}_0 - E_0 - \omega
- i\epsilon}} {\hat j}_x(-{\bf q}) | \phi_0 \rangle ],  \cr}
\tag {bl}
$$
\noindent and the notation was explained in section IV.B.2.
To study the particular case of the Meissner effect
let us consider the limit of a uniform
static transverse gauge potential that produces a magnetic field.
Then, we need to work at
$\omega=0$, and with ${\bf q} \rightarrow 0$.
This is non-trivial,
and care must be taken in the way in which the uniform limit is
approached. To understand how subtle this limit is, note that
the condition of transversality tells us that
$K({\bf q},0) = (1 - {{q^2_x}\over{{\bf q}^2}}) K({\bf q}^2)$
where $K({\bf q}^2)$ depends only on ${\bf q}^2$
(Schrieffer, 1988; section 8.3). Then, if the uniform limit is
obtained following the path
$q_y = 0, q_x \rightarrow 0$, the kernel $K({\bf q},0)$ cancels even
though $K({\bf q}^2)$ may be nonzero.
Then, it is clearly more convenient to take first $q_x=0$, and then $q_y
\rightarrow 0$ as the limiting process. With this approach, and following the
same steps that allowed us to derive Eq.(\call{az}) for the Drude weight,
we can express the superfluid weight as
$$
{{D_s}\over{2 \pi e^2}} = {{\langle - {\hat T} \rangle} \over{4N}}
- {{1}\over{N}} \sum_{n \neq 0}
{{|\langle \phi_0 | {\hat j}_x({ q_x}=0,q_y \rightarrow 0)
| \phi_n \rangle |^2}\over{E_n - E_0}},
\tag {bm}
$$
\noindent where the notation is the same as that used in section IV.B.2.
Then, in order to study $D_s$, or equivalently
the presence of a Meissner effect in the problem, we need to analyze
numerically the correlation between currents at a nonzero (but
vanishingly small) momentum (while for the Drude weight it was necessary
to study the particular case where ${\bf q}$ was strictly zero).
In practice, the computational effort to get both quantities
is basically the same.
This approach introduced by Scalapino, White and Zhang, 1992
has been recently applied to numerical studies of
both the one band Hubbard and the ${\rm t-J}$ models.
In the case of the ${\rm t-J}$ model, the superfluid weight has been
evaluated in the region where
indications of superconductivity were found (Fig.65,66),
i.e. density $\langle n \rangle \sim 1/2$, and close to
phase separation. A large peak was observed in $D_s$ on
a $4 \times 4$ cluster (Dagotto and Riera, 1993),
in the same region where
the pairing correlation functions suggested the presence of
superconductivity in the ground state.
Unfortunately, the minimum nonzero value of the
momentum in the y-direction on this small cluster is
$ q_y = \pi/2$, and thus the limit of ${\bf q} \rightarrow
0$ is difficult to approach smoothly.\footto{13}


Both the Drude and the superfluid weights can also be evaluated using
Quantum Monte Carlo methods. Although this algorithm does not allow the
calculation of $\omega$-dependent quantities, the particular case
of $\omega = 0$
can be studied in the imaginary time formalism using
a Matsubara frequency $\omega_m = 2 \pi T$. Then, the main
limitation of the method comes from the standard sign problem and
finite temperature effects. In Fig.70 results are shown for
an $8 \times 8$ cluster, working at ${\rm U/t=4}$, density
$\langle n \rangle = 0.72$, and temperature $T = t/6$. The result
indicates that the Drude weight is nonzero, in good agreement with the
exact diagonalization predictions of section IV.B.3 (Fig.44b).
Actually there is even quantitative agreement between the results
obtained
with both methods. Fig.44b
corresponding to a $4 \times 4$ cluster at zero temperature, predicted
that
$D/(2 \pi e^2) \sim 0.28$ at ${\rm U/t=4}$,
while from Fig.70b we found
$\Lambda_{xx}
({\bf q} = 0, \omega_m)$ between 0.0 and 0.10, which in combination
with the kinetic energy, makes a Monte Carlo prediction of
$D/(2 \pi e^2) \sim 0.26 - 0.31$. Thus, the Drude weight is only
weakly affected by finite size effects.
Finally, Fig.70b shows that the superfluid
density seem to vanish in this model. This result is in agreement with the
negative
conclusions about superconductivity obtained studying
pairing correlation functions in section IV.E.1.

Another criterion to search for superconducting phases is the ``flux
quantization''. In a normal state, the ground state energy is a periodic
function of the flux $\phi$ with period $\phi_0 = hc/e$, and it is minimized at
$\phi/\phi_0 = 0, \pm 1, \pm 2, ...$.
In a superconducting
phase, new stable states appear at $\phi/\phi_0 = \pm 1/2, \pm 3/2,
...$. The new unit of flux is $\phi_0/2$ due to the presence of pairs in
the ground state (``anomalous'' flux quantization).
The actual flux dependence of the energy can be
studied
on finite cluster using exact diagonalization and Quantum Monte Carlo
methods. The ${\rm t-J}$ model near phase separation shows anomalous
flux-quantization, in agreement with the conclusions based on pairing
correlations and a study of $D_s$ (Dagotto and Riera, 1993).
Also the one dimensional two band ${\rm Cu-O}$ model near phase
separation shows flux quantization on a ring (Sudbo et al., 1993) (Fig.71).
The Hubbard model has also been studied (Ferretti et al., 1992; Assaad and
Hanke, 1992).
However, care must be taken with this type of analysis since some
$non$-superconducting systems also show the presence of minima at
$\phi/\phi_0 = \pm 1/2, \pm 3/2, ...$. As example, consider two holes
in the ${\rm t-J} $ model (Poilblanc, 1991),
and CDW states on rings (Bogachek et al., 1990).

\vfill
\eject
\endit

\vskip 1cm

\noindent{\tt 5. $\dx2y2$ superconductivity}

\vskip 0.3cm

We close this review of properties of strongly correlated electrons
describing recent ideas that have induced considerable excitement among
experimentalists and theorists working on high-Tc superconductors. It
has been proposed, and
supported by several calculations, that the superconducting state
of the cuprates
has $\dx2y2$ symmetry, instead of the standard s-wave of the BCS theory
(see Monthoux, Balatsky and Pines, 1991; Bulut and Scalapino, 1991; and
references therein). The calculations are based on pairing mechanisms
that involve antiferromagnetic spin fluctuations.
Although early experiments seemed consistent with s-wave pairing, recent
results suggest that the pairing state is indeed highly anisotropic
giving support to these theoretical ideas. The
evidence comes from three sources: nuclear magnetic resonance (NMR) studies,
penetration depth measurements, and ARPES experiments:

\itemitem{$i)$} NMR experiments probe the local magnetic fields around an atom,
and allow measurements of the copper relaxation rates. Recent results by
Martindale et al. (1993) observed that this relaxation rate varies as
${\rm T^3}$ below the critical temperature, in agreement with the
predictions of some $\dx2y2$ models.

\itemitem{$ii)$} In an s-wave superconductor the penetration depth $\lambda$
(Sec.IV.E.4)
varies exponentially with temperature at small temperatures. This is a
direct consequence of the presence of a gap in the spectrum. However,
when nodes are present in the energy gap, and thus pairs can be broken
more easily, $\lambda$ is expected to change linearly with temperature
if the symmetry of the superconducting state is $\dx2y2$.
Hardy et al. (1993) have recently reported such a linear variation in
experiments carried out on clean ${\rm Y Ba_2 Cu_3 O_{6.95}}$
crystals in support of
d-wave pairing. However, note that previous
experiments in Bi2212 and thin films of YBCO found a ${\rm T^2}$
dependence of the penetration depth (see Beasley, 1993), and thus more
work is needed to clarify these experimental results.
Note also that
Wu et al. (1993) reported that for $\xnco$, $\lambda$ follows an
exponential temperature behavior as in an s-wave superconductor. This is in
agreement with some remarks made in this review about the differences between
electron
and hole doped materials in the behavior of the
resistivity $\rho$ with temperature (Sec. I.B), and the spin
correlations
(Sec. IV.A). In spite of what the one band Hubbard model may suggest, electron
and
hole doped compounds seem to behave differently from the experimental
point of view.

\itemitem{$iii)$} Shen et al.
(1993) have recently reported the presence of a strong anisotropy in the
superconducting gap of ${\rm Bi_2 Sr_2 Ca Cu_2 O_{8+ \delta}}$
crystals measured with ARPES techniques.
In some directions the gap is zero (within the experimental accuracy of
about 4 meV) compatible with $\dx2y2$ superconductivity.

However, note that there are neutron scattering
experiments (Mason et al., 1993) that urges caution about the
existence of a d-wave condensate in the cuprates. These authors observed
that
superconductivity does not induce anisotropy in the magnetic scattering
of $\x214x$, as would be expected from such a condensate.

Before these recent developments the presence of an attractive interaction in
the $\dx2y2$ channel appeared frequently in the theoretical analysis of holes
in
antiferromagnetic backgrounds (Miyake, Schmitt-Rink, and Varma, 1986;
Scalapino, Loh, and Hirsch, 1987; Gros, 1988; Chen et al., 1990). In
previous sections of this review, we have also found the presence of
hole attraction in the $\dx2y2$ channel both close to half-filling
(Poilblanc, Riera and Dagotto, 1993a; and references therein), and at
density $\langle n \rangle = 1/2$ (Dagotto and Riera, 1993). Although
the calculations used to obtain these results are approximate (finite
clusters and self-consistent equations), their common conclusions are
suggestive that indeed $\dx2y2$ is a concrete possibility in models of
correlated electrons presumed to describe the actual high-Tc materials.
In spite of these experimental and theoretical results, some theorists
remain sceptical. It
is believed that impurities can easily reduce the critical temperature
of a $\dx2y2$ superconductor. However, in the high-Tc cuprates such a
sensitivity has not been observed. Actually, the materials can be
easily prepared even by non-experts, under considerably less than perfect
conditions.
Thus, the study of the role of impurities
in a d-wave superconductor is an important issue to be addressed.
Needless to say, the physical properties of
$\dx2y2$ superconducting condensates are currently being investigated
by several experimental and theoretical
groups around the world, and a more detailed comparison with experiments
will clarify
whether the high-Tc superconductors are indeed $\dx2y2$ superconductors.
Surely, we will hear more about these interesting ideas in the near future!

\vskip 1.5cm

\centerline{\bf V. Conclusions}

\vskip 1cm

In this review we have attempted to summarize some of the results
obtained in the rapidly growing field of
computational techniques applied to models of strongly correlated
electrons. Some purely analytical methods and their predictions
have also been addressed. In addition,
we presented an overview of the current experimental
situation in high-Tc superconductors to provide the reader
with a summary of the main results, and its comparison with
computational calculations.
In the last few years, a remarkable level of maturity has been
reached in the computational studies of interacting electrons, with several
groups
independently arriving to similar conclusions as was shown in several
sections of this review.
It is becoming common practice to study models from as many
points of view as possible, including computational techniques,
perturbative or self-consistent calculations,
mean-field and variational approaches. It is clear that the complexity
of the problems require as much help as we can get, and thus the use of
numerical techniques is likely to keep on growing fast
in all areas of condensed matter theory.
The computational results are widely used as
benchmarks to test analytical approximations, especially in the
very difficult regime of strong correlations where there are no obvious
small parameters in the problem (for a good example see Sec. III. B. 3).
One cannot help but
think of this branch of theoretical physics as having common features with
experimental physics. It naturally provides a link between abstract
theoretical ideas, and the actual properties of a given model.

The main conclusions of this review are the following:

\itemitem{$\bullet$}
Regarding normal state properties,
a qualitative comparison between computational results and experiments for the
high
temperature superconducting materials was carried out. Remarkable
agreements
between theory and experiments have been observed, showing that some of
the ``anomalous'' properties of the copper oxides may have a simple
explanation through purely electronic models. In particular,
the magnetic susceptibility
in the ${\rm t-J}$ model and in the real cuprates behave similarly, both
showing deviations from a canonical Fermi liquid behavior which is caused by
the
presence of antiferromagnetic correlations.
The presence of a mid-infrared band in the optical conductivity
$\sigma(\omega)$ in Hubbard-like models is quite natural and related
to
the presence of an incoherent part in the hole spectral function
(although it is likely that holes trapped near Sr, or the
chains in the YBCO materials also appreciably contribute to this
feature in the experiments).
In the density of states $N(\omega)$, new states are observed
in the gap upon doping in the models of correlated electrons that have
been
analyzed, in agreement with experiments (again, other
possibilities, like the presence of
impurity bands, may also contribute to the
spectral weight in the gap).
In general, one gets the feeling that the standard pairing theory (perhaps in
an unusual non-s-wave channel) and Fermi liquid descriptions of the
normal state
may explain the features of the cuprates,
once the presence of antiferromagnetic correlations are taken into account, as
well
as the two dimensionality of the problem.
We are simply not used to
work with a material having competing orders. The
analytical tools are not well developed for this purpose, and thus the
help of computational techniques is crucial to obtain qualitative
information about the properties of a given model, and to check the accuracy of
the various mean-field and perturbative approaches described in the
literature,\footto{14} saving resources
and time to the condensed matter theory community.

\itemitem{$\bullet$}
New exciting ideas about the possible description of the cuprates
as $\dx2y2$ superconductors were briefly reviewed. It is reassuring that
the existence of hole attraction in this same channel was computationally
observed since the early numerical studies of the ${\rm
t-J}$ and Hubbard models,
providing a nice first-principles confirmation of the recently proposed
ideas based
on self-consistent approximations. Precisely, one of the main advantages
of numerical methods is that they allow a comparison between
abstract ``theories'' and the actual properties of the model under
consideration.

\itemitem{$\bullet$}
Thus far numerical studies are consistent with a quasiparticle (Fermi liquid)
description
of holes in antiferromagnetic backgrounds, in the sense that the wave
function renormalization $Z$ is nonzero. Then, the use of one
dimensional problems (where $Z=0$) as paradigms of their two dimensional
counterparts seems questionable. Unfortunately, for this particular problem
numerical techniques
have not yet reached a high enough level of accuracy to completely rule out
non-Fermi
liquid scenarios. More work is necessary to clarify this important
issue.

\itemitem{$\bullet$}
The development of better algorithms is crucial to improve the accuracy
of the numerical predictions for models of correlated electrons. In the
context of Lanczos, or exact diagonalization techniques, where the main
constraints come from the rapid growth of the Hilbert space with lattice
size, the use of
reduced basis sets, like in Quantum Chemistry problems, is a possible
direction to explore large clusters,
keeping the advantages of this method, especially the possibility of
studying response functions in real time.
With Monte Carlo algorithms efforts should be concentrated
on the alleviation of the sign problem, and the development of reliable
algorithms to study real-time dynamics.
Also the development of a Monte Carlo algorithm for the two dimensional
${\rm t-J}$ model is highly desirable.

\itemitem{$\bullet$}
It would be important to
explore other ``unusual'' properties of the cuprates like the behavior
of the Hall coefficient, and the resistivity as a function of temperature.
Not much work has been done on this front.
Also the paradoxical behavior of
the chemical potential with doping (section IV.C.1) does not seem to have
an explanation within two dimensional models of correlated electrons. It would
be
important to devote some effort to analyze the strange behavior of $\mu$ in the
cuprates.


\vskip 1.5cm

\centerline{\bf VI. Acknowledgments}
\vskip 0.5cm

The author specially thanks Adriana Moreo, Frank Marsiglio, Franco Nori, Bill
Putikka, and Richard
Scalettar for a careful reading of this manuscript, and useful
suggestions.
He also thanks
G. Aeppli, J. W. Allen, N. Bonesteel, S. L. Cooper, D. Dessau, H. Q. Ding, G.
Fano,
A. Fujimori,  P. Fulde, P. C. Hammel, P. Horsch, M. Imada,
T. Jolicoeur, S. Maekawa, A. Nazarenko, F. Ortolani, P. Prelovsek, J.
Riera, S. Sorella, and D. Tanner, for many useful comments about a
preliminary draft of this review.
The help of Linda Burns and Ken Ford in the preparation of the figures of
this review is acknowledged. This work was supported in part by
the Office of Naval Research through grant
ONR-N00014-93-1-0495.

\endpage

\footnotes

\singlespace

\footis{1} Unfortunately, any
finite transverse coupling or the addition of more holes induces matrix
elements
that can become negative. Thus this Monte Carlo approach cannot be
applied to more realistic cases without finding sign problems.

\footis{2}
It is worth noticing that results like those shown in
Fig.20 contain hundreds of poles, although to the eye
only a few peaks can be observed. To observe the individual poles
Dagotto et al. (1990b)
provided an example where
the $\delta$-functions have a small width $\epsilon = 0.01t$.

\footis{3} Present day supercomputers still do not allow an analysis
of the next interesting clusters which have $N=32$ and $36$ sites, for
the case of one hole in the ${\rm t-J}$ model.

\footis{4} In the continuum, $\Delta_B$
corresponds to the second derivative of the energy with respect to the
number of particles.

\footis{5} Note that the operator used to create the pair is not
strictly a spin singlet but a combination of singlet and triplet.
Nevertheless, it is supposed to have an overlap with the ground
state of two holes (which is a singlet) if the quantum numbers under
rotations and reflexions are properly selected. Note also that adding
the pairing correlation functions over all distances removes the
contribution of the triplets, and thus the superconducting
susceptibility is a pure singlet.

\footis{6} Note that the ground state of two holes
has zero momentum for all the clusters and couplings that Poilblanc,
Riera
and Dagotto considered,
and thus the systematic problem found in the one hole subspace related
with the change in the ground state momentum from cluster to cluster is absent
in this
subspace.

\footis{7} For theorists
it is interesting to notice that the d.c.
resistivity, $\rho_{dc}$, and the far-infrared resistivity, $\rho_{fir} =
\sigma^{-1}(\omega \rightarrow 0)$ obtained
from the optical conductivity, differ in this material in
about a factor two, perhaps due to the experimental uncertainties of
both methods. Then, care must be taken when detailed quantitative comparisons
between theory and experiment are attempted.

\footis{8} The help of Fabio Ortolani
in some difficult parts of the calculation is
acknowledged.

\footis{9} In deriving Eq.(\call{bb}) remember that
$\int^{\infty}_0 d\omega \delta(\omega) = 1/2$.

\footis{10} In the first studies of $\sigma(\omega)$ with periodic
boundary conditions (Moreo and Dagotto, 1990; and others)
it was claimed that the sum-rule Eq.(\call{bb}) was
not satisfied. This is not correct, and the
reason for this confusion was that the Drude peak
does not come out automatically from the numerical calculations, which
only provide the incoherent part. The weight at zero frequency needs to be
calculated separately from Eq.(\call{az}) as it is currently done in the
literature.

\footis{11} However, note that recent results by
King et al. (1993) also on $\xnco$, show a shift of
the chemical potential with electron doping as expected by band filling
scenarios, and no double peak structure in the results. Then, the
experimental situation is not quite clear.

\footis{12} The small weight inside the gap is caused by the tail of
the Lorentzians of width $\epsilon = 0.2t$ used to plot the
$\delta$-functions in Fig.56.

\footis{13}In the same region of parameter
space the Drude peak was very large suggesting that the resistivity in
the ground state near phase separation is zero for the ${\rm t-J}$ model.

\footis{14} In this respect computer studies of correlated electrons
plays a role similar to that of lattice gauge theory in the study of Quantum
Chromodynamics (which is a theory of strongly interacting quarks).

\endreferences

\vfill

\endpage

%
%
%
%
%
%

\references


Aeppli, G., et al., 1989, \lett {\bf 62}, 2052.

Aeppli, G., 1992a, Journal of Magnetism and Magnetic Materials {\bf
104-107}, 507.

Aeppli, G., 1992b, Lectures notes for E. Fermi Summer School,
Varenna, Italy.

Alexander M., et al., 1991, \phys {\bf 43}, 333.

Allen, J. W., et al., 1990, Phys. Rev. Letters {\bf 64}, 595.

Allen, J. W., 1991a, Physica {\bf B 171}, 175.

Allen, J. W., 1991b, Proceedings of the Adriatico Research Conference,
Trieste.

Almasan, C., and M. B. Maple, 1991, ``Chemistry of High
Temperature Superconductors'', ed. C. N. R. Rao. Singapore:
World Scientific.

Anderson, P. W., 1987, Science {\bf 235}, 1196.

Anderson, P. W., 1990a, \lett {\bf 64}, 1839.

Anderson, P. W., 1990b, \phys {\bf 42}, 2624.

Anderson, P. W., and J. R. Schrieffer, 1991, Phys. Today {\bf 44}, 55.

Anderson, R. O., et al., 1993, \lett {\bf 70}, 3163.

Assaad, F. F., and P. De Forcrand, 1990, in Quantum Simulations of Condensed
Matter Phenomena, eds. J. D. Doll and J. E. Gubernatis (World
Scientific, Singapore), p.1.

Assaad, F. F., and D. W\"urtz, 1991, \phys {\bf 44}, 2681.

Assaad, F. F., and W. Hanke, 1992, W\"urzburg preprint.

Auerbach, A., and B. Larson, 1991, Phys. Rev. Lett. {\bf 66}, 2262.

Bacci, S., E. Gagliano, R. Martin and J. Annet, 1991, \phys {\bf 44}, 7504.

Bacci, S., E. Gagliano, and F. Nori, 1991, Int. J. of Mod. Phys. {\bf B5}, 325.

Balseiro, C. A., A. G. Rojo, E. R. Gagliano, and B. Alascio, 1988, \phys {\bf
38}, 9315.

Barbieri, A., J. Riera, and A. P. Young, 1990, \phys {\bf 41}, 11697.

Barbour, I., et al., 1986, Nucl. Phys. {\bf B275}, 296.

Bardeen, J., L. N. Cooper, and J. R. Schrieffer, 1957, Phys. Rev. {\bf 108},
1175.

Barnes, T., and G. J. Daniell, 1988, \phys{\bf 37}, 3637.

Barnes, T., E. Dagotto, A. Moreo, and E. S. Swanson, 1989, \phys {\bf 40},
10977.

Barnes, T., and M. Kovarik, 1990, \phys{\bf 42}, 6159.

Barnes, T., 1991, Int. J. of Mod. Phys. {\bf C2}, 659.

Barnes, T., A. E. Jacobs, M. D. Kovarik, and W. G. Macready, 1992, \phys {\bf
45}, 256.

Barnes, T., et al., 1993, \phys {\bf 47}, 3196.

Batlogg, B., 1990, ``High Temperature Superconductivity: The Los
Alamos Symposium'', Addison-Wesley Publishing Co., Eds.
K. Bedell, D. Coffey, D. Meltzer, D. Pines, and J. R. Schrieffer.

Batlogg, B., 1991, Physics Today June, page 44.

Batlogg, B., H. Takagi, H. L. Kao, and J. Kwo, 1992,
``Electronic properties of High Tc Superconductors, The Normal and
the Superconducting State.'', Eds. Kuzmany et al., Springer-Verlag.

Batrouni, G. G. , and R. T. Scalettar, 1990, \phys {\bf 42}, 2282.

Beasley, M. R., 1993, Proceedings 2nd Int. Conf. on High-Temperature
Superconductivity,
Israel.

Bednorz, J. G., and K. A. M\"uller, 1986, Z. Phys.
{\bf B 64}, 189; 1988, Rev. Mod. Phys. {\bf 60}, 585.

B\'enard, P., Liang Chen, and A.-M. S. Tremblay, 1993, \phys {\bf 47}, 589.

Belinicher, V. I., and A. L. Chernyshev, 1993, \phys {\bf 47}, 390.

Bickers, N. E., D. J. Scalapino, and R. T. Scalettar, 1987, Int. J. Mod. Phys.
{\bf B1}, 687.

Bickers, N. E., D. J. Scalapino, adn S. R. White, 1989, \lett {\bf 62}, 961.

Binder, K., and D. Heerman, 1992, ``Monte Carlo Simulations in
Statistical Physics,'' Springer-Verlag series in Solid-State Science.

Birgeneau, R.  et al., 1988, \phys{\bf 38} 6614.

Birgeneau, R., 1990, Am. J. Phys. {\bf 58}, 28.

Blankenbecler, R., D. J. Scalapino, and
R. L. Sugar, 1981, Phys. Rev. {\bf D 24}, 2278.

Bogachek, E. N., et al., 1990, \phys {\bf 42}, 7614.

Bonca, J., P. Prelovsek, and I. Sega, 1989, \phys {\bf 39}, 7074.

Bonca, J., et al., 1989, Europhys. Lett. {\bf 10}, 87.

Bonca, J., et al., 1992, \lett {\bf 69}, 526.

Boninsegni, M., and E. Manousakis, 1991, \phys {\bf 43} 10353; 1992,
\phys {\bf 45}, 4877;
1992, \phys {\bf 46}, 560.

Boninsegni, M., and E. Manousakis, 1993, \phys {\bf 47}, 11897.

Brinkman, W., and T. M. Rice, 1970, \phys {\bf 39}, 6880.

Bulut, N. et al., 1990, \phys{\bf 41}, 1797.

Bulut, N., and D. Scalapino, 1991, \lett {\bf 67}, 2898.

Bulut, N., D. J. Scalapino, and R. T. Scalettar, 1992, \phys {\bf 45}, 5577.

Burns, G.,  1992, ``High-temperature superconductivity'',
Academic Press.

Campuzano, J. C., et al., 1990, \lett {\bf 64}, 2308.

Campuzano, J. C., et al., 1991, \phys{\bf 43}, 2788.

Chakravarty, S., 1990, in $High-Temperature$ $Superconductivity$, eds. K.
Bedell,
D. Coffey, D. Meltzer, D. Pines, and J. R. Schrieffer, Addison-Wesley,
p.136.

Chen, C. T., et al., 1991, \lett {\bf 66}, 104.

Chen, C.-X., and H.-B. Sch\"uttler, 1991, \phys {\bf 43}, 3771.

Chen, G. J., R. Joynt,
F. C. Zhang, and C. Gros, 1990, \phys {\bf 42}, 2662.

Chen, Y. C., and T. K. Lee, 1993, \phys {\bf 47}, 11548.

Chen, Y. C., 1993, private communication.

Chen, Y.-H., F. Wilczek, E. Witten, and B. I. Halperin, 1989, Int. J. Mod.
Phys. {\bf B 3}, 1001.

Cheong, S.-W., et al., 1991, \lett {\bf 67}, 1791.

Chernyshev, A., A. V. Dotsenko and O. P. Sushkov, 1993, preprint.

Cho, J. H., et al., 1992, \phys {\bf 46}, 3179.

Cho, J. H., F. C. Chou, and D. C. Johnston, 1993, \lett {\bf 70}, 222.

Claessen, R., et al., 1992, \lett {\bf 69}, 808.

Cooper, S. L., G. A. Thomas, J. Orenstein, D. Rapkine, A. Millis,
S-W. Cheong, A. S. Cooper, and Z. Fisk, 1990, \phys {\bf 41}, 11605.

Cooper, S. L., D. Reznik, A. Kotz, M. Karlow, R. Liu,
M. Klein, W. Lee, J. Giapintzakis, and D. M. Ginsberg, 1993, \phys
{\bf 47}, 8233.

Dagotto, E., and A. Moreo, 1985, Phys. Rev. {\bf D 31}, 865.

Dagotto, E., and A. Moreo, 1988, \phys {\bf 38}, 5087; 1991, {\bf 44}, 5396
(E).

Dagotto, E., A. Moreo, and T. Barnes, 1989, \phys {\bf 40}, 6721.

Dagotto, E., A. Moreo, R. L. Sugar, and D. Toussaint, 1990a, \phys {\bf 41},
811.

Dagotto, E., R. Joynt, A. Moreo, S. Bacci, and E.
Gagliano, 1990b, \phys {\bf 41}, 9049; 1990, \phys {\bf
41}, 2585.

Dagotto, E., J. Riera, and A. P. Young, 1990, \phys {\bf 42}, 2347.

Dagotto, E., and D. Poilblanc, 1990, \phys {\bf 42}, 7940.

Dagotto, E., and J. R. Schrieffer, 1991, \phys {\bf 43}, 8705.

Dagotto, E., 1991, Int. J. of Mod. Physics {\bf B5}, 907.

Dagotto, E., A. Moreo, F. Ortolani, J. Riera, and D. J. Scalapino, 1991, \lett
{\bf 67}, 1918.

Dagotto, E., J. Riera, and D. Scalapino, 1992, \phys {\bf 45}, 5744.

Dagotto, E., A. Moreo, F. Ortolani, J. Riera, and
D. Scalapino, 1992a, \phys {\bf 45}, 10107.

Dagotto, E., A. Moreo, F. Ortolani, D. Poilblanc, and
J. Riera, 1992b, \phys {\bf 45}, 10741.

Dagotto, E., F. Ortolani, and D. Scalapino, 1992, \phys {\bf 46}, 3183.

Dagotto, E., and J. Riera, 1992, \phys {\bf 46}, 12084.

Dagotto, E., 1992, ``Computational Approaches in Condensed-Matter Physics'',
Eds. S. Miyashita, M. Imada, and H. Takayama, Springer Verlag, page 84.

Dagotto, E., and J. Riera, 1993, \lett {\bf 70}, 682.

De Raedt, H., and W. von der Linden, 1992, \phys {\bf 45}, 8787.

Dessau, D., thesis, Dept. of Applied Physics, 1992, Stanford
University.

Di Castro, C., and M. Grilli, 1992, Physica Scripta, {\bf T 45}, 81.

Ding, H.-Q., 1992, Physica C {\bf 203}, 91.

Ding, H.-Q., G. Lang, and W. Goddard III, 1992, \phys {\bf 46}, 14317.

Ding, H.-Q., and W. Goddard III, 1993, \phys {\bf 47}, 1149.

Doll, G., J. Nicholls, M. Dresselhaus, A. Rao, J. Zhang,
G. Lehman, P. Eklund, G. Dresselhaus, and A. Strauss, 1988, \phys {\bf
38}, 8850;
and references therein.

Dopf, G., A. Muramatsu, and W. Hanke, 1990, \phys {\bf 41}, 9264.

Dopf, G., J. Wagner, P. Dieterich, A. Muramatsu, and W. Hanke, 1992, \lett
{\bf 68}, 2082.

Doucot, B., and X. G. Wen, 1989, \phys {\bf 40}, 2719.

Eder, R., K. W. Becker, and W. H. Stephan, 1990, Z. Phys. {\bf B 81}, 33.

Eder, R., 1992, \phys {\bf 45}, 319.

Eder, R., and P. Wr\'ober, 1992, Max-Planck (Stuttgart) preprint.

Eisaki, H., et al., 1992, \phys {\bf 45}, 12513.

Emery, V. J., 1986, Synthetic Metals {\bf 13}, 21.

Emery, V. J, 1987, \lett {\bf 58}, 2794.

Emery, V. J., and G. Reiter, 1988a, \phys {\bf 38}, 4547.

Emery, V. J., and G. Reiter, 1988b, \phys {\bf 38}, 11938.

Emery, V. J., and G. Reiter, 1990, \phys {\bf 41}, 7247.

Emery, V. J., S. A. Kivelson and H. Q. Lin, 1990, \lett {\bf 64},
475.

Emery, V. J., and S. A. Kivelson, 1993, preprint.

Eskes, H., and G. A. Sawatzky, 1988, \lett {\bf 61}, 1415.

Eskes, H., and G. A. Sawatzky, 1991, \phys {\bf 43}, 119.

Eskes, H., M. Meinders, and G. A. Sawatzky, 1991, \lett {\bf 67}, 1035.

Fahy, S. B., and D. R. Hamann, 1991, \phys {\bf 43}, 765;
1990, \lett {\bf 65}, 3437.

Fang, Y., A. Ruckenstein, E. Dagotto, and S. Schmitt-Rink, 1989, \phys {\bf
40}, 7406.

Fano, G., F. Ortolani, and F. Semeria, 1990, Int. J. Mod. Phys. B {\bf 3},
1845.

Fano, G., F. Ortolani, and A. Parola, 1990, \phys {\bf 42}, 6877.

Fano, G., F. Ortolani and A. Parola, 1992, \phys {\bf 46}, 1048.

Fehske, H., V. Waas, H. R\"oder, and H. B\"uttner, 1991, \phys {\bf 44}, 8473.

Feng, G., and S. R. White, 1992, \phys {\bf 46}, 8691.

Ferretti et al., 1992, \phys {\bf 45}, 5486.

Fetter, A., and J. Walecka, 1971, ``Quantum Theory of
Many-Particle
Systems'', Mc Graw-Hill Publishing Co..

Fink, J., et al., 1993, HTSC special issue of the Journal of
Electron Spectroscopy, in print.

Foster, C. M., K. Voss, T. Hagler, D. Mihailovic\',
A. Heeger, M. Eddy, W. Olsen and E. Smith, 1990, Solid State Comm. {\bf 76},
651.

Frick, M., P. Pattnaik, I. Morgenstern, D. Newns and
W. von der Linden, 1990, \phys {\bf 42}, 2665.

Fujimori, A., et al., 1992, \phys {\bf 46}, 9841.

Fujimori, A., 1992, JJAP Series 7, Mechanisms of
Superconductivity p.p. 125; 1992, J. Phys. Chem. Solids, Vol. 53, 1595.

Fukuda, Y., et al., 1989, Solid State Comm. {\bf 72}, 1183.

Fukuyama, H., S. Maekawa, and A. P. Malozemoff (eds.), 1989,
Springer Series in Solid State Sciences, Vol 89.

Furukawa, N. and M. Imada, 1991a, J. Phys. Soc. Jpn. {\bf 60}, 810.

Furukawa, N. and M. Imada, 1991b, J. Phys. Soc. Jpn. {\bf 60}, 3669.

Furukawa, N. and M. Imada, 1992, J. Phys. Soc. Jpn. {\bf 61},
3331.

Fulde, P., 1991, ``Electron Correlations in Molecules and Solids'',
Springer-Verlag series in Solid-State Sciences 100.

Fulde, P., and P. Horsch, 1993, Europhys. News {\bf 24}, 73.

Fulde, P., and P. Unger, 1993, \phys {\bf 47}, 8947.

Fye, R., M. Martins, and R. T. Scalettar, 1990, \phys {\bf 42}, 6809.

Fye, R., M. Martins, D. Scalapino, J. Wagner and W. Hanke, 1991,
\phys {\bf 44}, 6909.

Fye, R., M. Martins, D. Scalapino, J. Wagner and W. Hanke, 1992,
\phys{\bf 45}, 7311.

Gagliano, E., E. Dagotto, A. Moreo, and F. Alcaraz, 1986, Phys. Rev. {\bf B
34}, 1677; Erratum: \phys {\bf 35}, 5297 (1987).

Gagliano, E., and C. Balseiro, 1987, \lett {\bf 59}, 2999.

Gagliano, E., S. Bacci, and E. Dagotto, 1990, \phys{\bf 42}, 6222.

Gagliano, E., and S. Bacci, 1990, \phys {\bf 42}, 8772.

Gagliano, E., S. Bacci, and E. Dagotto, 1991, \phys{\bf 44}, 285.

Galan, J., and J. Verg\'es, 1991, \phys {\bf 44}, 10093.

Ghijsen, J., et al., 1988, \phys {\bf 38}, 11322.

Giamarchi, T., and A. Millis, 1992, \phys {\bf 46}, 9325.

Gomez-Santos, G., J. D. Joannopoulus, and J. W. Negele, 1989, Phys. Rev. B{\bf
39}, 4435.

Grilli, M., R. Raimondi, C. Castellani, C. Di Castro, and G. Kotliar, 1991,
\lett {\bf 67}, 259.

Gros, C., 1988, \phys {\bf 38}, 931.

Gubernatis, J. E., D. J. Scalapino, R. L. Sugar and
W. D. Toussaint, 1985, \phys {\bf 32}, 103.

Hamann, D. R., and S. B. Fahy, 1990, \phys {\bf 41}, 11352.

Hammel, P. C., et al., 1990, \phys {\bf 42}, 6781;
1991, Physica {\bf C 185}, 1095.

Hammel, P. C., E. Ahrens, A. Reyes, J. Thompson, Z. Fisk,
P. Canfield, J. Schirber and D. MacLaughlin, 1992, invited paper for workshop
on Phase Separation in Cuprate Superconductors, Erice, Italy.

Hardy, W. N., D. A. Bonn, D. C. Morgan, R. Liang and
K. Zhang, 1993, submitted to PRL.

Harshman, D. R., et al., 1988, \phys {\bf 38}, 852.

Harshman, D. R., et al., 1989, \lett {\bf 63}, 1187.

Harshman, D. R., and A. P. Millis, 1992, \phys {\bf 45}, 10684.

Hasegawa, Y., and D. Poilblanc, 1989, \phys {\bf 40}, 9035.

Hayden, S. M., et al., 1991a, \lett {\bf 66}, 821.

Hayden, S. M., et al., 1991b, \lett {\bf 67}, 3622.

Haydock, R., V. Heine, and M. Kelly, 1975, J. Phys. C {\bf 5}, 2845 (1972);
{\bf 8}, 2591.

Heeb, E. S., and T. M. Rice, 1993, Z. Phys. {\bf B 90}, 73.

Hellberg, C. S., and E. J. Mele, 1993, Univ. of Pennsylvania preprint.

Herr, S., K. Kamar\'as, C. Porter, M. Doss, D. Tanner, D.
Bonn, J. Greedan, C. Stager, and T. Timusk, 1987, \phys {\bf 36}, 733.

Hida, K., 1991, J. Phys. Soc. Jpn. {\bf 60}, 1347.

Hida, K., 1992, J. Phys. Soc. Jpn. {\bf 61}, 1013.

Hidaka, Y., 1989, and M. Suzuki, Nature {\bf 338}, 635.

Hirsch, J. E., 1985, \phys {\bf 31}, 4403.

Hirsch, J. E., S. Tang, E. Loh, Jr., and D. J. Scalapino, 1988, \lett {\bf
60}, 1668.

Hirsch, J. E., E. Loh, Jr., D. J. Scalapino, and S. Tang, 1989, \phys {\bf
39}, 243.

Hirsch, J., and S. Tang, 1989, \lett {\bf 62}, 591.

Hirsch, J., and F. Marsiglio, 1989, \phys {\bf 39}, 11515.

Horsch, P., W. Stephan, K. v.Szczepanski, M. Ziegler, and W. von der
Linden,
1989, Physica {\bf C 162-164}, 783.

Hubbard, J., 1963, Proc. R. Soc. London, Ser. A {\bf 276}, 238;
1964, {\bf 281}, 401.

Hybertsen, M., M. Schl\"uter, N. E. Christensen, 1989, \phys {\bf 39}, 9028.

Hybertsen, M., et al., 1990, \phys {\bf 41}, 11068.

Imada, M., and Y. Hatsugai, 1989, J. Phys. Soc. Jpn. {\bf 58},
3752.

Imada, M., 1990, J. Phys. Soc. Jpn. {\bf 59}, 4121.

Imada, M., 1991a, J. Phys. Soc. Jpn. {\bf 60}, 1877.

Imada, M., 1991b, J. Phys. Soc. Jpn. {\bf 60}, 2740.

Inui, M., S. Doniach, and M. Gabay, 1988, \phys {\bf 38}, 6631.

Inoue, J., and S. Maekawa, 1992,
Prog. of Theor. Physics, suppl. {\bf 108}, 313.

Jagla, E., K. Hallberg, and C. Balseiro, 1992, Bariloche preprint.

Jarrell, M., et al., 1991, \phys {\bf 43}, 1206.

Johnston, D. C., 1987, \phys {\bf 35}, 219.

Johnston, D. C., et al., 1988, Physica {\bf C}153-155, 572.

Johnston, D. C., 1989, Phys. Rev. Lett. {\bf 62}, 957.

Jorgensen, J. D., et al., 1988, \phys {\bf 38}, 11337.

Jorgensen, J. D., 1991, Physics Today {\bf 44}, June ,
page 34.

Kamar\'as, K., et al., 1990, \lett {\bf 64}, 84.

Kampf, A., and J. R. Schrieffer, 1990, \phys {\bf 42}, 7967.

Kanamori, J., 1963, Prog. Theor. Phys. {\bf 30}, 275.

Kane, C., P. Lee, and N. Read, 1989, \lett {\bf 39}, 6880.

Kawakami, N., and S.-K. Yang, 1991, \phys {\bf 44}, 7844.

Kaxiras, E. and E. Manousakis, 1988, \phys {\bf 38}, 866.

Keimer, B., et al., 1992, \phys {\bf 46}, 14034.

Kim, Y., A. Heeger, L. Acedo, G. Stucky, and
F. Wudl, 1987, \phys {\bf 36}, 7252.

King, D. M., et al., 1993, \lett {\bf 70}, 3159.

Kivelson, S., V. Emery, and H. Q. Lin, 1990, \phys {\bf 42}, 6523.

Knowles, P., and N. Handy, 1989, J. Chem. Phys. {\bf 91}, 2396.

Kogut, J., 1979, Rev. Mod. Phys. {\bf 51}, 659; 1983, Rev. Mod. Phys. {\bf
55}, 775.

Kohn, W., 1964, Phys. Rev. {\bf 133}, A171.

Koike et al., 1989, Physica C {\bf 159}, 105.

Kovarik, M., 1990, \phys {\bf 41}, 6889.

Lanczos, C., 1950, J. Res. Natl. Bur. Stand. {\bf 45}, 255.

Laughlin, R. B., 1988, Science {\bf 242}, 525;
1988, \lett {\bf 60}, 2677.

Lee, W. C., R. A. Klemm, and D. C. Johnston, 1989, Phys. Rev. Lett. {\bf 63},
1012.

Leung, P. W., Z. Liu, E. Manousakis, M. A. Novotny, and P. E.
Oppenheimer, 1992, \phys {\bf 46}, 11779.

Levin, K., Ju H. Kim, J. P. Lu, and Q. Si, 1992, Univ. of
Chicago preprint.

Littlewood, P. B., C. M. Varma, and E. Abrahams, 1987, \lett {\bf 60}, 379.

Liu, Z., and E. Manousakis, 1991, \phys {\bf 44}, 2414;
1992, \phys {\bf 45}, 2425.

Liu, Rong, et al., 1992a, \phys {\bf 45}, 5614.

Liu, Rong, et al., 1992b, \phys {\bf 46}, 11056.

Liu, L. Z., R. O. Anderson and J. W. Allen, 1991, Proceedings
of workshop on Fermiology of High Tc, Argonne National Lab, March 25-27.

Loh, E. Y., et al., 1990, \phys {\bf 41}, 9301.

Lu, J. P. et al., 1990, Phys. Rev. Lett. {\bf 65}, 2466.

Maekawa, S., and M. Sato (eds), 1991, Springer Series in Solid State Sciences,
Vol. 106.

Maldague, P., 1977, \phys {\bf 16} 2437.

Manousakis, E., 1991, Rev. Mod. Phys. {\bf 63}, 1.

Manousakis, E., 1992, \phys {\bf 45}, 7570.

Marsiglio, F., A. E. Ruckenstein, S. Schmitt-Rink,
and C. Varma, 1991, \phys {\bf 43}, 10882.

Martindale, J. A., et al., 1993, \phys {\bf 47}, 9155.

Martinez, G., and P. Horsch, 1991, \phys {\bf 44}, 317.

Mason, T. E., G. Aeppli, and H. A. Mook, 1992, \lett {\bf 68}, 1414.

Mason, T. E., et al., 1993, to appear in Phys. Rev. Letters.

Matsuyama et al., 1989, Physica (Amsterdam) {\bf 160C}, 567.

Mattis, D. C., 1981, editor, ``The theory of Magnetism I. Statistics and
Dynamics'', Springer Series in Solid-State Sciences, Vol.17.

Mila, F., 1988, \phys {\bf 38}, 11358.

Millis, A. J., H. Monien, and D. Pines, 1990, \phys {\bf 42},
167.

Millis, A., and S. N. Coppersmith, 1990, \phys {\bf 42}, 10807.

Millis, A., and B. Shraiman, 1992, \phys {\bf 46}, 14834.

Millis, A. J., and H. Monien, 1993, \lett {\bf 70}, 2810.

Miyake, K., S. Schmitt-Rink and C.M. Varma, 1986, Phys. Rev. B {\bf 34},
6554.

Monthoux, P., A. Balatsky, and D. Pines, 1991, \lett {\bf 67}, 3448.

Moreo, A., and E. Dagotto, 1990, \phys {\bf 42}, 4786.

Moreo, A., D. J. Scalapino, R. Sugar, S. White, and N.
Bickers, 1990a, \phys {\bf 41}, 2313.

Moreo, A., E. Dagotto, T. Jolicoeur and J. Riera, 1990b,
\phys {\bf 42}, 6283.

Moreo, A., E. Dagotto and D. Scalapino, 1991, \phys
{\bf 43}, 11442.

Moreo, A., and D. J. Scalapino, 1991, \lett {\bf 66}, 946.

Moreo, A., 1992a, \phys {\bf 45}, 4907.

Moreo, A., 1992b, \phys {\bf 45}, 5059.

Moreo, A., 1993a, FSU preprint, to appear in \phys.

Moreo, A., 1993b, unpublished.

Moreo, A., 1993c, FSU preprint.

Morgenstern, I., Th. Husslein, J. M. Singer, and H.-G. Matuttis,
1993, Regensburg preprint.

Mori, H., 1965, Prog. Theor. Phys. {\bf 33}, 423; 1965, {\bf 34}, 399.

Nagaoka, Y., 1966, Phys. Rev. {\bf 147}, 392.

Nagaosa, N., and P. A. Lee, 1990, \lett {\bf 64}, 2450.

Namatame, H., et al., 1990, \phys {\bf 41}, 7205.

Nieva, G., E. Osquiguil, J. Guimpel, M. Maenhoudt,
B. Wuyts, Y. Bruynseraede, M. Maple and I. Schuller, 1992, preprint.

Noack, R. M., R. T. Scalettar, N. Bulut, and D. J. Scalapino, 1992,
UCSB preprint.

Nori, F., E. Abrahams, and G. Zimanyi, 1990, \phys {\bf 41}, 7277;
1991, Int. J. of Mod. Phys. {\bf B 5}, 119.

Nori, F., and G. Zimanyi, 1991, Europhysics Lett. {\bf 16}, 397.

Nori, F., E. Gagliano, and S. Bacci, 1992, \lett {\bf 68}, 240.

N\"ucker, N., et al., 1987, Z. Phys. B{\bf 67}, 9.

Ogata, M., M. Luchini, S. Sorella, and F. Assaad, 1991, \lett {\bf 66}, 2388.

Ohta, Y., K. Tsutsui, W. Koshibae, T. Shimozato,
and S. Maekawa, 1992, \phys {\bf 46}, 14022.

Oitmaa, J., and D. D. Betts, 1978, Can. J. Phys. {\bf 56}, 897.

Olson, C. G., et al., 1989, Science {\bf 245}, 731.

Olson, C. G., et al., 1990, \phys {\bf 42}, 381.

Orenstein, J., G. A. Thomas, D. Rapkine, C. Bethea,
B. Levine, R. Cava, E. Reitman, and D. Johnson, Jr., 1987,
\phys{\bf 36}, 733.

Orenstein, J., G. A. Thomas, A. Millis, S. L. Cooper,
D. Rapkine, T. Timusk, L. Schneemeyer, and J. Waszczak, 1990,
\phys{\bf 42}, 6342.

Pettifor, D. G., D. L. Weaire (eds.): {\it The Recursion Method and Its
Applications}, Springer Ser. Solid-State Sci., Vol. 58 (Springer,
Berlin, Heidelberg 1985).

Poilblanc, D., and T. M. Rice, 1989, \phys {\bf 39}, 9749.

Poilblanc, D., and E. Dagotto, 1990, \phys {\bf 42}, 4861.


Poilblanc, D., and E. Dagotto, 1991, \phys {\bf 44}, 466.

Poilblanc, D., 1991, \phys {\bf 44}, 9562.

Poilblanc, D., H. J. Schulz, and T. Ziman, 1992, \phys {\bf 46},
6435.

Poilblanc, D., H. J. Schulz, and T. Ziman, 1993, \phys {\bf 47}, 3268.

Poilblanc, D., T. Ziman, H. J. Schulz and E. Dagotto, 1993,
FSU-SCRI preprint, to appear in Phys. Rev. B.

Poilblanc, D., J. Riera and E. Dagotto, 1993, FSU-SCRI
preprint.


Prelovsek, P., I. Sega, and J. Bonca, 1990, \phys {\bf 42}, 10706.

Prelovsek, P., et al., 1993, \phys {\bf 47}, 12224.

Prelovsek, P., and X. Zotos, 1993, \phys {\bf 47}, 5984.

Putikka, W. O., M. U. Luchini, and T. M. Rice, 1992,
\lett {\bf 68}, 538.

Putikka, W. O., M. U. Luchini, and M. Ogata, 1992, \lett {\bf 69}, 2288.

Putilin, S. N., E. V. Antipov, O. Chmaissem, and M. Marezio, 1993, Nature {\bf
362}, 226.


Ramsak, A., and P. Prelovsek, 1989, \phys {\bf 40}, 2239.

Randeria, M., N. Trivedi, A. Moreo, and R. T. Scalettar,
1992, \lett {\bf 69}, 2001.

Reedyk, M., T. Timusk, J. Xue, and J. Greedan, 1992, \phys {\bf 45},
7406.

Reger, J. D., and A. P. Young, 1988, \phys {\bf 37}, 5024.

Reiter, G. F., 1992, preprint.

Reyes, A. P., P. C. Hammel, E. T. Ahrens, J. D. Thompson, P. C.
Canfield, Z. Fisk, and J. E. Schirber, 1993, Proceedings of the ``Conference
on Spectroscopies in Novel Superconductors'', Santa Fe, NM.

Rice, T. M., and F. C. Zhang, 1989, \phys {\bf 39}, 815.

Rice, T. M., S. Gopalan, and M. Sigrist, 1993, ETH-TH/93-7 preprint.

Riera, J., and A. P. Young, 1989, \phys {\bf 39}, 9697.

Riera, J., and E. Dagotto, 1993a, \phys {\bf 47}, 15346.

Riera, J., and E. Dagotto, 1993b, Oak Ridge National Lab preprint.

Romberg, H., et al., 1990, \phys {\bf 42}, 8768.

Romero, D., C. Porter, D. Tanner, L. Forro, D. Mandrus,
L. Mihaly, G. Carr and G. Williams, 1992, Solid State Comm., submitted.

S\'a de Melo, C., and S. Doniach, 1990, \phys {\bf 41}, 6633.

Sachdev, S., 1989, \phys {\bf 39}, 12232.

Scalapino, D. J., E. Loh and J.E. Hirsch, 1986, Phys. Rev. B{\bf 34}, 8190

Scalapino, D. J., E. Loh and J.E. Hirsch, 1987, Phys. Rev. B{\bf 35}, 6694.

Scalapino, D. J., 1991, Physica C {\bf 185-189}, 104.

Scalapino, D. J., S. R. White, and S. C. Zhang, 1992, \lett
{\bf 68}, 2830.

Scalapino, D. J., S. R. White, and S. C. Zhang, 1993, UCSB preprint.

Scalapino, D. J., 1993, UCSB preprint.

Scalettar, R. T., et al., 1989, \phys {\bf 62}, 1407.

Scalettar, R. T., 1989, Physica {\bf C 162-164}, 313.

Scalettar, R. T., S. R. White, D. J. Scalapino, and
R. L. Sugar, 1991, \phys {\bf 44}, 770.

Schilling, A., et al., 1993, Nature {\bf 363}, 56.

Schlesinger, Z., R. Collins, F. Holtzberg, C. Feild,
S. Blanton, U. Welp, G. Crabtree, Y. Fang, and J. Liu, 1990, \lett {\bf 65},
801.

Schmitt-Rink, S., C. M. Varma, and A. E. Ruckenstein, 1988,
\lett {\bf 60}, 2793.

Schulz, H., 1990a, \lett {\bf 64}, 1445.

Schulz, H., 1990b, \lett {\bf 64}, 2831.

Schulz, H., 1991, Int. J. Mod. Phys. {\bf B}, 57.

Schrieffer, J. R. , 1988, ``Theory of Superconductivity'',
Addison-Wesley Publishing Co., Fourth printing.

Schrieffer, J. R., X.-G. Wen, and S.-C. Zhang, 1988, \lett {\bf 60},
944; 1989, \phys {\bf 39}, 11663.

Sch\"uttler, H.-B., and A. J. Fedro, 1989, J. Less Common Met. {\bf 149},
385.

Sega, I., and P. Prelovsek, 1990, \phys {\bf 42}, 892.

Serene, J. W., and D. W. Hess, 1991, \phys {\bf 44}, 3391.

Shastry, B., and B. Sutherland, 1990, \lett {\bf 65}, 243.

Shen, Z.-X., et al., 1987, \lett {\bf 36}, 8414.

Shen, Z.-X., et al., 1993, \lett {\bf 70}, 1553.

Shimada, M., M. Shimizu, J. Tanaka, I. Tanaka, and H.
Kojima, 1992a, Physica {\bf C} 193, 277.

Shimada, M., K. Mizuno, S. Miyamoto, M. Shimizu, and J.
Tanaka, 1992b, Physica {\bf C} 193, 353.

Shraiman, B., and E. Siggia, 1988a, \lett {\bf 61}, 467; 1989, \lett {\bf 62},
1564; 1989, \phys {\bf 40}, 9162.

Shraiman, B., and E. Siggia, 1988b, \lett{\bf 61}, 740.

Silver, R. N., et al., 1990, \lett{\bf 65}, 496.

Singh, R. R. P., and R. L. Glenister, 1992a, \phys {\bf 46}, 11871.

Singh, R. R. P., and R. L. Glenister, 1992b, \phys {\bf 46}, 14313.

Song, J., and J. F. Annett, 1992, Europhys. Lett. {\bf 18}, 549.

Sorella, S., S. Baroni, R. Car, and M. Parrinello, 1989,
Europhys. Lett. {\bf 8}, 663; 1988, S. Sorella et al., Int. J.
Mod. Phys. {\bf B1}, 993.

Sorella, S., 1991, Int. J. Mod. Phys. {\bf B5}, 937.

Sorella, S., 1992, \phys {\bf 46}, 11670.

Stafford, C. A., A. Millis, and B. Shastry, 1991, \phys {\bf 43},
13660.

Stafford, C. A., and A. Millis, 1993, preprint.

Stechel, E. B., and D. R. Jennison, 1988, \phys {\bf 38}, 4632.

Stephan, W., and P. Horsch, 1990, \phys {\bf 42}, 8736.

Stephan, W., and P. Horsch, 1991, \lett {\bf 66}, 2258.

Stephan, W., and P. Horsch, 1992, Int. J. Mod. Phys. {\bf B6}, 589.

Su, W. P., 1988, \phys {\bf 37}, 9904.

Sudbo, A., C. M. Varma, T. Giamarchi, E. B. Stechel, and R. T.
Scalettar, 1993, \lett {\bf 70}, 978.

Suzuki, T., et al., 1990, \phys {\bf 42}, 4263.

Svane, A., 1992, \lett {\bf 68}, 1900.

Takagi, H., S. Uchida, K. Kitazawa, and S. Tanaka, 1987, Jpn. J. Appl. Phys.
(Lett.) {\bf 26}, L123.

Takagi, H., et al., 1992, AT\&T preprint.

Takahashi, T., et al., 1990, Physica (Amsterdam) {\bf 170C}, 8414.

Takahashi, T., et al., 1991, Proceedings of Workshop on Fermiology of
High-Tc Superconductors, Argonne.

Tan, L., and J. Callaway, 1992, \phys {\bf 46}, 5499.

Tanamoto, T., H. Kohno, and H. Fukuyama, 1992, J. Phys. Soc. Jpn. {\bf 61},
1886.

Tanamoto, T., H. Kohno, and H. Fukuyama, 1993a, J. Phys. Soc. Jpn. {\bf 62},
717.

Tanamoto, T., H. Kohno, and H. Fukuyama, 1993b, J. Phys. Soc. Jpn. {\bf 62},
1455.

Tanner, D. B., and T. Timusk, 1992, ``Optical
Properties of high-temperature superconductors,''
{\it Physical Properties of High-Temperature
Superconductors III}, edited by Donald M. Ginsberg (World Scientific,
Singapore) pp 363--469.

Testardi, L. R., J. H. Wernick, and W. A. Roger, 1974,
Solid State Commun. {\bf 15}, 1.

Thomas, G. A., J. Orenstein, D. Rapkine, M. Capizzi,
A. Millis, R. Bhatt, L. Schneemeyer,
and J. Waszczak, 1988, \lett {\bf 61}, 1313.

Thomas, G. A., 1991, proceedings of the Thirty-Ninth Scottish
Universities Summer School in Physics, St. Andrews.

Thomas, G. A., D. Rapkine, S. Cooper, S-W. Cheong, A.
Cooper, L. Schneemeyer, and J. Waszczak, 1992, \phys {\bf 45}, 2474.

Tikofsky, A. M., R. B.  Laughlin, and Z. Zou, 1992, \lett {\bf
69}, 3670.

Timusk, T., and D. B. Tanner, 1989, ``Infrared
Properties of high-$T_c$ superconductors,''
{\it Physical Properties of High-Temperature
Superconductors I}, edited by Donald M. Ginsberg (World Scientific,
Singapore) pp. 339--407.

Tohyama, T., and S. Maekawa, 1991, J. Phys. Soc. Jpn. {\bf 60},
53.

Tohyama, T., and S. Maekawa, 1992, Physica {\bf C 191}, 193.

Thurston, T. et al., 1990, \lett {\bf 65}, 263.

Torrance, J. B., et al., 1989, \phys{\bf 40}, 8872.

Tranquada, J. M., et al., 1992, \phys {\bf 46}, 5561.

Troyer, M., H. Tsunetsugu, T. M. Rice, J. Riera, and E. Dagotto,
1993, to appear in \phys.

Trugman, S., 1988, \phys{\bf 37}, 1597.

Trugman, S., 1990a, \lett {\bf 65}, 500.

Trugman, S., 1990b, \phys {\bf 41}, 892.

Trugman, S., 1990c, \phys {\bf 42}, 6612.

Tsuei, C. C., A. Gupta, and G. Koren, 1989, Physica {\bf C 161},
415.

Uchida. S., T. Ido, H. Takagi, T. Arima, Y. Tokura and
S. Tajima, 1991, \phys {\bf 43}, 7942.

Varma, C. M., S. Schmitt-Rink, and E. Abrahams, 1987, Solid State Comm. {\bf
62}, 681.

Varma, C. M., P. B. Littlewood, S. Schmitt-Rink, E. Abrahams, and
A. Ruckenstein, 1989, \lett {\bf 63}, 1996.

Veal, B. W., et al., 1993, Argonne preprint.

Virosztek, A., and J. Ruvalds, 1990, \phys {\bf 42}, 4064.

von Szczepanski, K. J., P. Horsch, W. Stephan, and
M. Ziegler, 1990, \phys {\bf 41}, 2017.

Wagner, J., W. Hanke, and D. Scalapino, 1991, \phys {\bf 43}, 10517.

Watanabe, T., et al., 1991, \phys {\bf 44}, 5316.

Welp, U., et al., 1989, \lett {\bf 62}, 1908.

Wen, X.-G., F. Wilczek, and A. Zee, 1989, \phys {\bf 39}, 11413.

Wenzel, W., and K. G. Wilson, 1992, \lett {\bf 69}, 800.

White, S. R., D. J. Scalapino, R. L. Sugar, N. Bickers, and
R. Scalettar, 1989a, \phys {\bf 39}, 839.

White, S. R., D. J. Scalapino, R. L. Sugar, E. Y. Loh, J.
E. Gubernatis, and R. T. Scalettar, 1989b, \phys {\bf 40}, 506.

Wu, D. H., et al., 1993, \lett {\bf 70}, 85.

Wu, M. K., et al., 1987, \lett {\bf 58}, 908.

Yu, G., C. Lee, A. Heeger and S.-W. Cheong, 1992, Physica C203, 419.

Zaanen, J., G. A. Sawatzky, and J. W. Allen, 1985, \lett {\bf 55}, 418.

Zaanen, J., and A. M. Oles, 1988, \phys {\bf 37}, 9433.

Zhang, F. C., and T. M. Rice, 1988, \phys {\bf 37}, 3759.

Zhang, F. C., and T. M. Rice, 1990, \phys {\bf 41}, 7243.

Zhang, S., and M. H. Kalos, 1991, \lett {\bf 67}, 3074.

Zotos, Z., P. Prelovsek, and I. Sega, 1990, \phys {\bf 42}, 8445.

\endreferences

\vfill
\eject

\vskip 1cm

\singlespace

\bigskip
\centerline{\bf Figure Captions}
\medskip

\item{1} Crystal structure of $\x214x$ ({\bf T} phase)
(taken from Almasan and Maple, 1991).

\item{2} Phase diagram of $\x214x$ (from Keimer et al., 1992).

\item{3} Crystal structure of $\ybco$ (taken from Jorgensen, 1991).

\item{4} Phase diagram of $\ybco$ (taken from Burns, 1992;
Koike et al., 1989).

\item{5} Crystal structure of $\xnco$ ({\bf T'} phase)
(taken from Almasan and Maple, 1991).

\item{6} Phase diagram of $\xnco$ and $\x214x$ (together for better
comparison) taken from Almasan and Maple (1991).

\item{7} Temperature dependence of the in-plane resistivity
$\rho_{ab}$ measured on crystals of various cuprate superconductors.
$123 \ 90K$ denotes ${\rm YBCO}$ at the optimal composition ${\rm x=1}$,
while ${\rm 214}$ ${\rm LSCO}$ corresponds to $\x214x$ with ${\rm x = 0.15}$
(from Batlogg et al., 1992).

\item{8} In-plane resistivity $\rho_{ab}$ as a function of temperature
for several compositions of $\x214x$ (from Takagi et al., 1992)

\item{9} Bonding between a ${\rm Cu^{2+}}$ and two ${\rm O^{2-}}$ ions.
Only the d-electrons of Cu and the ${\rm p_x}$ and ${\rm p_y}$ orbitals
of the oxygens are
considered. The number in parentheses indicate the occupations of
the different levels in the undoped compound (from Fulde, 1991).

\item{10} Schematic band structure of the ${\rm Cu O_2}$ planes. ${\rm U_d}$
is the Coulombic repulsion at the copper ions, while $\Delta$ is the
difference in energy between copper and oxygen orbitals. The lower part of the
figure represents the one band Hubbard model that simulates the charge-transfer
gap by a Hubbard gap using an effective ${\rm U_{eff}}$.

\item{11} Mean value of the sign as a function of the density
for the one band Hubbard model on a
$4 \times 4$ cluster at ${\rm U/t=4}$. Results for two temperatures
are shown (from White et al., 1989b).

\item{12} Mean value of the energy $E$, for the one band Hubbard model
as a function of temperature $T$, on a $4 \times 4$ cluster working
at ${\rm U/t=4}$ and density $\langle n \rangle = 0.875$ (from Moreo, 1993a).

\item{13} Spin-spin correlation ${\rm C(r)}$ defined in the text,
corresponding
to the one band Hubbard model for
different values of the distance ${\rm r}$ on a $4 \times 4$ cluster, at
${\rm U/t = 4}$.
The full squares are exact results at zero temperature obtained using the
Lanczos
technique (Fano, Ortolani, and Parola, 1992).
The open squares are results obtained with the Quantum Monte
Carlo approach (Moreo, 1993a) at different temperatures (starting from
below $\beta=4$, $6$ and $12$).

\item{14a} Spin-spin correlation ${\rm C(r)}$ for the one band Hubbard
model at half-filling using Quantum Monte Carlo techniques at ${\rm U/t=4}$,
$\beta = 8/t$, and several cluster sizes (from Moreo, 1993b. See also
Moreo, 1992b).

\item{14b} Antiferromagnetic structure factor $S(\pi,\pi)$ as a
function of $\beta$ (inverse temperature) for a variety of lattice
sizes. These results were obtained with a Quantum Monte Carlo
algorithm (White et al., 1989b).

\item{15a} Spectrum of quasiparticles in the mean-field approximation used
to describe the SDW state at half-filling (see Schrieffer, Wen, and Zhang,
1988). The
lower band of negative energy states is populated, while the upper band
is empty.

\item{15b} Mean-field SDW gap as a function of the coupling ${\rm U/t}$.
(provided by A. Nazarenko, unpublished).

\item{16} Example of a hole moving in a N\'eel background to illustrate
the concept of strings. ``initial'' denotes
the site where the hole is initially injected, the circle is the hole
in its actual position after the hopping term acts three times,
and the dashed line is the path followed by the
hole. The double lines indicate links that are ``ferromagnetic'' i.e.
where magnetic energy is paid. The number of ferromagnetic links grows
like the length of the path.

\item{17} Energy of one hole with respect to the undoped system
$e_{1h}$,
as a function of the coupling ${\rm J}$ at ${\rm t=1}$. Results are
shown for the ${\rm t-J}$ model on a $ 4 \times 4$ cluster (from Dagotto
et al., 1990b), and for
the ${\rm t-J_z}$ model on a $8 \times 8$ cluster (from Barnes et al., 1989).

\item{18a} Energy of a hole with momentum ${\bf p}$ with respect to the
one hole ground state energy ($\Delta E$), for several values of the
coupling ${\rm J/t}$. The open circles correspond to ${\rm J/t=0.2}$,
the
triangles to ${\rm J/t=0.4}$, and the solid circles to ${\rm J/t=1.0}$
(from Dagotto et al., 1990b).

\item{18b} Bandwith ${\rm W}$ of the ${\rm t-J}$ model as a function of
${\rm J}$ (with ${\rm t=1}$) for different cluster sizes (from Poilblanc
et al., 1993).

\item{19} The hole dispersion curve, plotted along the direction
$\Gamma M X \Gamma$ in the Brillouin zone (see inset), for
a $32^2$ lattice, for ${\rm J=0.2}$ (taken from Liu and Manousakis,
1991).

\item{20} Spectral function of one hole in the ${\rm t-J}$ model at
${\bf p} = (\pi/2, \pi/2)$. a,b,c, and d, correspond to ${\rm J/t}$
equal to 1.0, 0.4, 0.2 and 0.0, respectively (from Dagotto et al., 1990b).

\item{21} $\AAA$ with ${\bf p} = (\pi/2, \pi/2)$ for ${\rm J=0.1}$ and
$\epsilon = 0.01$ (width of the $\delta$-functions), and several lattice
sizes (from Liu and Manousakis, 1991).

\item{22} Single hole spectral function for the ${\rm t-J}$ model at the
ground state momentum, for various cluster sizes and ${\rm J/t=0.3}$.
The actual values of the momenta are: ${\bf p}= (\pi/2, \pi/2)$ for the
cluster of 16 sites; ${\bf p}= (\pi, \pi/3)$ for 18 sites;
${\bf p}= (4\pi/5, 2\pi/5)$ for 20 sites; and
${\bf p}= (9\pi/13, 7\pi/13)$ for 26 sites (from Poilblanc et al., 1993).

\item{23} Spectral function of one hole in the ${\rm t-J}$ model at
${\rm J=0.2}$ for different values of ${\bf p}$ on a $4 \times 4$
cluster (Dagotto et al., 1990b).

\item{24a} Average hole-hole distance in the ground state of two holes
as a function
of the coupling constant for several clusters (from Poilblanc, Riera and
Dagotto, 1993).

\item{24b} Binding energy $\Delta_B$ of two holes in the ${\rm t-J}$
model as a function of the coupling. Open triangles denote results for
16 sites, open squares for 18 sites, full triangles for 20 sites, and
full squares for 26 sites (taken from Poilblanc, Riera and Dagotto,
1993a).
The points with the error bars joined by a dashed line are Green's
Function Monte Carlo results from Boninsegni and Manousakis (1993).

\item{25} Binding energy $\Delta_B$ as a function of ${\rm J_z/t}$ for
the ${\rm t-J_z}$ model and several cluster sizes. Starting from above,
the open squares denote results for ${\rm N = 16,20,26,36}$, and ${\rm
50}$ sites, respectively. The solid line is the binding energy
extrapolated
to the bulk limit, which is almost identical to the results in the
50-site cluster. These results were obtained with the ``truncation''
method described by Riera and Dagotto (1993a).

\item{26} Rough interpretation of the spin-charge separation in one
dimensional problems. The upper chain represents a hole injected in
a staggered spin background. The lower chain is the state obtained
after the hole hops three lattice spacings to the left. The
ferromagnetic link is still at the original position of the hole,
and thus the hole is not confined like in two dimensional problems
(contrast these results with Fig. 16).

\item{27a} Wave function renormalization $Z$ of one hole in the
${\rm t-J}$ model (for the actual definition see Eq.(3.16)).
The full squares denote results for a 16 sites cluster (Dagotto
and Schrieffer, 1991). Open squares are results for 18 sites, full
triangles for 20 sites,
and the open triangle corresponds to 26 sites (Poilblanc, Schulz,
and Ziman, 1993). The open circles joined by the dot-dashed line correspond
to results for the one band Hubbard model using ${\rm J=4t^2/U}$
and ${\rm t=1}$ (from Fano, Ortolani, and Parola, 1992). The
reader should note that the definition of $Z$ by these authors is
$Z = | \langle \psi^{gs}_{1h} | {\bar c_{{\bf p}\sigma} } |
\psi^{gs}_{0h} \rangle |^2$ i.e. different than that used in
Eq.(3.16).
With their definition $Z$ is restricted to the interval $[0,1/2]$.

\item{27b} $Z_{2h}$ as defined in the text for the ${\rm t-J}$ model
as a function of the coupling. The notation is as in Fig.27a.
(from Poilblanc, Riera, and Dagotto, 1993a).

\item{28} Spectral decomposition of the operator that creates a pair of
holes out of the undoped ground state (see text). Results are presented
for $\dx2y2$ symmetry (a), and extended s symmetry (b), on a $4 \times
4$ cluster, at ${\rm J/t=0.4}$ (from Dagotto, Riera and Young, 1990).
Similar results were obtained by Poilblanc, Riera and Dagotto, 1993a, on
larger clusters.

\item{29} Uniform magnetic susceptibility of $\x214x$ and $\ybco$ as a
function of temperature (from Johnston et al., 1988; and Johnston, 1989).
In the upper figure, (I) and (II) refer to different oxygen compositions
namely $y=0.0$ and $0.04$, respectively.

\item{30a} Uniform magnetic susceptibility as a function of density
for the one band Hubbard model
on a $4 \times 4$ cluster obtained using Quantum Monte
Carlo techniques, at temperature ${\rm T=t/4}$, and  ${\rm U/t=4}$.
(from Moreo, 1993a).

\item{30b} Same as figure 30.a but obtained at ${\rm U/t=10}$.

\item{30c} Temperature dependence of the uniform susceptibility obtained
with high temperature expansions (Singh and Glenister, 1992a) using the
${\rm t-J}$ model at ${\rm J/t=0.5}$ and several dopings.

\item{31} (a) Antiferromagnetic correlation length vs. temperature in
$\x214x$ (from Birgeneau et al., 1988); (b) Similar results but taking
into account the presence of incommensurate correlations. The solid line
is the fit, and the dashed line the experiments.

\item{32} $S = S({\bf Q}, \omega)$ (with ${\bf Q} = (\pi,\pi)$) as a
function of frequency at ${\rm J/t=0.4}$ for different doping fractions
on a $4 \times 4$ cluster. The hole doping is: a) x=0.125, b) x=0.25,
c) x=0.375, and d) x=0.50. The units in the vertical axis are arbitrary.
Results taken from Dagotto et al. (1992b).

\item{33} Inverse of the static spin structure factor as a function of
doping $\delta$ for a one band Hubbard model at ${\rm U/t=4}$ and
several clusters. The technique used is the Projector Monte Carlo.
Results are taken from Furukawa and Imada (1992).

\item{34} Neutron scattering experiment results for $\x214x$ at two
different ${\rm Sr}$ concentrations. For details see Cheong et al. (1991).

\item{35} Spin structure factor on an $8 \times 8$ cluster with ${\rm
U/t=4}$, $\beta = 6/t$ and several densities. The solid line is to guide
the
eyes. Results taken from Moreo et al. (1990a).

\item{36a} Equal-time spin correlation in momentum space on $10 \times
10$ lattices for 41 up and down spin fermions each. From Furukawa and
Imada (1992).

\item{36b} Spin structure factor as a function of momentum for the $4
\times 4$
cluster and different number of holes (2,4,7 and 11). From Moreo et al.
(1990b).

\item{37a} Optical conductivity of $\x214x$ at 300K vs. energy. Data are shown
parametric with the ${\rm Sr}$
concentration $x$, in the interval $0 \le x \le 0.34$ (from Uchida et
al., 1991).

\item{37b} Optical conductivity of ${\rm Nd_{2-x} Ce_x Cu O_{4-y}}$
(at room temperature) vs. energy,
parametric with Ce concentrations between 0.0 and 0.20 (from Uchida et
al., 1991).

\item{38a} Optical conductivity as a function of frequency of a Tc = 91K film
of
${\rm Y Ba_2 Cu_3 O_{7-\delta}}$ at several temperatures (from Kamar\'as
et al., 1990).

\item{38b} Optical conductivity of four crystals of
${\rm Y Ba_2 Cu_3 O_{7-\delta}}$ at 100K. $\delta$ ranges from $\sim 0.8$ for
the
lowest curve, to $\sim 0$ for the highest one (from Orenstein et al., 1990).
(see also Thomas et al., 1988).

\item{39a} Optical conductivity of $\s2212$ between 20 and 300K
(from Romero et al., 1993)

\item{39b} Frequency dependent conductivity of ${\rm Bi_2 Sr_2 Cu O_6}$
between 10 and 300K (from Romero et al., 1993)

\item{40a} Optical conductivity in the midinfrared region of
semi-insulating
${\rm Y Ba_2 Cu_3 O_{6+y}}$ (upper panel), ${\rm Nd_2 Cu
O_{4-y}}$ (center panel), and ${\rm La_2 Cu O_{4+y}}$ (lower panel). $E_J$
and $E_I$ are peaks discussed in the text (from Thomas et al., 1991).

\item{40b} Real part of the optical conductivity below 1eV in the $CuO_2$
planes [=a-axis] of single-domain  ${\rm Y Ba_2 Cu_3 O_{6+x}}$ (solid
lines),
compared with an estimate of the conductivity associated with the $CuO$
chains
[=$\sigma_b - \sigma_a$] (dotted lines). The dashed-dotted line is
the conductivity of $\s2212$ (from Cooper et al., 1993).

\item{41} Real part of the optical conductivity of the one band Hubbard
model at $U/t=10$ on a $4 \times 4$ cluster. The results are parametric
with hole doping $x$. $D$ denotes the Drude peak at zero frequency,
while
$MIR$ indicates the midinfrared band that is observed for doping
$x=0.125$. The $\delta$-functions appearing in the continued fraction
expansion have been given a large width $\epsilon=t$.

\item{42a} Optical conductivity of the $t-J$ model on a $4\times 4$
cluster evaluated by Stephan and Horsch (1990), for the case of one
hole, open boundary conditions, and several couplings ${\rm J/t}$. The
width of the deltas is $\epsilon = 0.1t$. The inset shows results for
$J=0.5$ with a higher resolution $\epsilon = 0.02t$.

\item{42b} Same as (a) but using periodic boundary conditions. These results
were obtained by Moreo and Dagotto (1990). The Drude peak is not shown,
only the ``incoherent'' part of the conductivity.

\item{43a} $\sigma_1(\omega)$ vs frequency for the three band Hubbard
model on a $2\times 2$ copper cluster (14 atoms total) with different
number of holes. The parameters in the Hamiltonian are $\Delta=4$,
$U_d=6$, $U_p=3$, and $t=1$. Periodic boundary conditions were used.
The drude peak is not shown (from Wagner, Hanke, and Scalapino, 1991).

\item{43b} Optical conductivity of the Kondo-Heisenberg model
on a $4 \times 4$ cluster with one hole. The Kondo spin exchange
coupling is $W=8$, and the nearest-neighbor oxygen-oxygen hole
transfer $t_{pp}= 4$. $c_h$ is 1/16, and the width of the delta
functions is $0.2t$ (for more details see Chen and Sch\"uttler, 1991).

\item{44a}  Kinetic energy per site of the Hubbard model
$K= \langle \phi_0 | (-{\hat T}) | \phi_0 \rangle / N$ on a
$4 \times 4$ cluster as a function of density $\langle n \rangle$ for
$U/t=4$ (full triangles), $U/t=8$ (full squares), and $U/t=20$ (full
pentagons). We also show results for a 10 site cluster at $U/t=100$
(full hexagons). The solid line without points corresponds to results
for $U/t=0$ in the bulk limit (from Dagotto et al, 1992b).

\item{44b} $D_n = D/(2 \pi e^2)$ vs density for various couplings $U/t$.
Full triangles, squares and pentagons denote results for $U/t=4$, $8$,
and $20$, respectively, on a $4 \times 4$ cluster. Open squares,
pentagons, and hexagons indicate results for a 10 site cluster at
$U/t=8$, $20$, and $100$, respectively. The solid line are exact results
at $U/t=0$ in the bulk limit.

\item{45a} Kinetic energy per site of the $t-J$ model
$K= \langle \phi_0 | (-{\hat T}) | \phi_0 \rangle / N$ on a
$4 \times 4$ cluster as a function of hole density $x$ for
$J/t=0.1$ (open squares), $J/t=0.4$ (full triangles), and $J/t=1$
(open triangles) (From Dagotto et al., 1992b).

\item{45b} $D_n = D/(2 \pi e^2)$ vs density for various couplings $J/t$.
The notation is as in Fig.45a.




\item{46} (a) Photoemission spectra calculated by Eskes and Sawatzky (1991)
using a ${\rm Cu_2 O_7}$ cluster and several ${\rm Cu}$ and ${\rm O}$
orbitals in the undoped limit, compared with experimental XPS
results (b) for ${\rm CuO}$ (Ghijsen et al., 1988).

\item{47} Schematic effect of doping for a charge transfer ${\rm Cu
O_2}$ plane.
Hole doping is expected to move the Fermi level into the charge-transfer
band, while electron doping is expected to move it into the upper
Hubbard
band. This would give an energy difference of the Fermi levels of
approximately 2eV (from Dessau, 1992).

\item{48} Photoemission spectra of $\xnco$ single crystals at three different
dopings. The position of $E_F$ is about 0.5eV above the top of the
valence band which implies that the spectral weight induced by
doping lies in the insulating gap (from Anderson et al., 1993).

\item{49} BIS spectrum (inverse photoemission) of ${\rm La_{2-x} Sr_x Ni O_4}$
(from
Fujimori, 1992; and Eisaki et al., 1992).

\item{50} (a) Density of states $N(\omega)$ obtained using the one band
Hubbard model on a $4 \times 4$ cluster at density $\langle n \rangle =
1$ (i.e. half-filling), and ${\rm U/t=8}$. The technique used is the Lanczos
method;
(b) Same as (a) but at density $\langle n
\rangle = 0.875$. Both figures taken from Dagotto, Ortolani, and
Scalapino (1992). The solid lines are the IPES spectrum, while the
dot-dashed lines denote PES results;
(c) Density of states obtained with
a recently developed technique to produce real-frequency results from the
Quantum Monte Carlo data
using a $4 \times 4$ cluster, ${\rm U/t=8}$,
densities $\langle n \rangle = 0.875$ (dashed line) and $\langle n
\rangle = 1$ (solid line),
and temperature ${\rm T = t/4}$.
(from Scalapino, 1991).

\item{51a} Snapshot of the ground state of Hubbard-like models on a
square lattice at large ${\rm U/t}$ and low hole doping.

\item{51b} Density of states of the ${\rm t-J}$ model at ${\rm J/t=0.4}$
and density $\langle n \rangle = 0.875$ (i.e. two holes on a $4 \times
4$ cluster). Results taken from Dagotto et al. (1992b). See also
Stephan and Horsch (1991).

\item{52} One particle excitation spectra for Cu and O orbitals in the
middle of a ${\rm Cu_4 O_{13}}$ cluster. (a) corresponds to $x=0.0$, and
(b) to $x=0.25$ for $\x214x$. The solid and dashed lines denote results
for Cu and O, respectively. The broadening of the Lorentzians is 0.4eV.
The results were obtained using exact
diagonalization techniques by Tohyama and Maekawa (1992).

\item{53a} Density $\langle n \rangle$ vs $\mu$ for the one band Hubbard
model at ${\rm U/t=4}$ using Quantum Monte Carlo on clusters of
different sizes (dots), and exact Lanczos results
for the $4 \times 4$ cluster (solid line). T denotes temperature.
Results taken from Moreo, Dagotto, and Scalapino (1991).
See also Dagotto et al. (1992b).

\item{53b} Total hole occupation number per elementary cell as a
function of the chemical potential $\mu$, for the parameters of the
three band Hamiltonian shown in the figure. Sixteen ${\rm Cu O_2}$ cells
are considered at two temperatures ($\beta = 3$ and $10$). The charge
transfer gap is clearly observed. Taken from
Dopf, Muramatsu, and Hanke (1990). See also Scalettar et al. (1991).

\item{54} ARPES obtained by Olson et al. (1990) in the normal state of
Bi2212 along the $\Gamma-Y$ edge (equivalent to scanning between
${\bf p}=(0,0)$ and $(\pi,\pi)$ in the notation of the square lattice).
The solid lines are fits assuming a Marginal Fermi Liquid behavior.
i.e. inverse hole lifetimes proportional to
$|E - E_F|$.
However, note the large background in the figures.

\item{55a} Momentum notation used for Bi2212. The solid lines denote the
standard convention for the square lattice with $\Gamma = (0,0)$,
$M = (\pi,\pi)$, $X = (\pi,0)$ and so on. The dashed line and the
characters in open letters denote the convention followed in Bi2212, caused
by the presence of the BiO planes.

\item{55b} The ARPES experimental band structure along various high symmetry
directions in Bi2212 (from Dessau, 1992). The momentum notation is that
of
Bi2212 i.e. the $Y$ (${\bar M}$ point in the figure corresponds to ${\bf
p}=(\pi,\pi)$ $({\bf p} = (\pi,0))$ in the notation of the square lattice.

\item{56} Mean-field spectral weight vs $\omega$ in units of $t$.
a) corresponds to momentum $(\pi,0)$, while b) is $(0,0)$. Lanczos
calculations
of $\AAA$ at $\langle n \rangle = 1$, using a $4 \times 4$ cluster and
${\rm U/t=8}$. c) has momentum $(\pi,0)$, and d) $(0,0)$.

\item{57} $\AAA$ obtained on a $4 \times 4$ cluster with ${\rm U/t=8}$
and $\langle n \rangle = 0.875$. The chemical potential is at $\mu \sim
-2.4t$. The IPES weight is shown as a solid line, and the PES as
dot-dashed. a) corresponds to momentum $(\pi,0)$; b) $(\pi/2,0)$;
c) $(0,0)$; d) $(\pi/2,\pi/2)$; e) $(\pi,\pi/2)$; and finally f) is
$(\pi,\pi)$. The arrows mark the peaks that can be labeled as
``quasiparticles'' (results taken from Dagotto, Ortolani and Scalapino, 1992).

\item{58} Fermi surface of models of correlated electrons according to
different studies: a) corresponds to the one band Hubbard model on a
$16 \times 16$ cluster using Quantum Monte Carlo (from Moreo et al.,
1990a); b) are results of Dagotto, Ortolani, and Scalapino (1992) on a
$4 \times 4$ cluster using exact diagonalization (one band Hubbard
model); c) are results
obtained by Stephan and Horsch (1991) using Lanczos techniques applied
to the ${\rm t-J}$ model on a 20 sites cluster; d) are results by Ding
(1993)
for the ${\rm t-J}$ model based on $\langle n_{\bf p} \rangle$. The
Fermi surfaces are sketched in all of the plots.

\item{59} Phase diagram of ${\rm La_2 Cu O_{4 + \delta}}$. The oxygen
poor
phase is described by the left set of data. Circles and diamonds are
results from two samples. The second phase having higher doping is
metallic,
and at low temperatures superconducting (not shown in the plot).
The region bounded by the two
curves and below $T_{ps}$ is inaccessible (from Reyes et al., 1993.
See also Hammel et al., 1992).

\item{60} (a) Line of phase separation in the two dimensional ${\rm t-J}$
model based on the high temperature expansions (Putikka, Luchini, and
Rice (1992) (solid line) and on the analysis of Emery, Kivelson and
Lin (1990) (points). The major difference between the two methods
occur near half-filling.
(b) For completeness, here we also show the
results for the same model in one dimension. The solid line denotes the high
temperature
series result, while the dashed one are finite cluster exact
diagonalization results by Ogata et al. (1991). Note the nice agreement
between the two techniques for the 1D chain.


\item{61} Projector Monte Carlo results for the two dimensional
one band Hubbard model at $U/t=4$, obtained by Furukawa and Imada (1992).
$\delta = 1-\langle n \rangle$,
and $\mu$ is the chemical potential. Results are obtained for
clusters of different sizes. They show that there are
no apparent discontinuities in the curve, and thus no phase separation
in agreement with the previous results by Moreo, Dagotto, and Scalapino (1991).

\item{62} The pair-field susceptibility (solid lines) and the
uncorrelated
pair-field susceptibility (dashed lines) vs temperature for
$\langle n \rangle = 0.875$, and different channels. The noninteracting
(${\rm U/t=0}$) $\dx2y2$ susceptibility is also in the upper right
box for comparison (from White et al., 1989a).

\item{63} Pair-pair correlation function for the one band Hubbard model
as a function of distance $r$ ($= \sqrt{ r_x^2 + r_y^2}$), obtained
using
Quantum Monte Carlo (Moreo, 1992b). The density is $\langle n \rangle = 0.85$.
(a) corresponds to $\dx2y2$ symmetry,
and (b) is extended-s.
The temperature and lattice sizes
are shown in the figure.


\item{64} Scaling of the s-wave pair correlation functions with the
system size N for the three band Hubbard model using
$\Delta = 3t$, $U_d = 5t$, $\beta = 8$, and 1.25
holes per unit cell. The four lines correspond to different pair
operators. All of them converge to zero in the bulk limit.
For more
details see Frick et al. (1990).

\item{65} (a) $\dx2y2$ superconducting susceptibility as a function of
${\rm J/t}$, at density $\langle n \rangle = 0.5$. (b) Pairing
correlation
function ${\rm C({\bf m})}$ as a function of distance, at $\langle n
\rangle = 0.5$, and ${\rm J/t=3}$. The full squares denote $\dx2y2$
correlations, while the open triangles are extended s correlations. (c)
Pairing-pairing correlation function ${\rm C({\bf m})}$ vs distance, at
density $\langle n \rangle = 0.5$. The open triangles, full squares, and
open squares, denote results for ${\rm J/t = 1.0}$, ${\rm 3.0}$,
and ${\rm 4.0}$, respectively. All the results were obtained on a $4
\times 4$ cluster using exact diagonalization techniques (Dagotto and
Riera, 1993).

\item{66a} Ground state energy of the ${\rm t-J-V}$ model on a $4 \times
4$ cluster at ${\rm V/t=10}$ and $\langle n \rangle = 0.5$,
as a function of ${\rm J/t}$. The results
clearly show three different regimes. The intermediate one (II) presents
electron pair formations. The solid lines
are variational results in the bulk limit discussed in Dagotto and
Riera (1992).

\item{66b} Superconducting susceptibility of the ${\rm t-J-V}$ model as
a function of ${\rm J/t}$, obtained at $\langle n \rangle = 0.5$ on
a $4 \times 4 $ cluster. From the left, the results correspond to
${\rm V/t=0.0}$, ${\rm 1.0}$, ${\rm 3.0}$, ${\rm 5.0}$, and ${\rm
10.0}$. Note the presence of a sharp peak for all values of ${\rm V/t}$
(results taken from Dagotto and Riera, 1992).

\item{67} Phase diagram of the one dimensional ${\rm t-J-V}$ model
obtained by exact diagonalization techniques at $\langle n \rangle =
0.5$. Contour lines of constant $K_{\rho}$, and constant spin gap
$\Delta$ are shown. $K_{\rho} > 1$ implies that the superconducting
correlations dominate in the ground state although they cannot develop
long-range order in this dimension
(taken from Troyer et al., 1993).

\item{68} (a) Superconducting susceptibility of the one
dimensional ${\rm t-J-V}$ model as a function of ${\rm J/t}$ for various
values of ${\rm V/t}$. Open circles, full squares, open squares, full
triangles, and open triangles, denote results for ${\rm V/t = 0.0, 1.0,
3.0}$,${\rm 5.0}$, and ${\rm 10.0}$, respectively. The peaks are located
immediately before the region of phase separation is reached. (b)
Pairing
correlation function ${\rm C({\bf m})}$ vs. distance for various values
of ${\rm V/t}$ (same convention as in (a)), and at the value of ${\rm
J/t}$ corresponding to the maximum in the susceptibility.

\item{69} (a) Schematic phase diagram of the two dimensional ${\rm t-J}$
model at zero temperature. ${\rm x = 1 - \langle n \rangle}$ is the hole
density. The meaning of the different phases is explained in the text (see
Dagotto and Riera, 1993). (b) Phase diagram of the one dimensional ${\rm
t-J}$
model taken from Ogata et al. (1991). n is the electronic density, and
the
contours corresponds to lines of constant $K_{\rho}$. In the region
$K_{\rho} > 1$, superconductivity dominates.

\item{70} (a) Quantum Monte Carlo results for the one band Hubbard model
obtained at the parameters
shown in the figure. The full dot denotes $-\langle k_x \rangle$ which
is the mean value of the kinetic energy per site divided by the dimension.
$\Delta_{xx}$ is the current-current correlation function in the
notation of Scalapino, White and Zhang, 1993. The Drude peak is obtained
from these results
as $D/(\pi e^2) = \langle -k_x \rangle - \Delta_{xx}({\bf q} = 0,
\omega_m \rightarrow 0)$. (b) Quantum Monte Carlo results at the same
parameters as (a) but now measuring the superfluid weight $D_s$ which is
obtained as $D_s/(\pi e^2) = \langle -k_x \rangle - \Delta_{xx}(q_x = 0,
q_y \rightarrow 0, \omega = 0)$ (also from Scalapino, White, and Zhang,
1993). The near cancellation of $D_s$ suggests that the one band Hubbard
model
does not superconduct for these parameters.

\item{71} Energy as a function of an external flux: (a) for the two band
Hubbard model in one dimension at different values of the
density-density interaction V. For more details see Sudbo et al. (1993).
(b) Energy of the ground state as a function of an external magnetic
flux $\phi$, for the two dimensional ${\rm t-J}$ model on a
$4 \times 4$ cluster at density $\langle n \rangle = 0.5$. Results are
presented with respect to the
energy at zero flux.
The full squares are results for ${\rm J/t=3.0}$,
while the open squares denote results for ${\rm J/t=4.0}$, i.e. inside
the phase separation region.





\endit